\documentclass[nofootinbib,notitlepage,superscriptaddress,showkeys,longbibliography]{revtex4-1}

\usepackage{amsmath}
\usepackage{amsfonts}
\usepackage{amssymb}
\usepackage{bbm}
\usepackage{microtype}
\usepackage{hyperref}
\usepackage{cleveref}
\usepackage{color}
\usepackage{float}
\restylefloat{table}

\usepackage[latin9]{inputenc}
\usepackage{array}
\usepackage{booktabs}
\usepackage{multirow}
\usepackage{breakurl}
\usepackage[justification=raggedright,singlelinecheck=false]{caption}
\usepackage{subcaption}
\usepackage{feynmp}

\usepackage{graphicx}
\usepackage{esint}

\newcommand{\Mp}{M_\mathrm{Pl}}
\newcommand{\dd}{\mathrm{d}}
\newcommand{\mn}{{\mu\nu}}

\makeatletter

 \@ifundefined{textcolor}{}
 {%
   \definecolor{BLACK}{gray}{0}
   \definecolor{WHITE}{gray}{1}
   \definecolor{RED}{rgb}{1,0,0}
   \definecolor{GREEN}{rgb}{0,1,0}
   \definecolor{BLUE}{rgb}{0,0,1}
   \definecolor{CYAN}{cmyk}{1,0,0,0}
   \definecolor{MAGENTA}{cmyk}{0,1,0,0}
   \definecolor{YELLOW}{cmyk}{0,0,1,0}
 }

\DeclareRobustCommand{\rcite}[1]{%
  \rcite@aux#1,\@nil{#1}%
}
\def\rcite@aux#1,#2\@nil#3{%
  \if\relax#2\relax
    Ref.~\cite{#3}%
  \else
    Refs.~\cite{#3}%
  \fi
}

\makeatother

\begin{document}

\title{Aller guten Dinge sind drei: Cosmology with three interacting spin-2
fields}

\author{Marvin Lüben}

\email{lueben@stud.uni-heidelberg.de}

\affiliation{Institut für Theoretische Physik, Ruprecht-Karls-Universität Heidelberg\\
 Philosophenweg 16, 69120 Heidelberg, Germany}

\author{Yashar Akrami}

\email{akrami@thphys.uni-heidelberg.de}

\affiliation{Institut für Theoretische Physik, Ruprecht-Karls-Universität Heidelberg\\
 Philosophenweg 16, 69120 Heidelberg, Germany}

\author{Luca Amendola}

\email{amendola@thphys.uni-heidelberg.de}

\affiliation{Institut für Theoretische Physik, Ruprecht-Karls-Universität Heidelberg\\
 Philosophenweg 16, 69120 Heidelberg, Germany}

\author{Adam R. Solomon}

\email{adamsol@physics.upenn.edu}

\affiliation{Institut für Theoretische Physik, Ruprecht-Karls-Universität Heidelberg\\
 Philosophenweg 16, 69120 Heidelberg, Germany}

\affiliation{Center for Particle Cosmology, Department of Physics and Astronomy,
University of Pennsylvania\\
 209 S. 33rd St., Philadelphia, PA 19104, USA}
\begin{abstract}
Theories of massive gravity with one or two dynamical metrics generically
lack stable and observationally-viable cosmological solutions that
are distinguishable from $\Lambda$CDM. We consider an extension to
trimetric gravity, with three interacting spin-2 fields which are
not plagued by the Boulware-Deser ghost. We systematically explore
every combination with two free parameters in search of background
cosmologies that are competitive with $\Lambda$CDM. For each case
we determine whether the expansion history satisfies viability criteria,
and whether or not it contains beyond-$\Lambda$CDM phenomenology.
Among the many models we consider, there
are only three cases that seem to be both viable and distinguishable
from standard cosmology. One of the models has only one free parameter and
displays a crossing from above to below the phantom divide. The other
two provide scaling behavior, although they contain future singularities that need to be studied in more detail.
These models possess interesting features that make them compelling targets for a full comparison to observations of both cosmological expansion history and structure formation.
\end{abstract}


\date{\today}

\maketitle
\tableofcontents{}

\section{Introduction}

The past half-decade has borne witness to a revolution in our understanding
of the physics of spin-2 fields. While it has been known for decades
that the unique theory describing a massless
spin-2 field is general relativity \cite{Gupta:1954zz,Weinberg:1965rz,Deser:1969wk,Boulware:1974sr,Feynman:1996kb},
it had similarly been a long-standing belief that massive and interacting
spin-2 fields were generically plagued by the nonlinear \emph{Boulware-Deser ghost} \cite{Boulware:1973my}, despite admitting a healthy linear
formulation \cite{Fierz:1939ix}. This story was turned on its head
when, building on earlier work in \rcite{ArkaniHamed:2002sp,Creminelli:2005qk},
de Rham, Gabadadze, and Tolley (dRGT) constructed a theory of a massive
graviton \cite{deRham:2010ik,deRham:2010kj} which has been shown
through a variety of methods to be free of the Boulware-Deser mode
\cite{Hassan:2011vm,Hassan:2011hr,deRham:2011rn,deRham:2011qq,Hassan:2011tf,Hassan:2011ea,Hassan:2012qv,Hinterbichler:2012cn}.

The breakthrough in massive gravity led to a corresponding advance
in theories of multiple interacting gravitons, or, equivalently, multiple
metrics. The dRGT construction contains two metrics, a spacetime metric
and a fixed reference metric which must be inserted by hand (typically
chosen to be that of Minkowski space). By promoting this fixed metric
to a dynamical one, one arrives at a theory of bimetric gravity (or
bigravity) which is also ghost-free \cite{Hassan:2011zd}.%
\footnote{Both massive gravity and bimetric gravity have deep histories describing
rich physics; for further details we refer the reader to the reviews
in \rcite{deRham:2014zqa,Hinterbichler:2011tt} on massive gravity,
and \rcite{Schmidt-May:2015vnx,Solomon:2015hja} on bigravity.%
} Theories describing multiple metrics can be trivially constructed
from here by coupling various pairs of metrics in the same manner
as in bigravity, using the ghost-free potential. These theories
also avoid the Boulware-Deser ghost \cite{Hinterbichler:2012cn},
up to certain conditions on which we elaborate below \cite{Nomura:2012xr,Scargill:2014wya,deRham:2015cha,deRham:2015rxa}.%
\footnote{In addition to these terms containing two metrics each, interactions
directly involving three or four different spin-2 fields can be written
using vielbeins \cite{Hinterbichler:2012cn}, but have no counterpart
in terms of metric interactions. Such interactions turn out to contain
the Boulware-Deser ghost precisely because they lack a metric-language
formulation \cite{deRham:2015cha,deRham:2015rxa}.%
}

With theoretically-consistent theories in hand, the next step is to
search for physical solutions. Given that multimetric theories are
fundamentally theories of massive gravitons in addition to a massless
one---generically a theory of $n$ metrics contains $n-1$ massive
gravitons and one massless one, of which matter couples to some combination%
\footnote{One might consider coupling matter to the massless graviton exclusively,
but this turns out to reintroduce the Boulware-Deser ghost \cite{Hassan:2012wr}.%
}---they modify general relativity predominantly at large distances,
i.e., they are \emph{infrared} modifications to gravity. As it turns out, general relativity has a well-known and significant problem in reconciling
theory and observation at cosmological distances (see, e.g., Ref.~\cite{Bull:2015stt}): the accelerating Universe \cite{Riess:1998cb,Perlmutter:1998np}, which naturally lends itself to solutions involving modifying gravity
on large scales \cite{Clifton:2011jh}. It is therefore entirely
natural to ask whether massive gravity or its multimetric generalizations
can solve this problem.

There are two immediately necessary (though not sufficient) criteria
for a modified-gravity theory to successfully address the accelerating
Universe. First, it needs to have cosmological solutions which self-accelerate,
i.e., which possess late-time acceleration in the absence of dark
energy. Second, it needs to have stable fluctuations about these self-accelerating
solutions. Unfortunately, this has proven rather difficult to achieve
in massive gravity and bigravity. In the simplest massive gravity
case, in which the reference metric is flat space, spatially-flat
and closed Friedmann-Lemaître-Robertson-Walker (FLRW) solutions do
not exist \cite{D'Amico:2011jj}. Solutions can be obtained by considering
open FLRW or more general reference metrics, but these solutions seem
to generically contain instabilities \cite{Gumrukcuoglu:2011ew,Gumrukcuoglu:2011zh,Vakili:2012tm,DeFelice:2012mx,Fasiello:2012rw,DeFelice:2013awa}.
In bigravity, the situation is slightly improved, as it is not difficult
to find FLRW solutions that agree with observations of the cosmic
expansion history \cite{Volkov:2011an,Comelli:2011zm,vonStrauss:2011mq,Akrami:2012vf,Akrami:2013pna,Konnig:2013gxa,Enander:2014xga}.
However, linear perturbations, studied extensively in \rcite{Comelli:2012db,Khosravi:2012rk,Berg:2012kn,Konnig:2014dna,Solomon:2014dua,Konnig:2014xva,Lagos:2014lca,Cusin:2014psa,Yamashita:2014cra,DeFelice:2014nja,Fasiello:2013woa,Enander:2015vja,Amendola:2015tua,Johnson:2015tfa,Konnig:2015lfa},
tend to contain either ghost or gradient instabilities. In each of
these cases there are potential ways out. In massive gravity, one
might consider large-scale inhomogeneities \cite{D'Amico:2011jj}.
In bigravity, cosmological solutions can be made stable back to arbitrarily
early times by taking one Planck mass to be much smaller than the
other \cite{Akrami:2015qga}, or by reintroducing a cosmological constant
which is much larger than the bimetric interaction parameter \cite{Konnig:2014dna}.
It is also possible that the gradient instability in bigravity is
cured at the nonlinear level \cite{Mortsell:2015exa} due to a version
of the Vainshtein screening mechanism \cite{Vainshtein:1972sx,Babichev:2013usa}.
However, there remains strong motivation to find a massive gravity
or multigravity theory with self-accelerating solutions that are linearly
stable at \emph{all} times.

One logical step in this direction is to inquire what happens cosmologically
if we have three, rather than two, interacting spin-2 fields. This
generalization has been discussed before in Refs.~\cite{Khosravi:2011zi,Tamanini:2013xia,Scargill:2014wya},
but all the cases studied in those references are pathological. Ref.~\cite{Khosravi:2011zi}
studied a theory of massive trimetric gravity where all three metrics
interact with each other directly, making a cycle of interactions;
as we will discuss in the next section, such theories are plagued
by the Boulware-Deser ghost \cite{Nomura:2012xr}. Ref.~\cite{Tamanini:2013xia}, on the
other hand, studied a multigravitational theory in terms of vierbeins
with no metric formulation. Such theories turned out later to also
suffer from the Boulware-Deser ghost. Ref.~\cite{Scargill:2014wya}
compares multimetric models in the metric and vierbein formalism,
but does not look at cosmological solutions, while Ref.~\cite{Baldacchino:2016jsz} examines maximally-symmetric solutions in multigravity.

In this paper, we consider the cosmologies of healthy theories of
trigravity and scan various models by examining a large number of
combinations of parameters in search for interesting background cosmological
solutions. Here, ``interesting'' means cosmological solutions that
are both viable and qualitatively different from those occurring in
bigravity. Appropriate perturbative analyses of the interesting models
will then be required to see whether any of them could be free from
instabilities; we leave this for future work.

We note that this paper examines a number of trimetric models in detail and is therefore fairly lengthy. The especially busy reader is directed to \cref{sec:novelpheno} for a summary of ``positive" results. An overview with an additional level of detail can be found in \cref{tab:T1-summary,tab:T2-summary}, where we present various cosmological viability criteria for all of the models studied.

\section{The theory of massive trigravity}

We begin by presenting the theory of massive trimetric gravity, or
trigravity, describing three interacting spin-2 fields in four dimensions.
We will work in the metric formulation of trigravity, in which the spin-2 fields are described by three metric-like tensors.

How should the three metrics couple to each other? When metrics interact,
the Boulware-Deser ghost looms as a nearly-inevitable pitfall, and
extreme caution must be taken to avoid it. Fortunately, as mentioned
in the previous secion, the interaction potentials which avoid this
ghost are known and have been formulated in the context of massive
gravity and bigravity. This provides an immediate
avenue for trimetric gravity: we could couple each pair of metrics
via the ghost-free potential, leading to a cycle of interactions. However,
such cycles turn out to be plagued by the Boulware-Deser ghost \cite{Nomura:2012xr,Scargill:2014wya,deRham:2015cha}.
Therefore we are forced to consider breaking the cycle into a line,
i.e., there must be one pair of metrics which do not directly interact
with each other.

Next we must consider how these metrics couple to matter. In the simpler
cases of massive gravity and bigravity, where there are two metrics
rather than three, the question of how matter couples was the source
of much discussion and debate \cite{Hassan:2011zd,Hassan:2012wr,Akrami:2013ffa,Tamanini:2013xia,Akrami:2014lja,Yamashita:2014fga,deRham:2014naa,deRham:2014naa,Hassan:2014gta,Enander:2014xga,Solomon:2014iwa,Schmidt-May:2014xla,deRham:2014fha,Gumrukcuoglu:2014xba,Heisenberg:2014rka,Gumrukcuoglu:2015nua,Hinterbichler:2015yaa,Heisenberg:2015iqa,Heisenberg:2015wja,Lagos:2015sya,Melville:2015dba},
leading to the conclusion that the Boulware-Deser ghost almost always
re-emerges if any matter field couples to more than one metric, or
if matter coupled to one metric interacts with matter coupled to another.%
\footnote{This conclusion can be partially avoided by coupling matter to a composite
metric of the form $g_{\mu\nu}^{\mathrm{eff}}=\alpha^{2}g_{\mu\nu}+2\alpha\beta g_{\mu\alpha}(\sqrt{g^{-1}f})^{\alpha}{}_{\nu}+\beta^{2}f_{\mu\nu}$,
although the Boulware-Deser ghost is present at high energies where
the decoupling limit is no longer a valid effective field theory \cite{deRham:2014naa}.%
} We will therefore take all matter to couple minimally to a single
metric, which we will call $g$. Because matter moves on geodesics
of this metric, we can interpret it as the physical metric describing
the geometry of spacetime, exactly like in general relativity. The
other two metrics, which we will denote as $f_{1}$ and $f_{2}$ (or
$f_{1,\mu\nu}$ and $f_{2,\mu\nu}$),%
\footnote{In this paper, commas do not denote spacetime derivatives.%
} couple only to each other or to $g$, and thus are responsible for
modifying gravity.

\begin{figure}
\centering \includegraphics[width=8cm]{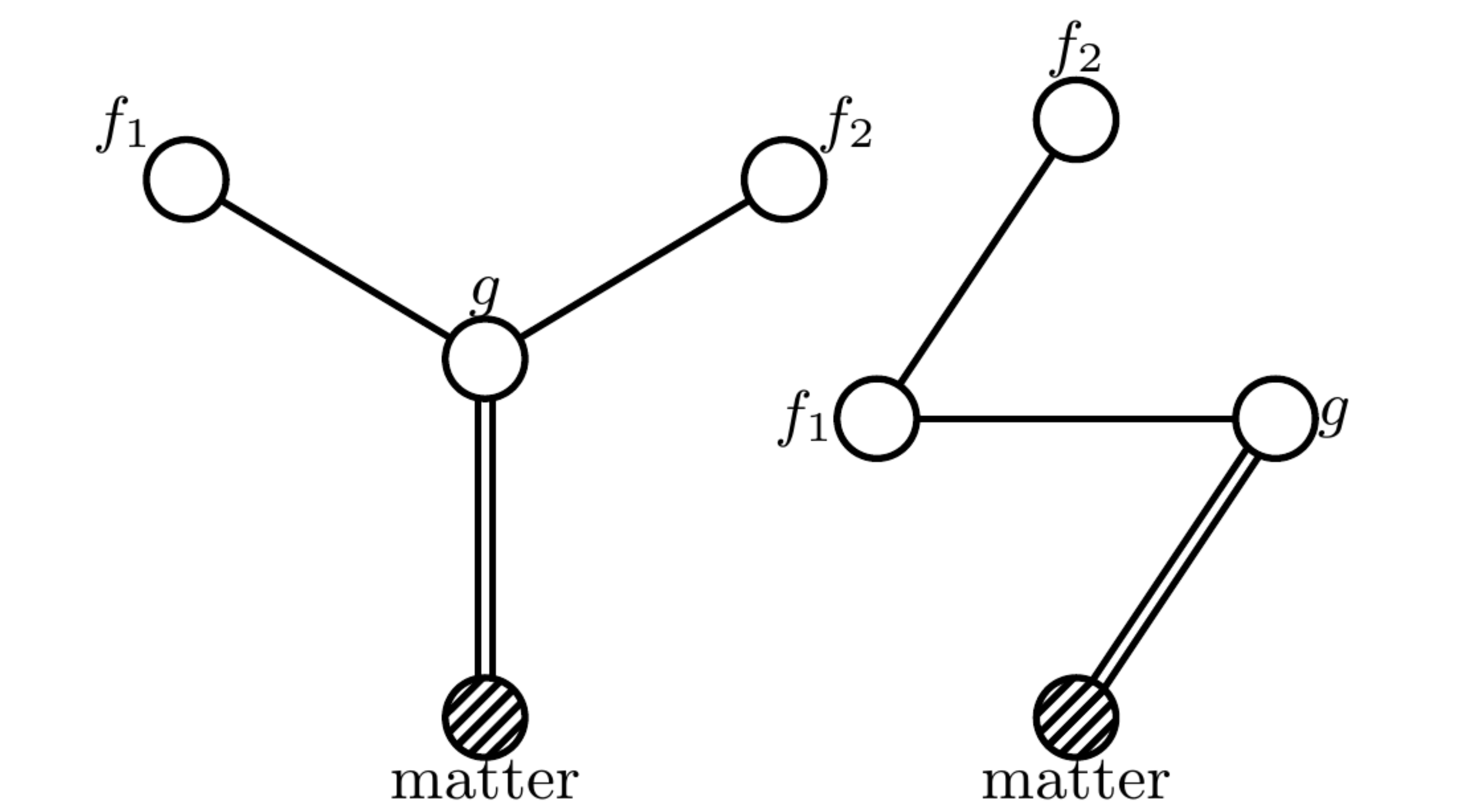} \protect\protect\caption{Visual depiction of star trigravity (left) and path trigravity (right).
The white circles and the single lines between them represent, respectively,
the three metrics and the interaction terms mixing them. The shaded
circles and the double lines represent the matter fields and their
couplings to the physical metric, $g$, respectively.}

\label{fig:star-path} 
\end{figure}

This leaves us with two different classes of ghost-free trimetric
theory. In the first, the metrics $f_{1}$ and $f_{2}$ both couple
to the physical metric $g$, but not to each other. We will call this
\emph{star trigravity}. The other possibility is to couple one of
the additional metrics, without loss of generality $f_{1}$, to each
of the other metrics, $g$ and $f_{2}$. In this theory, which we
call \emph{path trigravity}, there is no coupling between $g$ and
$f_{2}$. The two theories are depicted schematically in \cref{fig:star-path}.
In the rest of this section, we proceed with discussing both classes
of trigravity in full detail and generality, before moving on with
studying the background cosmology of the two theories in the next
sections.

\subsection{Star trigravity}

In star trigravity, $g_{\mu\nu}$ couples to $f_{1,\mu\nu}$, $f_{2,\mu\nu}$,
as well as all matter fields, $\Phi$. The action is given by 
\begin{align}
S & =-\frac{\Mp^{2}}{2}\int\mathrm{d}^{4}x\sqrt{-\det g}R(g)-\sum_{i=1}^{2}\frac{M_{i}^{2}}{2}\int\mathrm{d}^{4}x\sqrt{-\det f_{i}}R(f_{i})\nonumber \\
 & +\sum_{i=1}^{2}m_{i}^{2}\Mp^{2}\int\mathrm{d}^{4}x\sqrt{-\det g}\sum_{n=0}^{4}\beta_{i,n}e_{n}\Bigl(\sqrt{g^{-1}f_{i}}\Bigr)+\int\mathrm{d}^{4}x\sqrt{-\det g}\mathcal{L}_{\text{m}}(g,\Phi),\label{eq:star-action}
\end{align}
where $\sqrt{g^{-1}f_{i}}$ is the matrix square root of $g^{\mu\rho}f_{i,\rho\nu}$,
the $e_{n}$ are the elementary symmetric polynomials of the eigenvalues
of the square-root matrix, as presented in, e.g., Ref.~\cite{Hassan:2011zd},
and $\beta_{i,n}$ are the dimensionless coupling constants for the
interactions between $g_{\mu\nu}$ and $f_{i,\mu\nu}$. The first
index $i$ corresponds to the metric $f_{i,\mu\nu}$ involved in the
interaction with the physical metric $g_{\mu\nu}$, while the second
index $n$ specifies the order of the interaction and can take the
values $n=\{0,...,4\}$. $\Mp$ and $M_{i}$ are the Planck masses
and $R(g)$ and $R(f_{i})$ are the Ricci scalars for the metrics
$g_{\mu\nu}$ and $f_{i,\mu\nu}$, respectively. This theory is symmetric
under the interchange of the metrics $f_{1}$ and $f_{2}$, along
with their Planck masses and interaction parameters. The two mass
parameters $m_{i}^{2}$ can be absorbed into the $\beta_{i,n}$, so
that the $\beta_{i,n}$ will have dimensions of mass squared.

The two Planck masses of $f_{i,\mu\nu}$, $M_{i}$, are redundant
parameters and can be set equal to $\Mp$.%
\footnote{Though this rescaling does not change the physical solutions, one
has to be careful when considering certain limits of this theory.
This was demonstrated explicitly for bigravity in~\rcite{Akrami:2015qga},
and shown to be quite important for cosmological applications. We
expect an analogous story to hold in trigravity; this should be explored
in future work.%
} To see this, consider the rescaling $f_{i,\mu\nu}\rightarrow(\Mp/M_{i})^{2}f_{i,\mu\nu}$.
The Ricci scalars $R(f_{i})$ transform as $R(f_{i})\rightarrow(\Mp/M_{i})^{2}R(f_{i})$,
so the corresponding Einstein-Hilbert terms in the action become 
\begin{equation}
\frac{M_{i}^{2}}{2}\sqrt{-\det f_{i}}R(f_{i})\rightarrow\frac{\Mp^{2}}{2}\sqrt{-\det f_{i}}R(f_{i}).\label{eq:T1-EHT-transformation}
\end{equation}
In addition to the Einstein-Hilbert terms, the interaction terms in
the action also depend on $f_{i,\mu\nu}$. These transform as 
\begin{equation}
\sum_{n=0}^{4}\beta_{i,n}e_{n}\Bigl(\sqrt{g^{-1}f_{i}}\Bigr)\rightarrow\sum_{n=0}^{4}\beta_{i,n}e_{n}\left(\frac{\Mp}{M_{i}}\sqrt{g^{-1}f_{i}}\right)=\sum_{n=0}^{4}\beta_{i,n}\left(\frac{\Mp}{M_{i}}\right)^{n}e_{n}\Bigl(\sqrt{g^{-1}f_{i}}\Bigr),
\end{equation}
where in the last equality we used the scaling properties of the elementary
polynomials $e_{n}(\mathbb{X})$. Redefining the interaction couplings
as $\beta_{i,n}\rightarrow(M_{i}/\Mp)^{n}\beta_{i,n}$, we end up
with the original star trigravity action, but with $M_{1}=M_{2}=\Mp$.

Variation of the action \eqref{eq:star-action} with respect to $g_{\mu\nu}$
and $f_{i,\mu\nu}$ yields the modified Einstein equations for the
metrics (after absorbing $m_{i}^{2}$ into $\beta_{i,n}$ and seting
$M_{i}=\Mp$), 
\begin{flalign}
G_{\mu\nu}+\sum_{i=1}^{2}\sum_{n=0}^{3}(-1)^{n}\beta_{i,n}g_{\mu\lambda}Y_{(n)\nu}^{\lambda}\Bigl(\sqrt{g^{-1}f_{i}}\Bigr) & =\frac{1}{\Mp^{2}}T_{\mu\nu},\label{eq:T1-einstein-g}\\
G_{i,\mu\nu}+\sum_{n=0}^{3}(-1)^{n}\beta_{i,4-n}f_{i,\mu\lambda}Y_{(n)\nu}^{\lambda}\Bigl(\sqrt{f_{i}^{-1}g}\Bigr) & =0,\label{eq:T1-einstein-f}
\end{flalign}
where $G_{\mu\nu}$ and $G_{i,\mu\nu}$ are the Einstein tensors of
$g_{\mu\nu}$ and $f_{i,\mu\nu}$, respectively, and $T_{\mu\nu}$
is the stress-energy tensor defined with respect to $g_{\mu\nu}$
as $T_{\mu\nu}\equiv-\frac{2}{\sqrt{-\det g}}\frac{\delta\left(\sqrt{-\det g}\mathcal{L}_{\mathrm{m}}\right)}{\delta g^{\mu\nu}}$.
The matrices $Y_{(n)}(\mathbb{X})$ for a matrix $\mathbb{X}$ are
defined as 
\begin{align}
Y_{(0)}(\mathbb{X}) & \equiv\mathbb{I},\quad\nonumber \\
Y_{(1)}(\mathbb{X}) & \equiv\mathbb{X}-\mathbb{I}[\mathbb{X}],\quad\nonumber \\
Y_{(2)}(\mathbb{X}) & \equiv\mathbb{X}^{2}-\mathbb{X}[\mathbb{X}]+\frac{1}{2}\mathbb{I}\left([\mathbb{X}]^{2}-[\mathbb{X}^{2}]\right),\nonumber \\
Y_{(3)}(\mathbb{X}) & \equiv\mathbb{X}^{3}-\mathbb{X}^{2}[\mathbb{X}]+\frac{1}{2}\mathbb{X}\left([\mathbb{X}]^{2}-[\mathbb{X}^{2}]\right)-\frac{1}{6}\mathbb{I}\left([\mathbb{X}]^{3}-3[\mathbb{X}][\mathbb{X}^{2}]+2[\mathbb{X}^{3}]\right),\label{eq:matrixpolynomial}
\end{align}
where $\mathbb{I}$ is the identity matrix and $[...]$ is the trace
operator.

Let us now consider the divergence of the Einstein equations \eqref{eq:T1-einstein-g}
and \eqref{eq:T1-einstein-f}. The Einstein tensors satisfy the Bianchi
identities $\nabla^{\mu}G_{\mu\nu}=0$ and $\nabla_{i}^{\mu}G_{i,\mu\nu}=0$.
General covariance of the matter sector implies conservation of the
stress energy tensor, $\nabla^{\mu}T_{\mu\nu}=0$. Thus we are left
with the \emph{Bianchi constraints},%
\footnote{\label{footnote-sum}The sum of the three equations will vanish, i.e.,
one of the equations is redundant. Thus, this set of equations really
gives only two constraints.%
} 
\begin{flalign}
\nabla^{\mu}\sum_{i=1}^{2}\sum_{n=0}^{3}(-1)^{n}\beta_{i,n}g_{\mu\lambda}Y_{(n)\nu}^{\lambda}\Bigl(\sqrt{g^{-1}f_{i}}\Bigr)=0,\label{eq:T1-bianchi-g}\\
\nabla_{i}^{\mu}\sum_{n=0}^{3}(-1)^{n}\beta_{i,4-n}f_{i,\mu\lambda}Y_{(n)\nu}^{\lambda}\Bigl(\sqrt{f_{i}^{-1}g}\Bigr)=0,\label{eq:T1-bianchi-f}
\end{flalign}
where $\nabla^{\mu}$ is the $g$-metric covariant derivative raised
with respect to $g_{\mn}$, and $\nabla_{i}^{\mu}$ are the corresponding
operators for the $f_{i}$ metrics. These constraints arise from the
fact that the ghost-free potentials are invariant under combined diffeomorphisms
of the two metrics involved. They will be important in reducing some freedom in the cosmological solutions.

\subsection{Path trigravity}

In path trigravity, $g_{\mu\nu}$ couples directly to matter and to
one of the reference metrics, which we choose to be $f_{1,\mu\nu}$.
The latter couples in turn to $f_{2,\mu\nu}$. The action is therefore
given by 
\begin{flalign}
S= & -\frac{\Mp^{2}}{2}\int\mathrm{d}^{4}x\sqrt{-\det g}R(g)-\sum_{i=1}^{2}\frac{M_{i}^{2}}{2}\int\mathrm{d}^{4}x\sqrt{-\det f_{i}}R(f_{i})\nonumber \\
 & +m_{1}^{2}\Mp^{2}\int\mathrm{d}^{4}x\sqrt{-\det g}\sum_{n=0}^{4}\beta_{1,n}e_{n}\Bigl(\sqrt{g^{-1}f_{1}}\Bigr)+m_{2}^{2}\Mp^{2}\sqrt{-\det f_{1}}\sum_{n=0}^{4}\beta_{2,n}e_{n}\Bigl(\sqrt{f_{1}^{-1}f_{2}}\Bigr)\nonumber \\
 & +\int\mathrm{d}^{4}x\sqrt{-\det g}\mathcal{L}_{m}(g,\Phi),\label{eq:path-action}
\end{flalign}
with the same notations as in star trigravity, up to different definitions
of the interaction parameters. Here the parameters $\beta_{1,n}$
describe the interactions between the physical metric $g_{\mu\nu}$
and the metric $f_{1,\mu\nu}$, while the $\beta_{2,n}$ describe
the interactions between $f_{1,\mu\nu}$ and $f_{2,\mu\nu}$. In what
follows, the two mass parameters $m_{i}^{2}$ will again be absorbed
into the $\beta_{i,n}$.

Let us take a closer look at the $f_{i}$-metric Planck masses, $M_{i}$,
which, as discussed in the context of star trigravity, are redundant
parameters. Under the rescaling $f_{i,\mu\nu}\rightarrow(\Mp/M_{i})^{2}f_{i,\mu\nu}$,
the Ricci scalars for $f_{1}$ and $f_{2}$ transform as above. Therefore,
the Einstein-Hilbert terms transform as in \cref{eq:T1-EHT-transformation}.
However, the mass terms transform differently, 
\begin{align}
\sum_{n=0}^{4}\beta_{1,n}e_{n}\Bigl(\sqrt{g^{-1}f_{1}}\Bigr) & \rightarrow\sum_{n=0}^{4}\beta_{1,n}e_{n}\Bigl(\frac{\Mp}{M_{1}}\sqrt{g^{-1}f_{1}}\Bigr)=\sum_{n=0}^{4}\beta_{1,n}\left(\frac{\Mp}{M_{1}}\right)^{n}e_{n}\Bigl(\sqrt{g^{-1}f_{1}}\Bigr),\\
\sum_{n=0}^{4}\beta_{2,n}e_{n}\Bigl(\sqrt{f_{1}^{-1}f_{2}}\Bigr) & \rightarrow\sum_{n=0}^{4}\beta_{2,n}e_{n}\Bigl(\frac{M_{1}}{M_{2}}\sqrt{f_{1}^{-1}f_{2}}\Bigr)=\sum_{n=0}^{4}\beta_{2,n}\left(\frac{M_{1}}{M_{2}}\right)^{n}e_{n}\Bigl(\sqrt{f_{1}^{-1}f_{2}}\Bigr),
\end{align}
where we have again used the scaling properties of the elementary
symmetric polynomials $e_{n}(\mathbb{X})$. By redefining the interaction
parameters $\beta_{1,n}\rightarrow(M_{1}/\Mp)^{n}\beta_{1,n}$ and
$\beta_{2,n}\rightarrow(M_{2}/M_{1})^{n}\beta_{2,n}$ we end up with
the original path trigravity action, but with $M_{1}=M_{2}=\Mp$.
Therefore, we set $M_{i}=\Mp$ from now on.

Variation of the action \eqref{eq:path-action} with respect to $g_{\mu\nu}$
and $f_{i,\mu\nu}$ yields the modified Einstein equations for the
metrics, 
\begin{flalign}
 & G_{\mu\nu}+\sum_{n=0}^{3}(-1)^{n}\beta_{1,n}g_{\mu\lambda}Y_{(n)\nu}^{\lambda}\Bigl(\sqrt{g^{-1}f_{1}}\Bigr)=\frac{1}{\Mp^{2}}T_{\mu\nu},\label{eq:T2-einstein-g}\\
 & G_{1,\mu\nu}+\sum_{n=0}^{3}(-1)^{n}\beta_{1,4-n}f_{1,\mu\lambda}Y_{(n)\nu}^{\lambda}\Bigl(\sqrt{f_{1}^{-1}g}\Bigr)+\sum_{n=0}^{3}(-1)^{n}\beta_{2,n}f_{1,\mu\lambda}Y_{(n)\nu}^{\lambda}\Bigl(\sqrt{f_{1}^{-1}f_{2}}\Bigr)=0,\label{eq:T2-einstein-f1}\\
 & G_{2,\mu\nu}+\sum_{n=0}^{3}(-1)^{n}\beta_{2,4-n}f_{2,\mu\lambda}Y_{(n)\nu}^{\lambda}\Bigl(\sqrt{f_{2}^{-1}f_{1}}\Bigr)=0,\label{eq:T2-einstein-f2}
\end{flalign}
where $G_{\mu\nu}$ and $G_{i,\mu\nu}$ are the Einstein tensors for
$g_{\mu\nu}$ and $f_{i,\mu\nu}$, respectively. The matrices $Y_{(n)}$
are given by \cref{eq:matrixpolynomial} and $T_{\mu\nu}$ is the
stress-energy tensor defined with respect to the physical metric $g$.

Let us take the covariant derivative of the Einstein equations \eqref{eq:T2-einstein-g}--\eqref{eq:T2-einstein-f2}. The Bianchi identities for $g_{\mu\nu}$,
$f_{1,\mu\nu}$, and $f_{2,\mu\nu}$, and the covariant conservation
of the stress-energy tensor lead to the Bianchi constraints%
\footnote{See \cref{footnote-sum}.%
} 
\begin{flalign}
 & \nabla^{\mu}\sum_{n=0}^{3}(-1)^{n}\beta_{1,n}g_{\mu\lambda}Y_{(n)\nu}^{\lambda}\Bigl(\sqrt{g^{-1}f_{1}}\Bigr)=0,\label{eq:T2-bianchi-g}\\
 & \nabla_{1}^{\mu}\sum_{n=0}^{3}(-1)^{n}\beta_{1,4-n}f_{1,\mu\lambda}Y_{(n)\nu}^{\lambda}\Bigl(\sqrt{f_{1}^{-1}g}\Bigr)+\nabla_{1}^{\mu}\sum_{n=0}^{3}(-1)^{n}\beta_{2,n}f_{1,\mu\lambda}Y_{(n)\nu}^{\lambda}\Bigl(\sqrt{f_{1}^{-1}f_{2}}\Bigr)=0,\label{eq:T2-bianchi-f1}\\
 & \nabla_{2}^{\mu}\sum_{n=0}^{3}(-1)^{n}\beta_{2,4-n}f_{2,\mu\lambda}Y_{(n)\nu}^{\lambda}\Bigl(\sqrt{f_{2}^{-1}f_{1}}\Bigr)=0.\label{eq:T2-bianchi-f2}
\end{flalign}
As in star trigravity, these constraints will allow us to fix some otherwise-free variables.

\section{The background cosmology of trigravity}

After having introduced the theories of star and path trigravity,
we now turn to their cosmological solutions. We want to describe an
isotropic and homogeneous universe, so we choose all our metrics to
be of the FLRW form.%
\footnote{We follow the standard recipe as in bigravity, where both metrics
are usually taken to be of an FLRW form. One could in principle consider
cosmologies with some of the metrics being anisotropic or inhomogeneous.
In those cases, it is important to first investigate the consistency
of such choices. This has been done in, e.g., Ref.~\cite{Nersisyan:2015oha}
for bigravity. We leave a similar study for trigravity to future work.%
} This allows us to derive Friedmann equations for all the metrics.
After further massaging, we can analyze the solutions to the Friedmann
equations and find expressions for the matter density parameter and
the effective equation of state.

We first study the background equations of star trigravity and then
turn to the case of path trigravity, where we repeat the same procedure.
The results of this section are general and hold for any choices of
parameters.

\subsection{Star trigravity}

We assume that at the background level, the Universe is described
by spatially-flat FLRW metrics for $g_{\mu\nu}$, $f_{1,\mu\nu}$,
and $f_{2,\mu\nu}$, 
\begin{flalign}
\mathrm{d}s_{g}^{2} & =a^{2}(-\mathrm{d}\tau^{2}+\mathrm{d}\vec{x}^{2}),\label{eq:T1-g-metric}\\
\mathrm{d}s_{f_{i}}^{2} & =-N_{i}^{2}\mathrm{d}\tau^{2}+b_{i}^{2}\mathrm{d}\vec{x}^{2},\label{eq:T1-f-metric}
\end{flalign}
where $\tau$ is conformal time. The scale factor $a$ of $g_{\mu\nu}$
and the scale factors $b_{i}$ and lapses $N_{i}$ of $f_{i,\mu\nu}$
are functions of conformal time only. Since $g_{\mu\nu}$ is the physical
metric that minimally couples to matter, its scale factor $a(\tau)$
is the observable scale factor, and similarly the cosmic time $t$ measured by observers is given by $\dd t=a\dd\tau$. Plugging these ansätze in the Bianchi
constraints \eqref{eq:T1-bianchi-g} and \eqref{eq:T1-bianchi-f}
gives 
\begin{align}
 & \text{for \ensuremath{g_{\mn}}:} & \sum_{i=1}^{2}(a\dot{b}_{i}-\dot{a}N_{i})(\beta_{i,1}+2\beta_{i,2}r_{i}+\beta_{i,3}r_{i}^{2}) & =0,\label{eq:bianchi-g}\\
 & \text{for \ensuremath{f_{i,\mn}}:} & (a\dot{b}_{i}-\dot{a}N_{i})(\beta_{i,1}r_{i}^{-2}+2\beta_{i,2}r_{i}^{-1}+\beta_{i,3}) & =0,\label{eq:bianchi-f}
\end{align}
where an overdot denotes a derivative with respect to conformal time
$\tau$. The ratios of the scale factors of the physical and reference
metrics, 
\begin{flalign}
r_{i}\equiv\frac{b_{i}}{a},
\end{flalign}
will be of major importance in the cosmological solutions. We will
use the Bianchi constraints to fix the $f$-metric lapses as%
\footnote{Recall from above that out of the three Bianchi constraints, two are
independent. In each case we can choose either the \textit{dynamical}
branch, fixing one of the lapses, or the \textit{algebraic} branch,
fixing one of the $r_{i}$. These correspond to setting to zero either
the first term in the parentheses of \cref{eq:bianchi-g,eq:bianchi-f}
or the second, respectively. In general, there are four possibilities
to solve the Bianchi constraints in star trigravity: taking the dynamical
branch for both constraints, the algebraic branch for both constraints,
or mixing the dynamical branch for one and the algebraic branch for
the other. This is a novel feature of trigravity; in bigravity such
mixed branches are not possible. In bigravity, the algebraic branch
reproduces general relativity with a cosmological constant at the
background level, as we have a fixed solution for $r$, which, when
plugged back into the Friedmann equations, generates a constant term~\cite{vonStrauss:2011mq,Comelli:2011zm}.
However, these solutions possess perturbations with vanishing kinetic
terms \cite{Comelli:2012db}, signalling an infinitely strong coupling,
and moreover, are plagued by instabilities in the tensor sector \cite{Cusin:2015tmf}.
Whether this is the case also in trigravity needs investigation, and
we leave it for future work.\label{foot:bianchi}%
} 
\begin{equation}
N_{i}=\frac{\dot{b}_{i}}{\dot{a}}a.\label{eq:star-dynam}
\end{equation}
We additionally define the conformal-time Hubble parameter for each
metric as $\mathcal{H}\equiv\frac{\dot{a}}{a}$ and $\mathcal{H}_{i}\equiv\frac{\dot{b}_{i}}{b_{i}}$.
These quantities are related via 
\begin{flalign}
\mathcal{H}_{i}=\mathcal{H}+\frac{\dot{r}_{i}}{r_{i}}.
\end{flalign}

Let us now turn to the Einstein field equations \eqref{eq:T1-einstein-g}
and \eqref{eq:T1-einstein-f}. Inserting our ansätze for $g_{\mu\nu}$
and $f_{i,\mu\nu}$ into the $0$-$0$ components of the equations,
we obtain the three Friedmann equations, 
\begin{flalign}
3\mathcal{H}^{2}-\sum_{i=1}^{2}a^{2}\Bigr[\beta_{i,0}+3\beta_{i,1}r_{i}+3\beta_{i,2}r_{i}^{2}+\beta_{i,3}r_{i}^{3}\Bigl] & =\frac{a^{2}\rho_{\text{m}}}{\Mp^{2}},\\
3\mathcal{H}_{i}^{2}-N_{i}^{2}\Bigr[\beta_{i,1}r_{i}^{-3}+3\beta_{i,2}r_{i}^{-2}+3\beta_{i,3}r_{i}^{-1}+\beta_{i,4}\Bigl] & =0,
\end{flalign}
where we have assumed a perfect fluid source with $\rho_{\text{m}}=-T^{0}{}_{0}$.
Using \cref{eq:star-dynam} we can write the $f$-metric lapses
as $N_{i}=\frac{\mathcal{H}_{i}}{\mathcal{H}}r_{i}a$, and the Friedmann
equations for $f_{i}$ therefore become 
\begin{flalign}
3\mathcal{H}^{2}-a^{2}\Bigr[\beta_{i,1}r_{i}^{-1}+3\beta_{i,2}+3\beta_{i,3}r_{i}+\beta_{i,4}r_{i}^{2}\Bigl] & =0.\label{eq:T1-f-friedmann}
\end{flalign}
The spatial components of the $g$-metric Einstein equation yield
\begin{flalign}
2\mathcal{\dot{H}}+\mathcal{H}^{2}=-\frac{a^{2}p_{\text{m}}}{\Mp^{2}}+a^{2}\sum_{i=1}^{2}\Bigl[\beta_{i,0}+\beta_{i,1}\bigl(\frac{N_{i}}{a}+2r_{i}\bigr)+\beta_{i,2}\bigl(2\frac{N_{i}}{a}+r_{i}\bigr)r_{i}+\beta_{i,3}\frac{N_{i}}{a}r_{i}^{2}\Bigr],
\end{flalign}
where $T^{i}{}_{j}=p_{\text{m}}\delta^{i}{}_{j}$ for a perfect fluid.
We rewrite the lapses as $N_{i}=a(r_{i}+\dot{r}_{i}\mathcal{H}^{-1})=a(r_{i}+r_{i}^{\prime})$,
where $^{\prime}$ denotes a derivative with respect to the number
of $e$-foldings $N\equiv\ln a$.%
\footnote{From now on, we will work only in terms of $N$ as our time variable.
A conformal-time derivative of a quantity $X$ can be transformed
into a derivative with respect to $N$ as 
\begin{flalign}
\dot{X}=\frac{\mathrm{d}}{\mathrm{d}\tau}X=\frac{\mathrm{d}a}{\mathrm{d}\tau}\frac{\mathrm{d}}{\mathrm{d}a}X=\frac{\dot{a}}{a}\frac{\mathrm{d}}{\mathrm{d}\ln a}=\mathcal{H}X^{\prime},
\end{flalign}
as long as $\mathcal{H}\ne0$. %
} That yields 
\begin{flalign}
2\mathcal{HH^{\prime}}+\mathcal{H}^{2}=-\frac{a^{2}p_{\text{m}}}{\Mp^{2}}+a^{2}\sum_{i=1}^{2}\Bigl[\beta_{i,0}+\beta_{i,1}(r_{i}^{\prime}+3r_{i})+\beta_{i,2}(2r_{i}^{\prime}+3r_{i})r_{i}+\beta_{i,3}(r_{i}^{\prime}+r_{i})r_{i}^{2}\Bigr].\label{eq:T1-g-friedmann}
\end{flalign}

As the Friedmann \cref{eq:T1-f-friedmann,eq:T1-g-friedmann} suggest,
the dynamics of trimetric cosmology are captured by the scale-factor ratios $r_{1}$ and $r_{2}$. Thus,
we need to find an expression for $r_{i}^{\prime}$ in order to be
able to analyze the background cosmology of star trigravity. We start
by subtracting \cref{eq:T1-f-friedmann} with $i=2$ from \cref{eq:T1-f-friedmann}
with $i=1$ to obtain 
\begin{flalign}
\beta_{1,1}r_{1}^{-1}+3\beta_{1,2}+3\beta_{1,3}r_{1}+\beta_{1,4}r_{1}^{2}=\beta_{2,1}r_{2}^{-1}+3\beta_{2,2}+3\beta_{2,3}r_{2}+\beta_{2,4}r_{2}^{2}.\label{eq:T1-scalefactorrelation}
\end{flalign}
With this equation, it is possible to relate the two ratios of the
scale factors $r_{1}$ and $r_{2}$. It is a cubic polynomial in $r_{1}$
and $r_{2}$, and therefore always has analytic solutions for
$r_{1}$ as a function of $r_{2}$, and vice versa, though of course there is more than one solution in general.%
\footnote{\label{ftnt:excep-star}The only exception is the model with $\beta_{1,n}=0=\beta_{2,n}\,\forall n\ne2$,
i.e., with only $\beta_{1,2}$ and $\beta_{2,2}$ being nonzero. In
that case \cref{eq:T1-scalefactorrelation} reduces to $\beta_{1,2}=\beta_{2,2}$,
but does not give a relation between $r_{1}$ and $r_{2}$. %
} For every solution, one has to therefore check whether it leads to viable cosmologies.
For the star trigravity models discussed
in this paper it turns out that the different solutions are redundant
at the level of the Friedmann equations and the models' phase space.

Taking the derivative of \cref{eq:T1-scalefactorrelation} with
respect to $N$ and rearranging the whole expression give 
\begin{flalign}
r_{2}^{\prime}=\frac{-\beta_{1,1}r_{1}^{-2}+3\beta_{1,3}+2\beta_{1,4}r_{1}}{-\beta_{2,1}r_{2}^{-2}+3\beta_{2,3}+2\beta_{2,4}r_{2}}r_{1}^{\prime}\equiv D_{\text{ST}}r_{1}^{\prime},\label{eq:T1-derivative-rel}
\end{flalign}
where we use $r_{2}^{\prime}=D_{\text{ST}}r_{1}^{\prime}$ as a short-hand
notation. With these two equations, it is possible to reduce the dimension
of the phase space from $2$ to $1$, which simplifies the analysis
significantly.%
\footnote{\label{footnote:star-derivative-relation}One can use \cref{eq:T1-derivative-rel} only when the denominator does not vanish.
				If it vanishes, then the relation between the derivatives of the two scale factor ratios does not hold anymore.
				However, this situation does not occur in the $1+1$-parameter models of star trigravity discussed in this paper.
				}
Combining the Friedmann \cref{eq:T1-f-friedmann}
with $i=1$ and \cref{eq:T1-g-friedmann} gives an algebraic equation
for $r_{1}$ and $r_{2}$, 
\begin{flalign}
\beta_{1,3}r_{1}^{3}+(3\beta_{1,2}-\beta_{1,4})r_{1}^{2}+3(\beta_{1,1}-\beta_{1,3})r_{1}+(\beta_{1,0}-3\beta_{1,2})-\beta_{1,1}r_{1}^{-1}\nonumber \\
+\beta_{2,3}r_{2}^{3}+3\beta_{2,2}r_{2}^{2}+3\beta_{2,1}r_{2}+\beta_{2,0}+\frac{\rho_{\text{m}}}{\Mp^{2}} & =0,\label{eq:T1-quartic}
\end{flalign}
and the same for $1\leftrightarrow2$ exchanged. Taking the derivative
with respect to $N$, specializing to pressureless dust with $p_{\text{m}}=0$
obeying the continuity equation 
\begin{flalign}
\rho_{\text{m}}^{\prime}+3\rho_{\text{m}}=0,\label{eq:continuity}
\end{flalign}
and using \cref{eq:T1-derivative-rel} to rewrite $r_{2}^{\prime}$
in terms of $r_{1}^{\prime}$, yields a differential equation for $r_{1}$,
\begin{flalign}
r_{1}^{\prime}=\frac{3\rho_{\text{m}}/\Mp^{2}}{3\beta_{1,3}r_{1}^{2}+2(3\beta_{1,2}-\beta_{1,4})r_{1}+3(\beta_{1,1}-\beta_{1,3})+\beta_{1,1}r_{1}^{-2}+3\Bigr[\beta_{2,3}r_{2}^{2}+2\beta_{2,2}r_{2}+\beta_{2,1}\Bigl]D_{\text{ST}}}.\label{eq:T1-derivative}
\end{flalign}
Since exchanging $1\leftrightarrow2$ in this equation yields the
same result, we need an expression for the density $\rho_{\text{m}}$
that is symmetric under $1\leftrightarrow2$. In order to find such an expression,
we add \cref{eq:T1-f-friedmann} for $i=1$ and the one for $i=2$,
and combine the resulting equation with \cref{eq:T1-g-friedmann}.
We obtain 
\begin{flalign}
\frac{\rho_{\text{m}}}{\Mp^{2}}=\sum_{i=1}^{2}\Bigr[-\beta_{i,3}r_{i}^{3}+\Bigl(\frac{\beta_{i,4}}{2}-3\beta_{i,2}\Bigr)r_{i}^{2}+3\Bigl(\frac{\beta_{i,3}}{2}-\beta_{i,1}\Bigr)r_{i}-\beta_{i,0}+\frac{3}{2}\beta_{i,2}+\frac{\beta_{i,1}}{2}r_{i}^{-1}\Bigl].\label{eq:T1-matter-dens}
\end{flalign}
With these equations we can analyze the phase space.

In order to check the cosmological viability of a model, we will make
use of the matter density parameter $\Omega_{\text{m}}$ defined as
\begin{flalign}
\Omega_{\text{m}}\equiv\frac{a^{2}\rho_{\text{m}}}{3\mathcal{H}^{2}\Mp^{2}},\label{eq:matter-density-param}
\end{flalign}
where the matter density follows $\rho_{\text{m}}\propto a^{-3}$.
Using \cref{eq:T1-f-friedmann,eq:T1-matter-dens} to rewrite $\rho_{\text{m}}$
and $\mathcal{H}^{2}$ in terms of $r_{i}$ we obtain 
\begin{flalign}
\Omega_{\text{m}}=\frac{\sum_{i=1}^{2}\Bigr[-\beta_{i,3}r_{i}^{3}+\left(\frac{\beta_{i,4}}{2}-3\beta_{i,2}\right)r_{i}^{2}+3\left(\frac{\beta_{i,3}}{2}-\beta_{i,1}\right)r_{i}-\beta_{i,0}+\frac{3}{2}\beta_{i,2}+\frac{\beta_{i,1}}{2}r_{i}^{-1}\Bigl]}{\frac{1}{2}\sum_{i=1}^{2}\Bigr[\beta_{i,1}r_{i}^{-1}+3\beta_{i,2}+3\beta_{i,3}r_{i}+\beta_{i,4}r_{i}^{2}\Bigl]}.\label{eq:T1-matter-dens-param}
\end{flalign}
We can also define the modified-gravity energy density parameter as
$\Omega_{\text{mg}}\equiv1-\Omega_{\text{m}}$ since we are working
in flat space without curvature terms. Note that we additionally do
not consider radiation here as we are interested in observations at
low redshifts. However, we could easily add a radiation component
to the pressureless matter and it would qualitatively not change any
of the conclusions below.

The effective equation of state of a fluid consisting of different
constituents is defined as 
\begin{flalign}
p=w_{\text{eff}}\rho,
\end{flalign}
with $p$ the total pressure and $\rho$ the total energy density.
We can then rewrite the Friedmann \cref{eq:T1-g-friedmann} as $3\mathcal{H}^{2}=\frac{1}{\Mp^{2}}a^{2}\rho$,
and the acceleration \cref{eq:T1-g-friedmann} as $2\mathcal{HH^{\prime}}+\mathcal{H}^{2}=\frac{a^{2}p}{\Mp^{2}}$,
yielding 
\begin{flalign}
w_{\text{eff}}=-\frac{1}{3}\Bigl(1+2\frac{\mathcal{H^{\prime}}}{\mathcal{H}}\Bigr).\label{eq:weff}
\end{flalign}
Using \cref{eq:T1-f-friedmann,eq:T1-g-friedmann}, the effective
equation of state in star trigravity reads 
\begin{flalign}
w_{\text{eff}}=-\frac{\sum_{i=1}^{2}\Bigl[\beta_{i,0}+\beta_{i,1}(r_{i}^{\prime}+3r_{i})+\beta_{i,2}(2r_{i}^{\prime}+3r_{i})r_{i}+\beta_{i,3}(r_{i}^{\prime}+r_{i})r_{i}^{2}\Bigr]}{\frac{1}{2}\sum_{i=1}^{2}\Bigr[\beta_{i,1}r_{i}^{-1}+3\beta_{i,2}+3\beta_{i,3}r_{i}+\beta_{i,4}r_{i}^{2}\Bigr]}.\label{eq:T1-eos}
\end{flalign}

\subsection{Path trigravity}

Let us now repeat the procedure of the previous subsection for path
trigravity. We assume the metrics $g_{\mu\nu}$, $f_{1,\mu\nu}$,
and $f_{2,\mu\nu}$ to be of the spatially-flat FLRW form 
\begin{flalign}
\mathrm{d}s_{g}^{2} & =a^{2}(-\mathrm{d}\tau^{2}+\mathrm{d}\vec{x}^{2}),\\
\mathrm{d}s_{f_{i}}^{2} & =-N_{i}^{2}\mathrm{d}\tau^{2}+b_{i}^{2}\mathrm{d}\vec{x}^{2},
\end{flalign}
where the scale factors $a$ and $b_{i}$ of $g_{\mu\nu}$ and $f_{i,\mu\nu}$,
respectively, as well as the lapses $N_{i}$ of $f_{i,\mu\nu}$, are
all functions of conformal time $\tau$ only. As $g_{\mu\nu}$ is
the physical metric that couples to matter, its scale factor $a(\tau)$
plays the same role as in general relativity, and in particular is the same scale factor as usually deduced from observations. The path trigravity Bianchi constraints
\eqref{eq:T2-bianchi-g}--\eqref{eq:T2-bianchi-f2} simplify to 
\begin{align}
 &  &  & \text{for \ensuremath{g_{\mn}}:} & (a\dot{b}_{1}-\dot{a}N_{1})(\beta_{1,1}+2\beta_{1,2}r_{1}+\beta_{1,3}r_{1}^{2})=0,\\
 &  &  & \text{for \ensuremath{f_{1,\mn}}:} & (N_{1}\dot{a}-a\dot{b}_{1})(\beta_{1,1}r_{1}^{-2}+2\beta_{1,2}r_{1}^{-1}+\beta_{1,3})-(N_{1}\dot{b}_{2}-N_{2}\dot{b}_{1})(\beta_{2,1}+2\beta_{2,2}r_{2}+\beta_{2,3}r_{2}^{2})=0,\\
 &  &  & \text{for \ensuremath{f_{2,\mn}}:} & (N_{1}\dot{b}_{2}-N_{2}\dot{b}_{1})(\beta_{2,1}r_{2}^{-2}+2\beta_{2,2}r_{2}^{-1}+\beta_{2,3})=0,
\end{align}
where overdot again denotes a derivative with respect to conformal
time $\tau$. The quantities 
\begin{flalign}
r_{1}\equiv\frac{b_{1}}{a}\ \ ,\ \ r_{2}\equiv\frac{b_{2}}{b_{1}}
\end{flalign}
are the ratios of the scale factors of $f_{1}$ and $g$, and $f_{2}$
and $f_{1}$, respectively. Note the different definition here compared
to star trigravity. We use these Bianchi constraints to fix the lapses
as 
\begin{equation}
N_{i}=\frac{\dot{b}_{i}}{\dot{a}}a,
\end{equation}
but we note that other solutions are also possible, in principle.%
\footnote{See \cref{foot:bianchi} for star trigravity. The same statements
are true for path trigravity.%
} Similarly to star trigravity, we define the conformal-time Hubble
parameter for the metrics as $\mathcal{H}\equiv\frac{\dot{a}}{a}$
and $\mathcal{H}_{i}\equiv\frac{\dot{b}_{i}}{b_{i}}$. These quantities
are related via 
\begin{flalign}
\mathcal{H}_{1}=\mathcal{H}+\frac{\dot{r}_{1}}{r_{1}}\ ,\ \mathcal{H}_{2}=\mathcal{H}_{1}+\frac{\dot{r}_{2}}{r_{2}}.
\end{flalign}

Let us turn back to the Einstein field equations \eqref{eq:T2-einstein-g}--\eqref{eq:T2-einstein-f2}
and insert the ansätze for the metrics into the $0$-$0$ components
of the equations. We arrive at the Friedmann equations for the metrics,
\begin{flalign}
3\mathcal{H}^{2}-a^{2}\Bigr[\beta_{1,0}+3\beta_{1,1}r_{1}+3\beta_{1,2}r_{1}^{2}+\beta_{1,3}r_{1}^{3}\Bigl] & =\frac{a^{2}\rho_{\text{m}}}{\Mp^{2}},\label{eq:T2-g-friedmann}\\
3\mathcal{H}_{1}^{2}-N_{1}^{2}\Bigr[\beta_{1,1}r_{1}^{-3}+3\beta_{1,2}r_{1}^{-2}+3\beta_{1,3}r_{1}^{-1}+\beta_{1,4}\Bigl]\nonumber \\
-N_{1}^{2}\Bigr[\beta_{2,0}+3\beta_{2,1}r_{2}+3\beta_{2,2}r_{2}^{2}+\beta_{2,3}r_{2}^{3}\Bigl] & =0,\\
3\mathcal{H}_{2}^{2}-N_{2}^{2}\Bigr[\beta_{2,1}r_{2}^{-3}+3\beta_{2,2}r_{2}^{-2}+3\beta_{2,3}r_{2}^{-1}+\beta_{2,4}\Bigl] & =0,
\end{flalign}
where we have assumed a perfect fluid source with $\rho_{\text{m}}=-T^{0}{}_{0}$.
The Bianchi constraints on the lapses can be rewritten as $N_{1}=\frac{\mathcal{H}_{1}}{\mathcal{H}}r_{1}a$ and $N_{2}=\frac{\mathcal{H}_{2}}{\mathcal{H}}r_{1}r_{2}a$.
The $f_{i}$-metric Friedmann equations then become 
\begin{flalign}
3\mathcal{H}^{2}-a^{2}\Bigr[\beta_{1,1}r_{1}^{-1}+3\beta_{1,2}+3\beta_{1,3}r_{1}+\beta_{1,4}r_{1}^{2}\Bigl]\nonumber \\
-a^{2}r_{1}^{2}\Bigr[\beta_{2,0}+3\beta_{2,1}r_{2}+3\beta_{2,2}r_{2}^{2}+\beta_{2,3}r_{2}^{3}\Bigl] & =0,\label{eq:T2-f1-friedmann}\\
3\mathcal{H}^{2}-a^{2}r_{1}^2\Bigr[\beta_{2,1}r_{2}^{-1}+3\beta_{2,2}+3\beta_{2,3}r_{2}+\beta_{2,4}r_{2}^{2}\Bigl] & =0.\label{eq:T2-f2-friedmann}
\end{flalign}
If we plug in the ansätze for the metrics into the $i$-$i$ components
of the $g$-metric Einstein field equations, we obtain 
\begin{flalign}
2\mathcal{\dot{H}}+\mathcal{H}^{2}=-\frac{a^{2}p_{\text{m}}}{\Mp^{2}}+a^{2}\Bigl[\beta_{1,0}+\beta_{1,1}\bigl(\frac{N_{1}}{a}+2r_{1}\bigr)+\beta_{1,2}\bigl(2\frac{N_{1}}{a}+r_{1}\bigr)r_{1}+\beta_{1,3}\frac{N_{1}}{a}r_{1}^{2}\Bigr],
\end{flalign}
where we have assumed $T^{i}{}_{j}=p_{\text{m}}\delta^{i}{}_{j}$.
Rewriting the lapse $N_{1}$ as $N_{1}=a(r_{1}+r_{1}^{\prime})$, the equation
reads 
\begin{flalign}
2\mathcal{HH^{\prime}}+\mathcal{H}^{2}=-\frac{a^{2}p_{\text{m}}}{\Mp^{2}}+a^{2}\Bigl[\beta_{1,0}+\beta_{1,1}(r_{1}^{\prime}+3r_{1})+\beta_{1,2}(2r_{1}^{\prime}+3r_{1})r_{1}+\beta_{1,3}(r_{1}^{\prime}+r_{1})r_{1}^{2}\Bigr].\label{eq:T2-g-friedmann-2}
\end{flalign}

As \cref{eq:T2-g-friedmann,eq:T2-f1-friedmann,eq:T2-f2-friedmann,eq:T2-g-friedmann-2}
suggest, the cosmology of this path trigravity model depends on the
dynamics of $r_{1}$ and $r_{2}$. Thus we need an expression for
$r_{1}^{\prime}$ and $r_{2}^{\prime}$. We start with subtracting
\cref{eq:T2-f1-friedmann} from \cref{eq:T2-f2-friedmann} to find
\begin{flalign}
\Bigr[\beta_{1,1}r_{1}^{-1}+3\beta_{1,2}+3\beta_{1,3}r_{1}+\beta_{1,4}r_{1}^{2}\Bigl]+r_{1}^{2}\Bigr[-\beta_{2,1}r_{2}^{-1}+(\beta_{2,0}-3\beta_{2,2})+3(\beta_{2,1}-\beta_{2,3})r_{2}+(3\beta_{2,2}-\beta_{2,4})r_{2}^{2}+\beta_{2,3}r_{2}^{3}\Bigl]=0.\label{eq:T2-scalefactorrelation}
\end{flalign}
This equation will allow us to analytically rewrite $r_{2}$ in terms
of $r_{1}$ for any choices of the parameters.%
\footnote{\label{ftnt:excep-path}This is not true for models with at least
$\beta_{2,1}\ne0$ and $\beta_{2,3}\ne0$ because the polynomial is
quartic in $r_{2}$ in this case, and an analytic solution is not
guaranteed to exist.%
} Let us now take the derivative of \cref{eq:T2-scalefactorrelation}
with respect to $N$ and rearrange the whole expression; we obtain
\begin{flalign}
r_{2}^{\prime}=\frac{\beta_{1,1}r_{1}^{-3}-3\beta_{1,3}r_{1}^{-1}-2\beta_{1,4}+2\left(\beta_{2,1}r_{2}^{-1}+(3\beta_{2,2}-\beta_{2,0})+3(\beta_{2,3}-\beta_{2,1})r_{2}+(\beta_{2,4}-3\beta_{2,2})r_{2}^{2}-\beta_{2,3}r_{2}^{3}\right)}{\beta_{2,1}r_{2}^{-2}+3(\beta_{2,1}-\beta_{2,3})+2(3\beta_{2,2}-\beta_{2,4})r_{2}+3\beta_{2,3}r_2^2}\frac{r_{1}^{\prime}}{r_1}\equiv D_{\text{PT}}r_{1}^{\prime},\label{eq:T2-derivative-rel}
\end{flalign}
where we use $r_{2}^{\prime}=D_{\text{PT}}r_{1}^{\prime}$ as a short-hand
notation.%
\footnote{\label{footnote:path-derivative-relation}See \cref{footnote:star-derivative-relation}.
				For path trigravity, we expect this situation to occur in the $1+1$-parameter models with $\beta_{2,3}\ne0$.
				In those cases, the denominator reduces to $3\beta_{2,3}(1-r_2^2)$, and therefore we cannot use \cref{eq:T2-derivative-rel} whenever $r_2^2=1$.}
Combining the Friedmann \cref{eq:T2-g-friedmann,eq:T2-f1-friedmann}
gives an algebraic equation for $r_{1}$ and $r_{2}$,
\begin{flalign}
\beta_{1,3}r_{1}^{3}+(3\beta_{1,2}-\beta_{1,4})r_{1}^{2}+3(\beta_{1,1}-\beta_{1,3})r_{1}+(\beta_{1,0}-3\beta_{1,2})-\beta_{1,1}r_{1}^{-1}\nonumber \\
-r_{1}^{2}\Bigr[\beta_{2,0}+3\beta_{2,1}r_{2}+3\beta_{2,2}r_{2}^{2}+\beta_{2,3}r_{2}^{3}\Bigl]+\frac{\rho_{\text{m}}}{\Mp^{2}} & =0.\label{eq:T2-quartic}
\end{flalign}
This equation can be interpreted as defining $\rho_{\text{m}}$ as
a function of $r_{1}$ and $r_{2}$. However, the matter density also
obeys $\rho_{\text{m}}\propto a^{-3}$. Taking the derivative with
respect to $N$, specializing to pressureless dust with $p_{\text{m}}=0$
obeying the continuity equation 
\begin{flalign}
\rho_{\text{m}}^{\prime}+3\rho_{\text{m}}=0,
\end{flalign}
and using \cref{eq:T2-derivative-rel} to rewrite $r_{2}^{\prime}$ in terms of $r_{1}^{\prime}$,
yield the differential equation 
\begin{flalign}
r_{1}^{\prime}=3\frac{\rho_{\text{m}}}{\Mp^{2}}\Bigr\{ & \Bigr[3\beta_{1,3}r_{1}^{2}+2(3\beta_{1,2}-\beta_{1,4})r_{1}+3(\beta_{1,1}-\beta_{1,3})+\beta_{1,1}r_{1}^{-2}\Bigl]\nonumber \\
 & -2r_{1}\Bigr[\beta_{2,0}+3\beta_{2,1}r_{2}+3\beta_{2,2}r_{2}^{2}+\beta_{2,3}r_{2}^{3}\Bigl]-3r_{1}^{2}\Bigr[\beta_{2,1}+2\beta_{2,2}r_{2}+\beta_{2,3}r_{2}^{2}\Bigl]D_{\text{PT}}\Bigl\}^{-1}\label{eq:T2-derivative}
\end{flalign}
for $r_{1}$, with $\rho_{\text{m}}$ given by \cref{eq:T2-quartic}.

We now derive an expression for the matter density parameter $\Omega_{\text{m}}$
that is defined according to \cref{eq:matter-density-param}. Plugging
in \cref{eq:T2-f1-friedmann,eq:T2-quartic} yields the path trigravity
matter density parameter 
\begin{flalign}
\Omega_{\text{m}}= & \Biggl\{-\Bigl[\beta_{1,3}r_{1}^{3}+(3\beta_{1,2}-\beta_{1,4})r_{1}^{2}+3(\beta_{1,1}-\beta_{1,3})r_{1}+(\beta_{1,0}-3\beta_{1,2})-\beta_{1,1}r_{1}^{-1}\Bigl]\nonumber \\
 & +r_{1}^{2}\Bigr[\beta_{2,0}+3\beta_{2,1}r_{2}+3\beta_{2,2}r_{2}^{2}+\beta_{2,3}r_{2}^{3}\Bigr]\Biggr\}\times\nonumber \\
 & \Biggl\{\Bigr[\beta_{1,1}r_{1}^{-1}+3\beta_{1,2}+3\beta_{1,3}r_{1}+\beta_{1,4}r_{1}^{2}\Bigl]+r_{1}^{2}\Bigr[\beta_{2,0}+3\beta_{2,1}r_{2}+3\beta_{2,2}r_{2}^{2}+\beta_{2,3}r_{2}^{3}\Bigl]\Biggl\}^{-1}.\label{eq:T2-matter-dens-param}
\end{flalign}
The effective modified-gravity density parameter is again given by
$\Omega_{\text{mg}}=1-\Omega_{\text{m}}$ because we are working in
flat space such that curvature terms are absent; we in addition neglect
radiation as we are interested only in the low-redshift regime, i.e.,
late times. In order to find an expression for the effective equation
of state $w_{\text{eff}}$, we make use of \cref{eq:weff}. Plugging
in \cref{eq:T2-f1-friedmann,eq:T2-g-friedmann-2}, the effective
equation of state in path trigravity is 
\begin{flalign}
w_{\text{eff}}=-\frac{\beta_{1,0}+\beta_{1,1}(r_{1}^{\prime}+3r_{1})+\beta_{1,2}(2r_{1}^{\prime}+3r_{1})r_{1}+\beta_{1,3}(r_{1}^{\prime}+r_{1})r_{1}^{2}}{\Bigr[\beta_{1,1}r_{1}^{-1}+3\beta_{1,2}+3\beta_{1,3}r_{1}+\beta_{1,4}r_{1}^{2}\Bigl]+r_{1}^{2}\Bigr[\beta_{2,0}+3\beta_{2,1}r_{2}+3\beta_{2,2}r_{2}^{2}+\beta_{2,3}r_{2}^{3}\Bigl]}.\label{eq:T2-eos}
\end{flalign}

\section{The cosmology of $1+1$-parameter models}

\label{sec:cos-11params}

After having introduced the cosmological background equations for
star and path trigravity, we now analyze their $1+1$-parameter models.
That means we consider models with only one interaction parameter
being non-zero for each interaction potential, such that the models
discussed here are of a $\beta_{1,n}\beta_{2,m}$ form. These are the simplest non-trivial trigravity models that one can construct, i.e., models with the minimum number of free parameters which may give new phenomenology compared to general relativity and bigravity. In order to keep the analysis as
simple as possible, and to respect the Occam's razor guiding principle
in building cosmological models, we therefore only analyze these $1+1$-parameter models in the present paper, i.e., we adhere to the minimal
versions of trigravity. Furthermore, we only consider
models with vanishing $\beta_{i,0}$ and $\beta_{i,4}$. As we
study only the $1+1$-parameter models, cases where only one of the $\beta_{i,0}$
or $\beta_{i,4}$ is turned on will effectively be equivalent to bigravity
and general relativity, and cases where both are turned on will effectively
be equivalent to three independent copies of general relativity.

As we will see later explicitly for both star and path trigravity theories, for the $1+1$-parameter models studied in this paper only the
ratio of the two interaction parameters appears in the phase-space equations, and we therefore define the ratio 
\begin{equation}
B_{mn}\equiv\frac{\beta_{2,n}}{\beta_{1,m}},\label{eq:Bmn}
\end{equation}
which we will use to characterize different cosmological solutions
of the models.

In the following sections, we will analyze the phase space of the
different models and study the behavior of $\Omega_{\text{m}}$ and
$w_{\text{eff}}$ as functions of $r_{i}$. Therefore, we will need
to find the fixed points $r_{i}^{\text{fix}}$, defined as solutions
to the equations 
\begin{equation}
\left.r_{i}^{\prime}\right|_{r_{i}^{\text{fix}}}=0.
\end{equation}
These fixed points will identify different branches of solutions for $r_i$,%
\footnote{A fixed point $r_1^\text{fix}$ identifies different branches only if both $\left.r_{1}^{\prime}\right|_{r_{1}^{\text{fix}}}=0$ and $\left.r_{2}^{\prime}\right|_{r_{1}^{\text{fix}}}=0$ are satisfied, due to the caveats discussed in \cref{footnote:star-derivative-relation,footnote:path-derivative-relation}.}
and will additionally
be the initial or the final values for $r_{i}$ depending on the
sign of $r_{i}^{\prime}$. From \cref{eq:T1-matter-dens-param}
for star trigravity and from \cref{eq:T2-matter-dens-param} for
path trigravity we find $\left.\Omega_{\text{m}}\right|_{r_{i}^{\text{fix}}}=0$
for the fixed points with $r_{i}^{\text{fix}}\ne0$. Therefore, a non-vanishing fixed
point $r_{i}^{\text{fix}}$ can only be a final value.

In addition, there are models where 
\begin{flalign}
r_{i}^{\prime}\rightarrow\pm\infty\ \ \ \text{when}\ \ \ r_{i}\rightarrow r_{i}^{\text{sing}}
\end{flalign}
for some $r_{i}^{\text{sing}}$ that we call a {\it singular} point. Singular
points also separate branches from each other; they cannot be crossed
as $r_{i}^{\prime}$ changes its sign.

In principle, it is always possible to rewrite $r_{2}$ in terms of $r_{1}$ analytically, or the other way around, with the help of \cref{eq:T1-scalefactorrelation} for star and \cref{eq:T2-scalefactorrelation} for path trigravity.
	We will use these for the analysis of the models.%
		\footnote{\label{ftnt:excep-model}As mentioned in \cref{ftnt:excep-star,ftnt:excep-path}, there are models where this is not possible. The only $1+1$-parameter model where this rewriting is not possible is the $\beta_{1,2}\beta_{2,2}$ model of star trigravity.
		}
	Additionally, more than one root exist and it is not clear which one will correspond to a viable solution to the Friedmann equations.
	The roots can take complex values depending on the values of $r_{i}$ and $\beta_{i,n}$, but this does not rule out those roots.
	Plugging in the roots into the Friedmann equations has to lead to real values.
	For the $1+1$-parameter star trigravity models, there is only one root, while for the $1+1$-parameter path trigravity models, there do exist more roots.

Since trigravity should account for the late-time acceleration of
the Universe, we are interested only in the low-redshift regime. In
particular, we do not include radiation. We will therefore analyze
only the phenomenology of the models after matter-radiation equality
($N\approx-8$); the models developed in this paper can therefore not describe
earlier stages of the Universe.

Finally, we will distinguish between models with three different phenomenologies: 
\begin{itemize}
\item \textit{Standard phenomenology:} The model follows standard background
cosmology, i.e., that of the $\Lambda$CDM model, or it mimics
viable bigravity models at the background level as discussed in, e.g.,
Ref.~\cite{Konnig:2013gxa}. This means that the matter density parameter
satisfies $\Omega_{\text{m}}^{\text{init}}=1$, where $\Omega_{\text{m}}^{\text{init}}$
is the initial value of $\Omega_{\text{m}}$, and vanishes in the
infinite future such that the Universe approaches a de Sitter point.
The effective equation of state evolves from $w_{\text{eff}}^{\text{init}}=0$
during matter domination to $w_{\text{eff}}^{\text{fin}}=-1$ at late
times. We already know that this phenomenology describes the background
cosmology properly and one can therefore perform a statistical analysis
to find the best-fit parameters of the model (similarly to what has
been done in Ref.~\cite{Akrami:2012vf} for bigravity). 
\item \textit{New phenomenology:} One can think of various alternatives
to the standard phenomenology. Examples are a non-vanishing fraction
of dark energy during matter domination, i.e., what is called {\it early
dark energy} (see, e.g., Refs.~\cite{Doran:2006kp,Pettorino:2013ia}
and references therein), a non-vanishing matter density parameter
in the infinite future ({\it scaling solutions}) (see, e.g., Refs.~\cite{Gomes:2013ema,Gomes:2015dhl}),
or a {\it phantom} equation of state $w_{\text{eff}}<-1$ at late times~\cite{Caldwell:1999ew,Caldwell:2003vq}. 
\item \textit{Unviable phenomenology:} Models with unviable phenomenologies
are not able to describe our universe. This is the case, for example,
for models which have a matter density parameter with values $\Omega_{\text{m}}\notin[0,1]$,
do not lead to an accelerating universe at late times (i.e., with
$w_{\text{eff}}^{\text{late}}>-1/3$), or lead to an accelerated expansion
of the Universe even during early times (i.e., with $w_{\text{eff}}^{\text{early}}<-1/3$).
Singularities in the past or an increasing $\Omega_{\text{m}}$ in
time are other examples of unviable phenomenologies. 
\end{itemize}
For a model with new phenomenology, we will solve the differential equations
\eqref{eq:T1-derivative} for star and \eqref{eq:T2-derivative}
for path trigravity numerically in terms of the time variable $N$.
The value $N=0$ corresponds to today. After having found the evolution
of $r_{i}$, we can determine the evolution of $\Omega_{\text{m}}$
and $w_{\text{eff}}$ as functions of $N$. For the initial condition,
we will set the value of the ratio of the scale factors today, i.e.
we will fix $r_{i,0}\equiv r_{i}(N=0)$ such that the model produces a
present-time matter density parameter $\Omega_{\text{m},0}\approx0.3$,
consistent with the current observational constraints. However, this is
just a rough estimate based on the constraints on a $\Lambda$CDM-like
model. In order to find out whether a model can describe our universe,
one needs to compare the model's predictions to the data in a careful
and consistent statistical way; this is beyond the scope of the present
paper, and we leave it for future work.

\subsection{Star trigravity}

The procedure is as follows. We first simplify \cref{eq:T1-scalefactorrelation,eq:T1-derivative-rel},
relating the two scale factor ratios $r_{i}$ by specifying the $\beta_{i,n}$,
and rewrite these equations in terms of $B_{mn}$ defined by \cref{eq:Bmn}.
We then simplify the differential \cref{eq:T1-derivative} for $r_{i}$
and read off the fixed and singular points. As the final step, we
simplify \cref{eq:T1-matter-dens-param} for the matter density
parameter $\Omega_{\text{m}}$ and \cref{eq:T1-eos} for the effective
equation of state $w_{\text{eff}}$.

In what follows, we apply this procedure to all $1+1$-parameter models
of star gravity. Therefore, noting that in star gravity $\beta_{1,n}\beta_{2,m}$
models are equivalent to $\beta_{1,m}\beta_{2,n}$ models, the models
we consider here are $\beta_{1,1}\beta_{2,1}$, $\beta_{1,1}\beta_{2,2}$,
$\beta_{1,1}\beta_{2,3}$, $\beta_{1,2}\beta_{2,2}$, $\beta_{1,2}\beta_{2,3}$,
and $\beta_{1,3}\beta_{2,3}$.

\subsubsection{The $\beta_{1,1}\beta_{2,1}$ model}

For the $\beta_{1,1}\beta_{2,1}$ model, the relations between the
two scale factor ratios and their time derivatives, \cref{eq:T1-scalefactorrelation,eq:T1-derivative-rel}, simplify to 
\begin{flalign}
r_{2}=B_{11}r_{1}\ ,\ r_{2}^{\prime}=B_{11}r_{1}^{\prime},\label{eq:T1-M11-scalefactorrelation}
\end{flalign}
and the derivative of $r_{1}$ with respect to $N$, i.e., \cref{eq:T1-derivative},
simplifies to 
\begin{flalign}
r_{1}^{\prime} & =3\frac{r_{1}-3(B_{11}^{2}+1)r_{1}^{3}}{1+3(B_{11}^{2}+1)r_{1}^{2}},
\end{flalign}
from which we can read off the fixed point as 
\begin{flalign}
r_{1}^{\text{fix}}=\frac{1}{\sqrt{3(B_{11}^{2}+1)}}.
\end{flalign}
This fixed point exists for any values of $B_{11}$, i.e., the qualitative
behavior of the model is independent of the numerical value of $B_{11}$.
The matter density parameter \eqref{eq:T1-matter-dens-param} is given
by 
\begin{equation}
\Omega_{\text{m}}=1-3(B_{11}^{2}+1)r_{1}^{2},
\end{equation}
and the effective equation of state \eqref{eq:T1-eos} simplifies
to 
\begin{flalign}
w_{\text{eff}}=-(B_{11}^{2}+1)(r_{1}^{\prime}+3r_{1})r_{1}.
\end{flalign}

\begin{figure}
\begin{centering}
\includegraphics[width=0.4\linewidth]{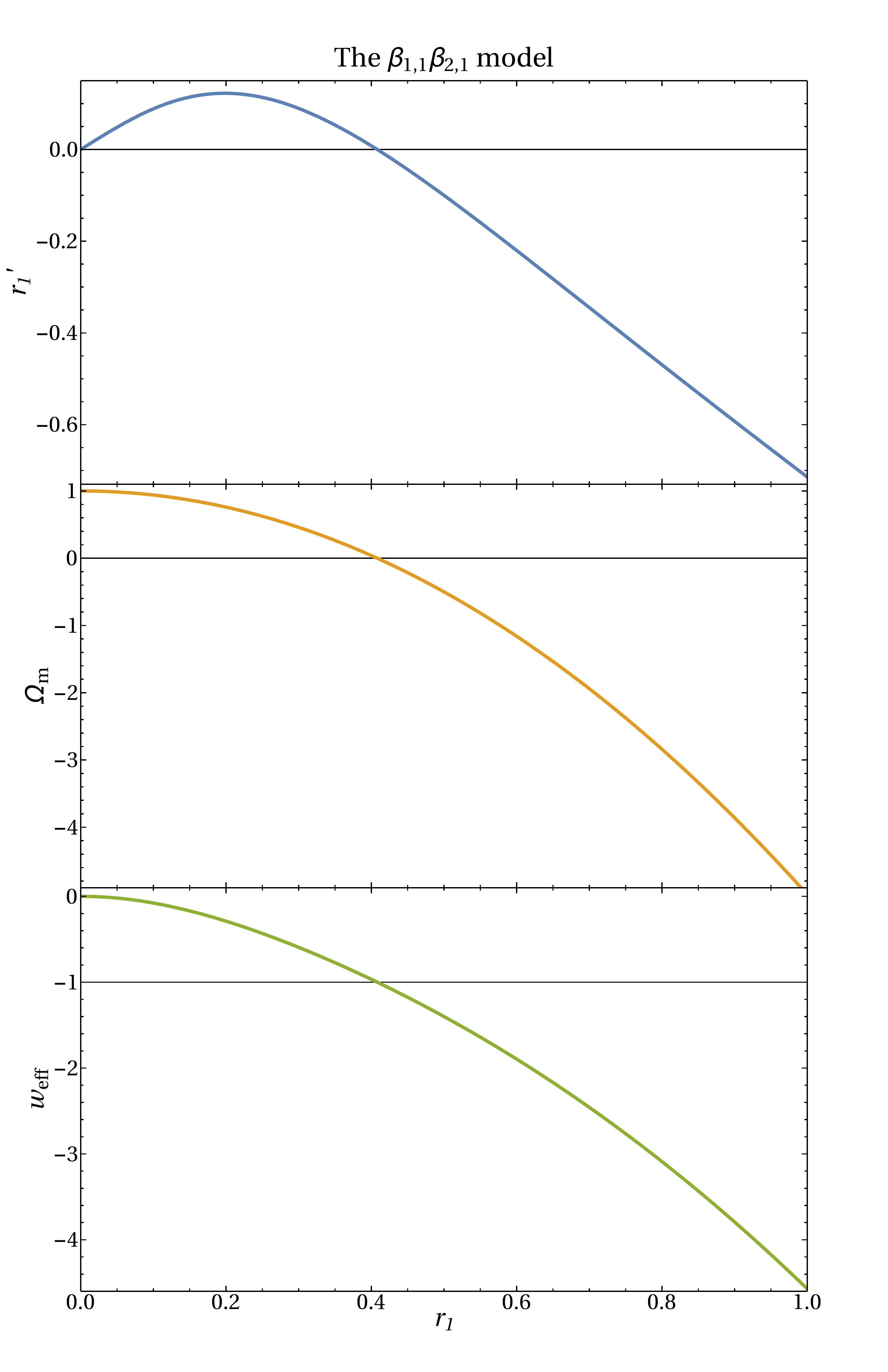} 
\par\end{centering}

\protect\protect\caption{\label{fig:T1-11}Evolution of $r_{1}^{\prime}$, the matter density
parameter $\Omega_{\text{m}}$, and the equation of state $w_{\text{eff}}$
as functions of $r_{1}$ for the $\beta_{1,1}\beta_{2,1}$ model of
star trigravity with $B_{11}=1$. In this and subsequent figures, we remind the reader that $r_i$ effectively stands in for time, as it monotonically increases or decreases throughout cosmological evolution. Of course, whether $r_i$ increases or decreases with time can be determined from the sign of $r_i'$.}
\end{figure}

The phase space, matter density parameter, and effective equation of
state of the $\beta_{1,1}\beta_{2,1}$ model are presented in \cref{fig:T1-11}
for the representative value $B_{11}=1$. The general behavior is
similar to the $\beta_{1}$ model in bigravity (see Ref. \cite{Konnig:2013gxa}),
with a finite branch over the range $[0,r_{1}^{\text{fix}}]$ and
an infinite branch over $[r_{1}^{\text{fix}},\infty]$ with $r_{1}^{\text{fix}}$
being the final value of $r_{1}$ for both branches. The behavior
of $\Omega_{\text{m}}$ and $w_{\text{eff}}$ indicates that the infinite
branch is not viable as the matter density parameter is always negative and
the effective equation of state is always phantom. The finite branch behaves
well as there is a matter-dominated past with $w_{\text{eff}}^{\text{init}}=0$,
and $w_{\text{eff}}$ evolves towards a de Sitter point with $\Omega_{\text{m}}^{\text{fin}}=0$
and $w_{\text{eff}}^{\text{fin}}=-1$ as in standard cosmology.

Indeed, the $g_{\mu\nu}$ Friedmann \cref{eq:T1-g-friedmann} can
be transformed into the corresponding equation in bigravity using
\cref{eq:T1-M11-scalefactorrelation}, 
\begin{flalign}
3\mathcal{H}^{2}=\frac{a^{2}\rho_{\text{m}}}{\Mp^{2}}+3a^{2}(\beta_{1,1}r_{1}+\beta_{2,1}r_{2})=\frac{a^{2}\rho_{\text{m}}}{\Mp^{2}}+3a^{2}\underbrace{\beta_{1,1}(B_{11}^{2}+1)r_{1}}_{\equiv\beta_{1}r},
\end{flalign}
where $\beta_{1}$ and $r$ are, respectively, the interaction parameter
and ratio of the scale factors in $\beta_{1}$ bigravity (see Ref.
\cite{Konnig:2013gxa} for the notation). That means that, at the
background level, the $\beta_{1,1}\beta_{2,1}$ model is completely
equivalent to the $\beta_{1}$ model of bigravity. We leave it for
future work to analyze whether this equivalence still holds at the
level of linear perturbations.

\subsubsection{The $\beta_{1,1}\beta_{2,2}$ and $\beta_{1,2}\beta_{2,3}$ models}

According to \cref{eq:T1-scalefactorrelation}, $r_{1}$ in the
$\beta_{1,1}\beta_{2,2}$ model is non-dynamical and given by $r_{1}=\frac{1}{3B_{12}}$.
Therefore we express everything in terms of $r_{2}$, with the derivative
\begin{flalign}
r_{2}^{\prime}=\frac{3-3r_{2}^{2}-B_{12}^{-2}}{2r_{2}}.
\end{flalign}
We can read off the fixed point as 
\begin{flalign}
r_{2}^{\text{fix}}=\sqrt{1-\frac{1}{3B_{12}^{2}}}.
\end{flalign}
Since $r_{2}^{\text{fix}}$ has to be a real number, we have to distinguish
between three qualitatively different cases: (a) $B_{12}>1/\sqrt{3}$,
(b) $B_{12}=1/\sqrt{3}$, and (c) $B_{12}<1/\sqrt{3}$.
For case (a) there is one fixed point, $r_{2}^{\text{fix}}$, and one singular point, $0$, for case (b) there is only one fixed point, $0$, and for case (c)
there are no fixed points, but it has one singular point, $0$. The matter density parameter is 
\begin{flalign}
	\Omega_{\text{m}}=1-r_{2}^{2}-\frac{1}{3B_{12}^{2}},\label{eq:T1-M12-matter-density}
\end{flalign}
while the effective equation of state is independent of $r_{2}$ and
$B_{12}$; it is in fact a constant: $w_{\text{eff}}=-1$ at all times. This is enough to rule the model out as an effective equation of state of $w_{\text{eff}}=-1$ at all times would lead to an accelerated expansion at all times, which clearly contradicts observations. In addition, we can see from \cref{eq:T1-M12-matter-density} that for cases (b) and (c), i.e., for $B_{12}\le1/\sqrt{3}$, the matter density parameter is negative, i.e., $\Omega_\text{m}\le0$, during the entire evolution, which additionally excludes those cases. The $\beta_{1,1}\beta_{2,2}$ model is therefore not viable and we do not present its phase space here.

The $\beta_{1,2}\beta_{2,3}$ model is completely analogous to the
model discussed here if we replace $r_{2}$ by $r_{1}$. In this model,
$r_{2}$ is non-dynamical and given by $r_{2}=B_{23}^{-1}$. Thus,
this model is ruled out as well because of the same arguments as in
the $\beta_{1,1}\beta_{2,2}$ model.

\subsubsection{The $\beta_{1,1}\beta_{2,3}$ model}

Here, the scale factor ratios $r_{1}$ and $r_{2}$ are related via
$r_{2}=\frac{1}{3B_{13}r_{1}}$, which is the simplified form of \cref{eq:T1-scalefactorrelation}.
Plugging this into \cref{eq:T1-derivative} gives 
\begin{flalign}
r_{1}^{\prime} & =\frac{r_{1}-27B_{13}r_{1}^{3}+81B_{13}^{2}r_{1}^{5}}{1-9B_{13}^{2}r_{1}^{2}-27B_{13}^{2}r_{1}^{4}},\label{eq:T1-13-r}
\end{flalign}
which allows us to find the fixed and singular points 
\begin{flalign}
r_{1}^{\text{fix}\pm} & =\frac{1}{3\sqrt{2}}\sqrt{3\pm B_{13}^{-1}\sqrt{9B_{13}^{2}-4}},\label{eq:T1-r-crit-13}\\
r_{1}^{\text{sing}} & =\frac{1}{2\sqrt{3}}\sqrt{-3+B_{13}^{-1}\sqrt{3(3B_{13}^{2}+4)}}.\label{eq:T1-r-sing-13}
\end{flalign}
From this we find three cases for the model: (a) $B_{13}>2/3$,
(b) $B_{13}=2/3$, and (c) $B_{13}<2/3$. These should
be analyzed separately as the qualitative behavior of the phase space
is different in each case. For case (a) there are two fixed points
and one singular point, while for case (b) there is only one fixed
point. Case (c) admits only one singular point, and no fixed point.
Before analyzing the three cases one by one, we obtain the simplified
form of the matter density parameter and the effective equation of
state: 
\begin{flalign}
\Omega_{\text{m}} & =1-3r_{1}^{2}-\frac{1}{27B_{13}^{2}r_{1}^{2}},\\
w_{\text{eff}} & =-\Bigl[\Bigl(1-\frac{1}{27B_{13}^{2}}r_{1}^{-4}\Bigr)r_{1}^{\prime}+3r_{1}+\frac{1}{27B_{13}^{2}}r_{1}^{-3}\Bigr]r_{1}.
\end{flalign}

\begin{figure}
\begin{centering}
\includegraphics[width=0.4\linewidth]{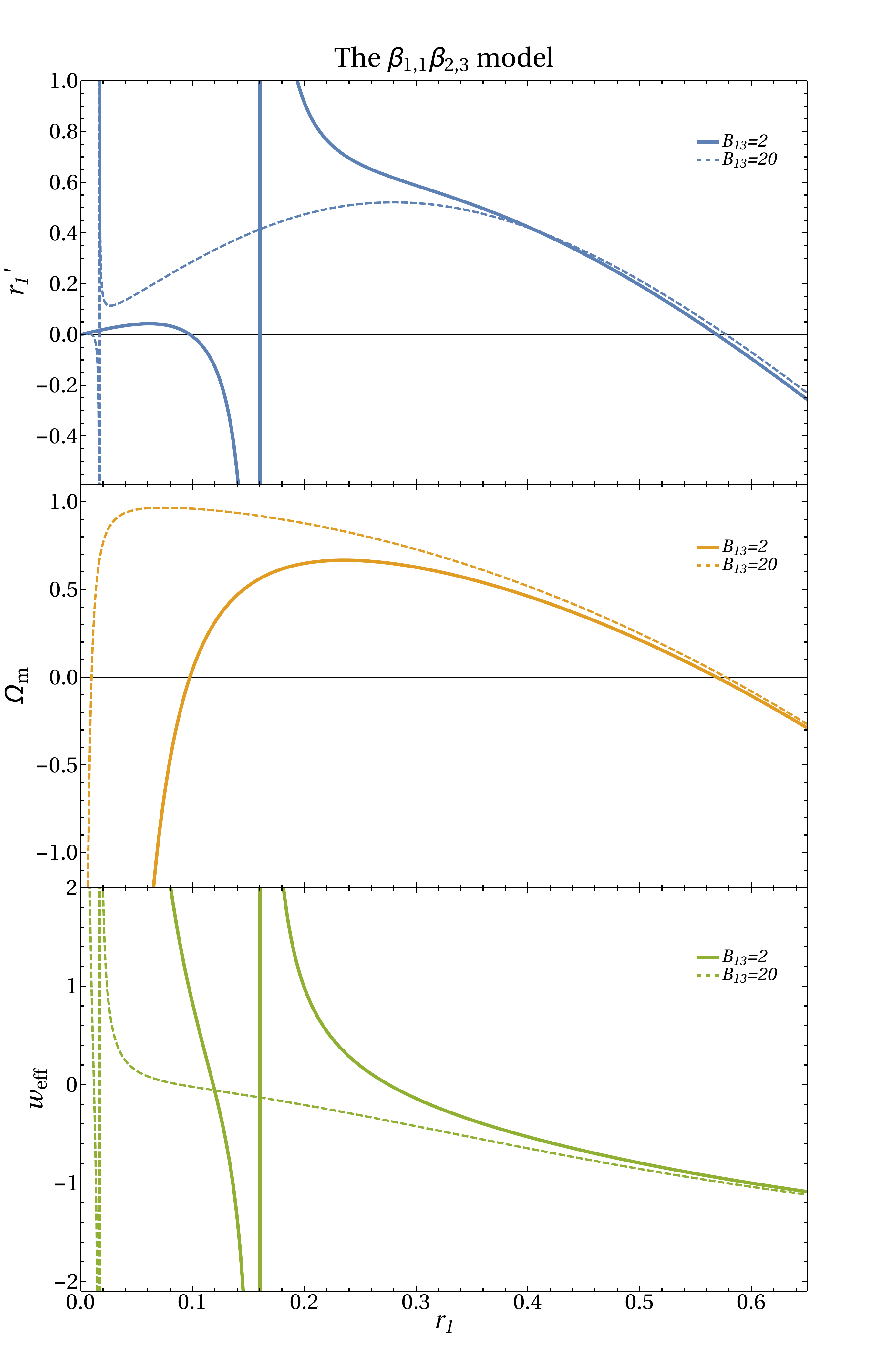} \includegraphics[width=0.4\linewidth]{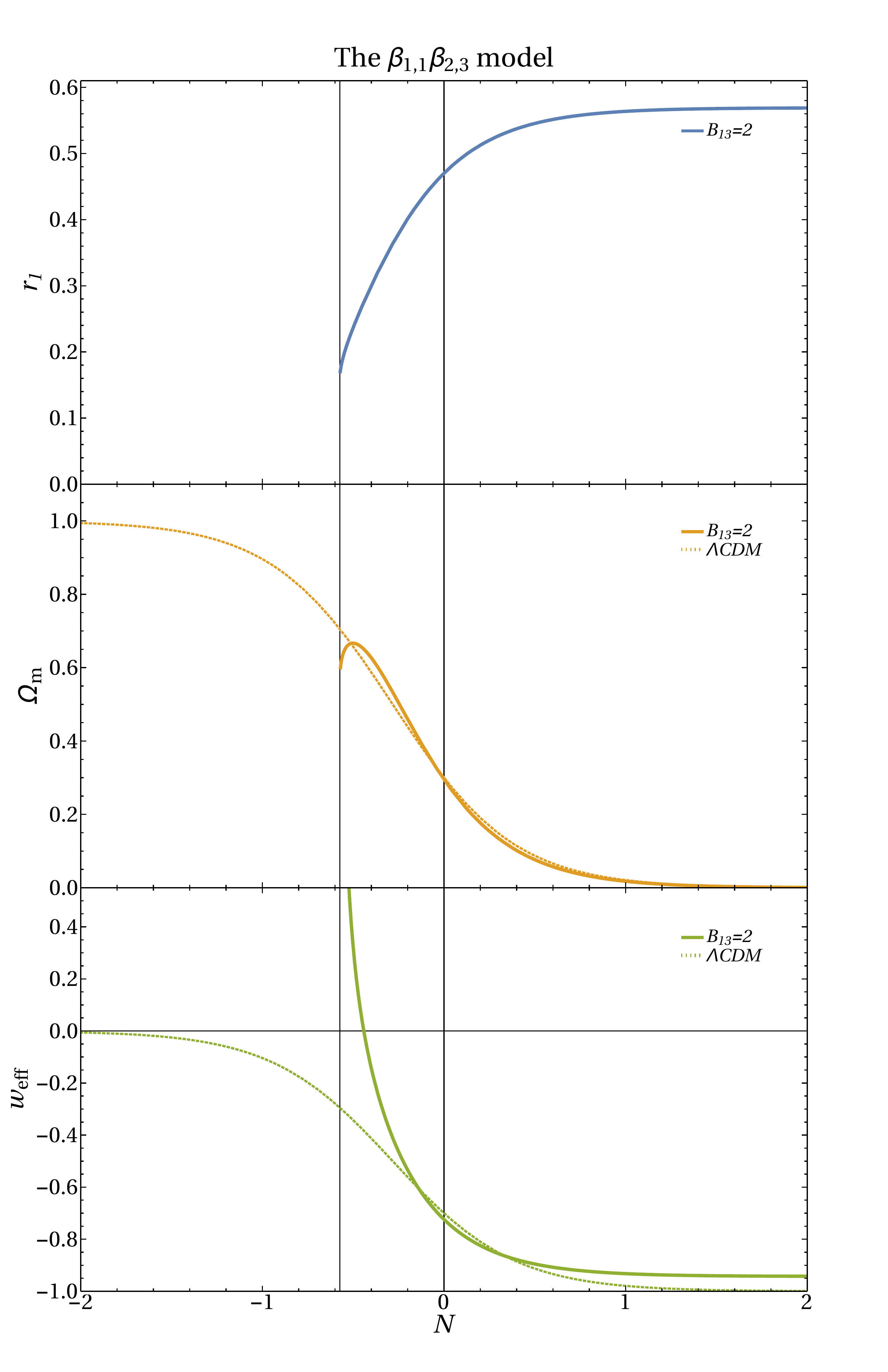} 
\par\end{centering}

\protect\protect\caption{\label{fig:T1-13a} Left panel: Evolution of $r_{1}'$, the matter
density parameter $\Omega_{\text{m}}$, and the effective equation
of state $w_{\text{eff}}$ in terms of $r_{1}$ for the $\beta_{1,1}\beta_{2,3}$
model of star trigravity with $B_{13}=\{2,20\}$. Right panel: Time
evolution of $r_{1}$, $\Omega_{\text{m}}$, and $w_{\text{eff}}$
for the finite branch $[r_{1}^{\text{sing}},r_{1}^{\text{fix}+}]$
of the $\beta_{1,1}\beta_{2,3}$ model with $B_{13}=2$, together
with the time evolution of $\Omega_{\text{m}}$ and $w_{\text{eff}}$
for standard $\Lambda$CDM cosmology. The right vertical line shows
$N=0$, i.e., today, while the left vertical line represents the would-be
initial value of $N$ for $B_{13}=2$, i.e., the would-be initial
size of the Universe.}
\end{figure}

$\bullet$ \textbf{Case (a):} The quantities $r_{1}'$, $\Omega_{\text{m}}$,
and $w_{\text{eff}}$ as functions of $r_{1}$ are shown in \cref{fig:T1-13a}
(left panel) for two examples of $B_{13}=2,20$. We can see from
the figure that there are four branches in each case. The finite branch
$[0,r_{1}^{\text{fix}-}]$ and the infinite branch $[r_{1}^{\text{fix}+},\infty]$
are ruled out because $\Omega_{\text{m}}<0$ on those branches. On
the branch $[r_{1}^{\text{fix}-},r_{1}^{\text{sing}}]$, the matter
density is positive, $\Omega_{\text{m}}>0$, and decreases with time,
but the effective equation of state is phantom during matter domination
and is positive when the matter density parameter vanishes. This branch
is therefore ruled out.

We are thus left with only the finite branch $[r_{1}^{\text{sing}},r_{1}^{\text{fix}+}]$.
The scale factor ratio increases from $r_{1}^{\text{sing}}$, where
the matter density parameter takes its initial value 
\begin{flalign}
\Omega_{\text{m}}^{\text{init}}=2+\frac{8}{9B_{13}^{2}-3B_{13}\sqrt{3(3B_{13}^{2}+4)}}<1,
\end{flalign}
to the de Sitter point $r_{1}^{\text{fix}+}$ with a vanishing matter
density parameter $\Omega_{\text{m}}^{\text{fin}}=0$.
$\Omega_{\text{m}}$ first increases to the maximum value 
\begin{flalign}
	\Omega_{\text{m}}^{\text{max}}\equiv1-\frac{2}{3B_{13}}\label{eq:T1-M13-maxmatterdensity}
\end{flalign}
at $r_{1}^{\text{max}}=\frac{1}{3\sqrt{B_{13}}}$ and then decreases
towards the de Sitter point. At early times, the effective equation
of state is singular.

In order to analyze this branch further, we integrate \cref{eq:T1-13-r}
w.r.t $N\equiv\ln a$ to calculate the evolution of the matter density
parameter and the effective equation of state. As the initial condition
we set $r_{1,0}\equiv r_{1}(N=0)=0.47$ for $B_{13}=2$, in order
to achieve a present-time matter density parameter of $\Omega_{\text{m},0}\approx0.3$ (consistent with observational measurements).
The results for $B_{13}=2$ are presented in \cref{fig:T1-13a}
(right panel) together with the time evolution of $\Omega_{\text{m}}$
and $w_{\text{eff}}$ for $\Lambda$CDM with $\Omega_{\text{m,0}}=0.3$,
for comparison.

Since $r_{1}$ starts to evolve from its singular value, there is
only a finite number of $e$-foldings in the past. In our $B_{13}=2$
example, the initial number of $e$-foldings is $N_{0}\approx-0.57$.
That would imply that the Universe had started to evolve with a finite
size $a_{0}=e^{N_{0}}$ such that there would be no big bang. Although
we have not presened in the figure, our analysis of the behavior of
the model for $B_{13}=20$ compared to $B_{13}=2$ indicates that
increasing the value of $B_{13}$ might help to push the singularity
back in time and gain larger numbers of $e$-foldings. The maximum
value of the matter density parameter will then be closer to $1$.
For example, if we choose $B_{13}=700000$, the initial number of
$e$-foldings will be $N_{0}\approx-5.3$. For larger values of $B_{13}$
numerical instability leads to problems. However, a value $B_{13}\gg1$
seems to be unnatural.

$\bullet$ \textbf{Cases (b) and (c):} From \cref{eq:T1-M13-maxmatterdensity} we can read off that any value $B_{13}\le2/3$ will lead to a negative matter density parameter, $\Omega_\text{m}\le0$.
Thus these cases do not lead to a viable phenomenology.
Therefore, we do not present the phase space for these cases.

\subsubsection{The $\beta_{1,2}\beta_{2,2}$ model}

Since in this case \cref{eq:T1-scalefactorrelation} does not yield
a relation between the two scale factor ratios $r_{1}$ and $r_{2}$,
and, additionally, \cref{eq:T1-derivative-rel} is not applicable,
the procedure for finding the phase space for the $\beta_{1,2}\beta_{2,2}$
model is rather different. First, we notice that \cref{eq:T1-scalefactorrelation}
results in $\beta_{1,2}=\beta_{2,2}\equiv\beta$, and therefore, the interaction
parameter ratio is fixed to $B_{22}=1$, and there is no free parameter
left at the level of the phase space. The Friedmann \cref{eq:T1-g-friedmann,eq:T1-f-friedmann}
then read 
\begin{flalign}
3\mathcal{H}^{2}= & 3a^{2}\beta(r_{1}^{2}+r_{2}^{2})+a^{2}\frac{\rho_{\text{m}}}{M_{\text{g}}^{2}},\\
3\mathcal{H}^{2}= & 3a^{2}\beta.\label{eq:T1-M22-fi-friedmann}
\end{flalign}
By subtracting these equations we find 
\begin{flalign}
\frac{\rho_{\text{m}}}{M_{\text{g}}^{2}}=3\beta(1-r_{1}^{2}-r_{2}^{2}),\label{eq:T1-M22-matter-dens}
\end{flalign}
which is in agreement with \cref{eq:T1-matter-dens}. Taking the
derivative of \cref{eq:T1-quartic} with respect to the number of $e$-foldings
$N$ yields 
\begin{flalign}
r_{1}r_{1}^{\prime}+r_{2}r_{2}^{\prime}=\frac{3}{2}(1-r_{1}^{2}-r_{2}^{2}),
\end{flalign}
after plugging in \cref{eq:continuity,eq:T1-M22-matter-dens}. According
to \cref{eq:T1-matter-dens-param}, the matter density parameter
reads 
\begin{flalign}
\Omega_{\text{m}}=1-r_{1}^{2}-r_{2}^{2}.
\end{flalign}

Looking at these equations it seems that the phase space for this
model is $2$-dimensional, which will make the dynamical analysis
more complicated. Let us however try to find a $1$-dimensional phase
space for the model, by changing our dynamical variables. We define
the variable $r$ as $r^{2}\equiv r_{1}^{2}+r_{2}^{2}$, in terms
of which the equations above can be written as 
\begin{flalign}
3\mathcal{H}^{2}= & 3a^{2}\beta r^{2}+a^{2}\frac{\rho_{\text{m}}}{M_{\text{g}}^{2}},\\
r^{\prime}= & \frac{3}{2}(r^{-1}-r),\\
\Omega_{\text{m}}= & 1-r^{2}.
\end{flalign}
We can therefore see that the dynamics of the model are completely
captured in terms of our new variable $r$, which has a $1$-dimensional
phase space, with the fixed and singular points $r^{\text{fix}}=1$
and $r^{\text{sing}}=0$, respectively. What is missing is an
expression for the effective equation of state. In order to find it
we start with \cref{{eq:T1-M22-fi-friedmann}} and take its derivative
with respect to $N$, which yields $\mathcal{H}^{\prime}=a\sqrt{\beta}$ as
$a^{\prime}=a$. Plugging this into \cref{eq:weff} together with
\cref{eq:T1-M22-fi-friedmann} we find a constant effective equation
of state, independent of $r$: $w_{\text{eff}}=-1$.

Since the effecive equation of state is constant with $w_\text{eff}=-1$ at all times, this model is clearly ruled out.
Thus we do not present the phase space of this model.

\subsubsection{The $\beta_{1,3}\beta_{2,3}$ model}

In this model, the unique relation between the two scale factor ratios
is $r_{2}=B_{33}^{-1}r_{1}$, such that \cref{eq:T1-derivative}
becomes 
\begin{flalign}
r_{1}^{\prime} & =\frac{3r_{1}-(B_{33}^{-2}+1)r_{1}^{3}}{(B_{33}^{-2}+1)r_{1}^{2}-1},
\end{flalign}
allowing us to read off the fixed and the singular points 
\begin{flalign}
r_{1}^{\text{fix}} & =\frac{\sqrt{3}B_{33}}{\sqrt{B_{33}^{2}+1}},\\
r_{1}^{\text{sing}} & =\frac{1}{\sqrt{B_{33}^{-2}+1}}.
\end{flalign}
We see that there are no cases to be distinguished. The matter density
parameter and the effective equation of state are 
\begin{flalign}
\Omega_{\text{m}} & =1-\frac{B_{33}^{-2}+1}{3}r_{1}^{2},\\
w_{\text{eff}} & =-\frac{1}{3}(B_{33}^{-2}+1)(r_{1}^{\prime}+r_{1})r_{1}.
\end{flalign}

\begin{figure}
\centering{}\includegraphics[width=0.4\linewidth]{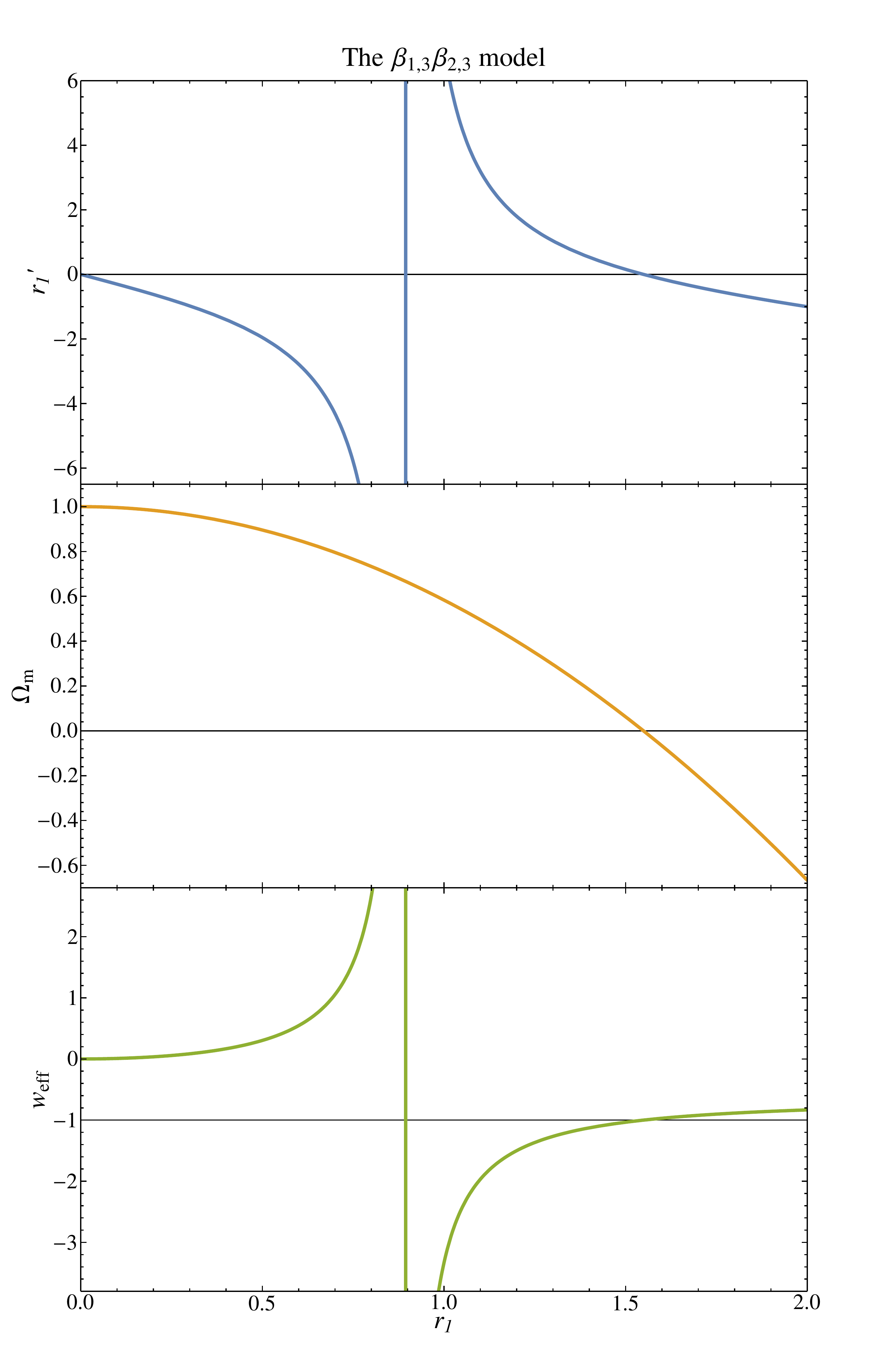}
\protect\protect\caption{\label{fig:T1-33}Evolution of $r_{1}'$, the matter density parameter
$\Omega_{\text{m}}$, and the effective equation of state $w_{\text{eff}}$
in terms of $r_{1}$ for the $\beta_{1,3}\beta_{2,3}$ model of star
trigravity with an interaction parameter ratio of $B_{33}=2$. }
\end{figure}

Looking at \cref{fig:T1-33}, for the parameter choice of $B_{33}=2$
without loss of generality, we can identify three branches. The finite
branch $[0,r_{1}^{\text{sing}}]$ is not viable as $\Omega_{\mathrm{m}}$
increases with time and $w_{\text{eff}}$ is always positive, i.e.,
the branch does not give a late-time acceleration. The infinite branch
$[r_{1}^{\text{fix}},\infty]$ is also ruled out as $\Omega_{\text{m}}<0$ always. Finally, on the intermediate finite branch $[r_{1}^{\text{sing}},r_{1}^{\text{fix}}]$,
the matter density parameter decreases to its final value $\Omega_{\text{m}}^{\text{fin}}=0$,
but its initial value is $\Omega_{\mathrm{m}}^{\text{init}}=2/3$. Since, additionally,
the effective equation of state is phantom at all times, this branch
is also ruled out.

\subsubsection{Summary}

We now summarize the phenomenology of the $1+1$-parameter models of
star trigravity. This is presented in \cref{tab:T1-summary}, where
we briefly describe the behavior of $\Omega_{\text{m}}$ and $w_{\text{eff}}$
for different models, their cases and branches. As $\Omega_{\text{m}}<0$
for all infinite branches of $1+1$-parameter star trigravity models,
there are no viable such branches, and we therefore do not mention
them in the table in order to keep things simple.

A $\checkmark$ in the matter density parameter column means that
it decreases from $\Omega_{\text{m}}^{\text{init}}=1$ to $\Omega_{\text{m}}^{\text{fin}}=0$
monotonically, i.e., as in the standard $\Lambda$CDM model. A $\checkmark$
in the effective equation of state column means that $w_{\text{eff}}=0$
in the matter-dominated epoch (at early times) and $w_{\text{eff}}=-1$
at late times, again similarly to $\Lambda$CDM. Otherwise, if $\Omega_{\text{m}}$
and/or $w_{\text{eff}}$ do not behave as in standard cosmology, we
briefly describe their behavior, and point out whether/why the phenomenology
of the model/branch is new or unviable.

\begin{table}
\begin{tabular}{lllp{5cm}p{5cm}l}
\toprule[0.5mm]
  & & & \multicolumn{2}{c}{{\bf Viability criterion}} & \tabularnewline
\cmidrule{4-5}
 {\bf Model}\footnote{As mentioned previously, in star trigravity $\beta_{1,n}\beta_{2,m}$ and $\beta_{1,m}\beta_{2,n}$ are the same models, as star trigravity is symmetric under exchanging $f_1$ and $f_2$ (along with the interaction parameters and Planck masses). Additionally, the $\beta_{1,1}\beta_{2,2}$ and $\beta_{1,2}\beta_{2,3}$ models are completely equivalent.} &  {\bf Case} & {\bf Branch} & $\Omega_{\text{m}}$ & $w_{\text{eff}}$ & {\bf Phenomenology} \tabularnewline
\midrule[0.5mm]
$\beta_{1,1}\beta_{2,1}$ & & Finite & $\checkmark$ & $\checkmark$ & Standard\tabularnewline
\midrule[0.3mm]
\multirow{4}{*}{$\beta_{1,1}\beta_{2,2}$} & $B_{12}>1/\sqrt{3}$  & Finite & $\Omega_{\text{m}}^{\text{init}}<1$ & Constant $w_{\text{eff}}$: $w_{\text{eff}}=-1$ & Unviable\tabularnewline
\cmidrule{2-6} 
 & $B_{12}=1/\sqrt{3}$ & Finite & $\Omega_\text{m}<0$ & Constant $w_{\text{eff}}$: $w_{\text{eff}}=-1$ & Unviable\tabularnewline
 \cmidrule{2-6} 
 & $B_{12}<1/\sqrt{3}$ & & $\Omega_\text{m}<0$ & Constant $w_{\text{eff}}$: $w_{\text{eff}}=-1$ & Unviable\tabularnewline
\midrule[0.3mm]
\multirow{8}{*}{$\beta_{1,1}\beta_{2,3}$} & \multirow{6}{*}{$B_{13}>2/3$ } & $[0,r_{1}^{\text{fix}-}]$  & $\Omega_\text{m}<0$ & $w_\text{eff}>0$ & Unviable\tabularnewline
\cmidrule{3-6} 
 &  & $[r_{1}^{\text{fix}-},r_{1}^{\text{sing}}]$ & $\Omega_{\text{m}}^{\text{init}}<1$  & Phantom at early and positive at late times & Unviable\tabularnewline
 \cmidrule{3-6} 
 &  & $[r_{1}^{\text{sing}},r_{1}^{\text{fix}+}]$ & First increases and then decreases to zero  & Positive at early times & New\tabularnewline
 \cmidrule{2-6} 
 & $B_{13}=2/3$ & Finite & $\Omega_\text{m}<0$ & $w_\text{eff}>0$ & Unviable\tabularnewline
 \cmidrule{2-6} 
 & $B_{13}<2/3$ & Finite & $\Omega_\text{m}<0$ & $w_\text{eff}>0$ & Unviable\tabularnewline
\midrule[0.3mm]
$\beta_{1,2}\beta_{2,2}$ & & Finite & $\checkmark$ & Constant $w_{\text{eff}}$: $w_\text{eff}=-1$ & Unviable\tabularnewline
\midrule[0.3mm]
\multirow{2}{*}{$\beta_{1,3}\beta_{2,3}$ } & & $[0,r_{1}^{\text{sing}}]$ & Increases in time & $w_\text{eff}>0$ & Unviable\tabularnewline
\cmidrule{3-6} 
 &  & $[r_{1}^{\text{sing}},r_{1}^{\text{fix}}]$ & $\Omega_\mathrm{m}^\mathrm{init}=2/3$ & Phantom at all times & Unviable\tabularnewline
\bottomrule[0.5mm]
\end{tabular}\protect\caption{ An overview of the cosmological viability of different $1+1$-parameter models in star trigravity.
We consider different branches for different cases in each
model. We do not present infinite branches as for all of them $\Omega_{\text{m}}<0$, making the models unviable. The only viable models are the finite branch of the $\beta_{1,1}\beta_{2,1}$
model, with a phenomenology similar to that of the $\beta_1$ model of bigravity at the background level, and the intermediate finite branch of the $\beta_{1,1}\beta_{2,3}$
model with a new phenomenology.}
\label{tab:T1-summary} 
\end{table}

The table shows that we are left with only two models that are not
ruled out by our analysis. The finite branch of the $\beta_{1,1}\beta_{2,1}$
model behaves exactly like the $\beta_{1}$ model of bigravity at
the background level. Therefore we already know that it has a viable
background cosmology. A next natural step could then be to study linear
perturbations for the model to see if the gradient instabilities,
which are present in the finite branch of $\beta_{1}$ bigravity,
are absent in the $\beta_{1,1}\beta_{2,1}$ model. The finite branch
$[r_{1}^{\text{sing}},r_{1}^{\text{fix}+}]$ of the $\beta_{1,1}\beta_{2,3}$
model behaves differently from any bigravity models, and therefore
gives rise to a new phenomenology. Although it seems to be difficult
to achieve a viable cosmology with this model, it is not necessarily
ruled out. By choosing a very large value for $B_{13}$ one can make
this model describe the late-time evolution of the Universe, i.e.,
after the matter-radiation equality. However, such a large value of
$B_{13}$ seems unnatural.

\subsection{Path trigravity}

We now repeat the procedure of the previous section for path trigravity.
We first simplify \cref{eq:T2-scalefactorrelation,eq:T2-derivative-rel},
relating the two scale factor ratios $r_{i}$ by specifying the $\beta_{i,n}$,
and rewrite these equations in terms of $B_{mn}$ defined by \cref{eq:Bmn}.
We then simplify the differential \cref{eq:T2-derivative} for $r_{1}$
and read off the fixed and singular points. As the final step, we
simplify \cref{eq:T2-matter-dens-param} for the matter density
parameter $\Omega_{\text{m}}$ and \cref{eq:T2-eos} for the effective
equation of state $w_{\text{eff}}$. We apply the procedure to all
possible $1+1$-parameter models of path trigravity, one by one; these
are the nine models $\beta_{1,1}\beta_{2,1}$, $\beta_{1,1}\beta_{2,2}$,
$\beta_{1,1}\beta_{2,3}$, $\beta_{1,2}\beta_{2,1}$, $\beta_{1,2}\beta_{2,2}$,
$\beta_{1,2}\beta_{2,3}$, $\beta_{1,3}\beta_{2,1}$, $\beta_{1,3}\beta_{2,2}$,
and $\beta_{1,3}\beta_{2,3}$.

\subsubsection{The $\beta_{1,1}\beta_{2,1}$ model}
					For this model, the relations between the two scale factor ratios $r_{1}$ and $r_{2}$, as well as their time derivatives \cref{eq:T2-derivative-rel}, simplify to  
					\begin{flalign}
						r_{2}^{\pm}=\frac{\pm\sqrt{12B_{11}^{2}r_{1}^{6}+1}-1}{6B_{11}r_{1}^{3}},\ \ r_{2}^{\prime}=\frac{r_2^2+2B_{11}r_1^3 r_2-6B_{11}r_1^3 r_2^3}{B_{11}r_1^4(1+3r_2^2)}r_{1}^{\prime},
					\end{flalign}
					where $r_{2}^{\pm}$ are the two roots of \cref{eq:T2-scalefactorrelation}.
					The root $r_{2}^{-}$ does not yield consistent results because $\Omega_\text{m}>1$ always.
					Thus, we will focus only on $r_{2}^{+}$ to rewrite $r_{2}$ in terms of $r_{1}$.
					The differential \cref{eq:T2-derivative} for $r_{1}$ simplifies to
					\begin{flalign}
						r_{1}^{\prime}=\frac{3r_1(1-3r_1^2+3B_{11} r_1^3 r_2)(1+3r_2^2)}{1-12B_{11} r_1^3 r_2+3r_1^2(1+3r_2^2)}.\label{eq:T2-11-derivative}
					\end{flalign}
					We can read off the fixed points
					\begin{flalign}
						r_{1}^{\text{fix}}=0,\ \ r_{1}^{\text{fix}\pm}=\frac{\sqrt{3\pm\sqrt{9-4B_{11}^2}}}{\sqrt{2}B_{11}},
					\end{flalign}
					which means that we should distinguish between the qualitatively different cases (a) $B_{11}>3/2$, (b) $B_{11}=3/2$, and (c) $B_{11}<3/2$.
					The matter density parameter and the effective equation of state are given by 
					\begin{flalign}
						\Omega_{\text{m}} & =1-\frac{3r_{1}^{2}}{1+3B_{11}r_{1}^{3}r_{2}},\\
						w_{\text{eff}} & =-\frac{(r_{1}^{\prime}+3r_{1})r_{1}}{1+3B_{11}r_{1}^{3}r_{2}}.
					\end{flalign}

					\begin{figure}
						\centering{}
						\includegraphics[width=0.4\linewidth]{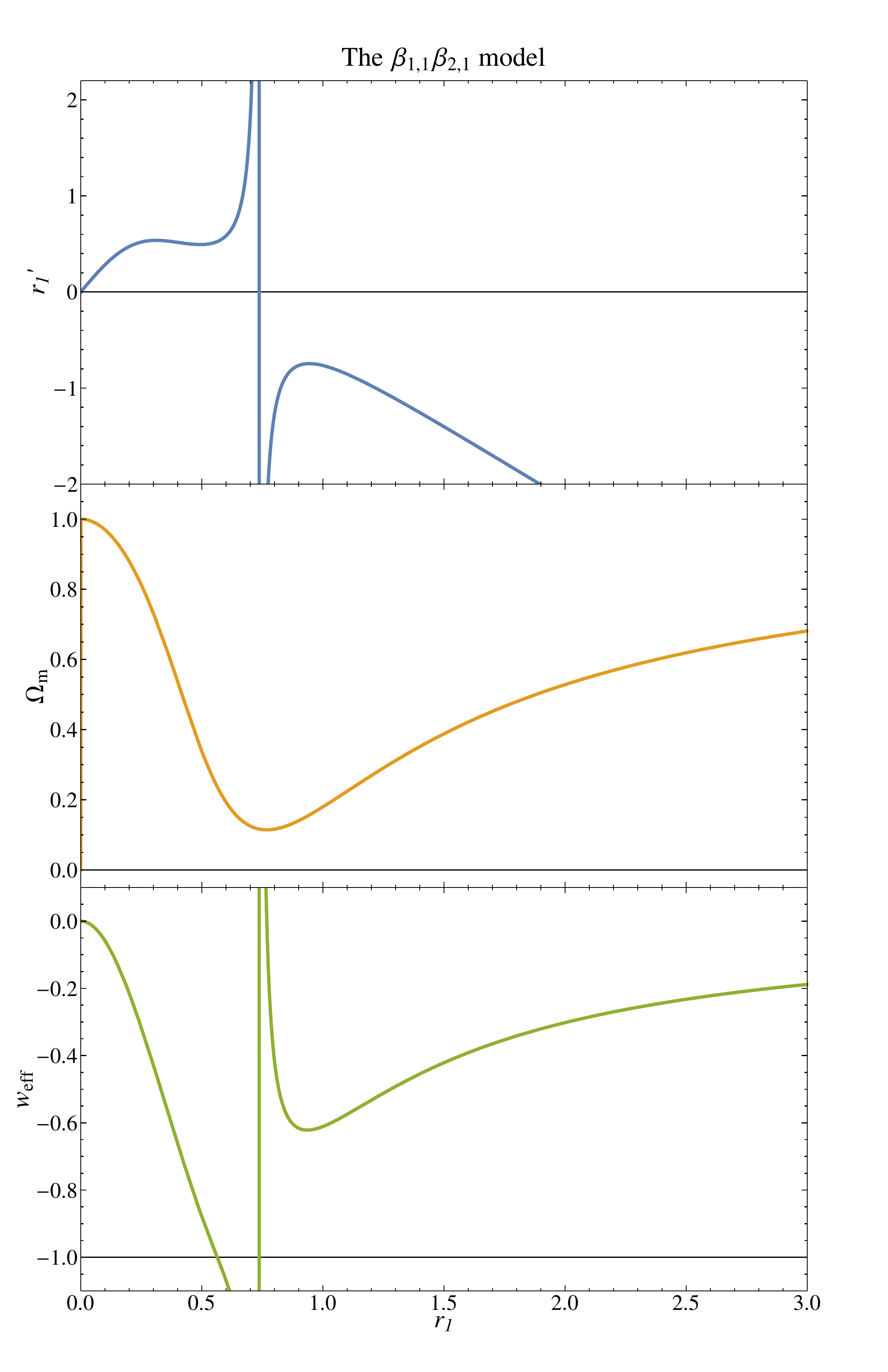}
						\caption{\label{fig:T2-11a}Evolution of $r_{1}^\prime$, $\Omega_{\text{m}}$ and $w_{\text{eff}}$ as functions of $r_{1}$ for the $\beta_{1,1}\beta_{2,1}$ model of path trigravity with ratio of the interaction parameters $B_{11}=1.8$ representing case (a). 
							}
					\end{figure}

$\bullet$ \textbf{Case (a):} As presented in \cref{fig:T2-11a},
there is a finite branch, as well as an infinite one, which are separated by a singular point $r_{1}^{\text{sing}}$.
Let us first discuss the finite branch.
The matter density parameter decreases from $\Omega_{\text{m}}^{\text{init}}=1$ to a finite final value $\Omega_{\text{m}}^{\text{fin}}>0$, i.e., there is no de Sitter point in the infinite future.
Such so-called scaling solutions, as mentioned in the beginning of \cref{sec:cos-11params}, are the ones with matter and dark energy density parameters approaching constant non-vanishing values in the future.
The exact expression for $\Omega_{\text{m}}^{\text{fin}}=\left.\Omega_{\text{m}}\right|_{r_{1}=r_{1}^{\text{sing}}}$ is quite lengthy, but can be found analytically.
In fact, the final value of $\Omega_{\text{m}}$ depends on the value of $B_{11}$ such that $\Omega_{\text{m}}^{\text{fin}}$ increases as $B_{11}$ increases.
This places an upper bound on the value of $B_{11}$ if we want a certain final value for the matter density parameter.
For the case of $\Omega_{\text{m}}^{\text{fin}}\le0.3$, this upper limit is $B_{11}\approx2.8$.
Thus, we are left with $B_{11}\in(2/3,2.8)$ in order to have a scaling solution with $\Omega_{\text{m}}^{\text{fin}}\le0.3$.
The effective equation of state starts from $0$ and then decreases with time, becoming phantom at late times.

					\begin{figure}
						\centering{}
						\includegraphics[width=0.4\linewidth]{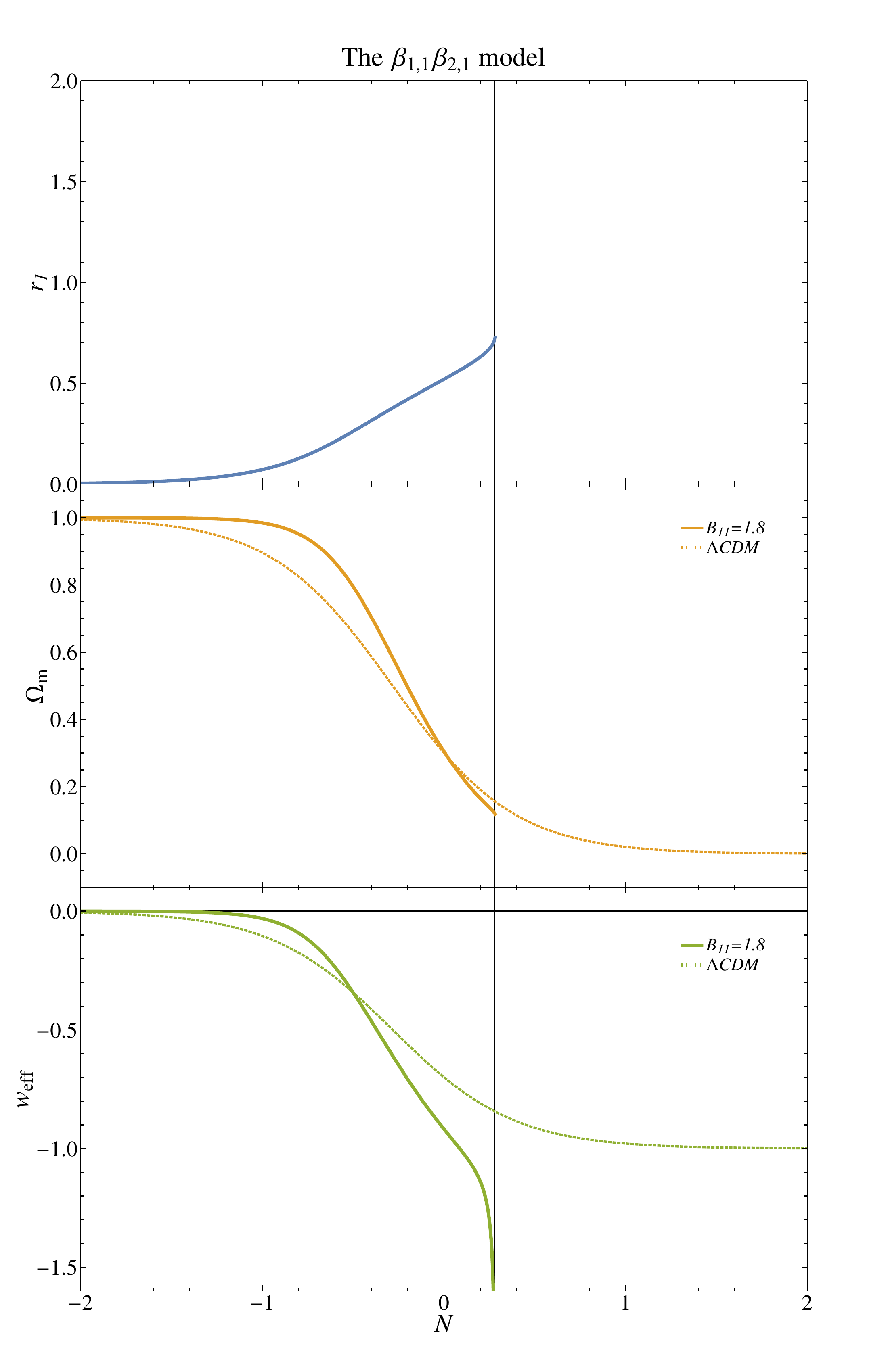}
						\includegraphics[width=0.4\linewidth]{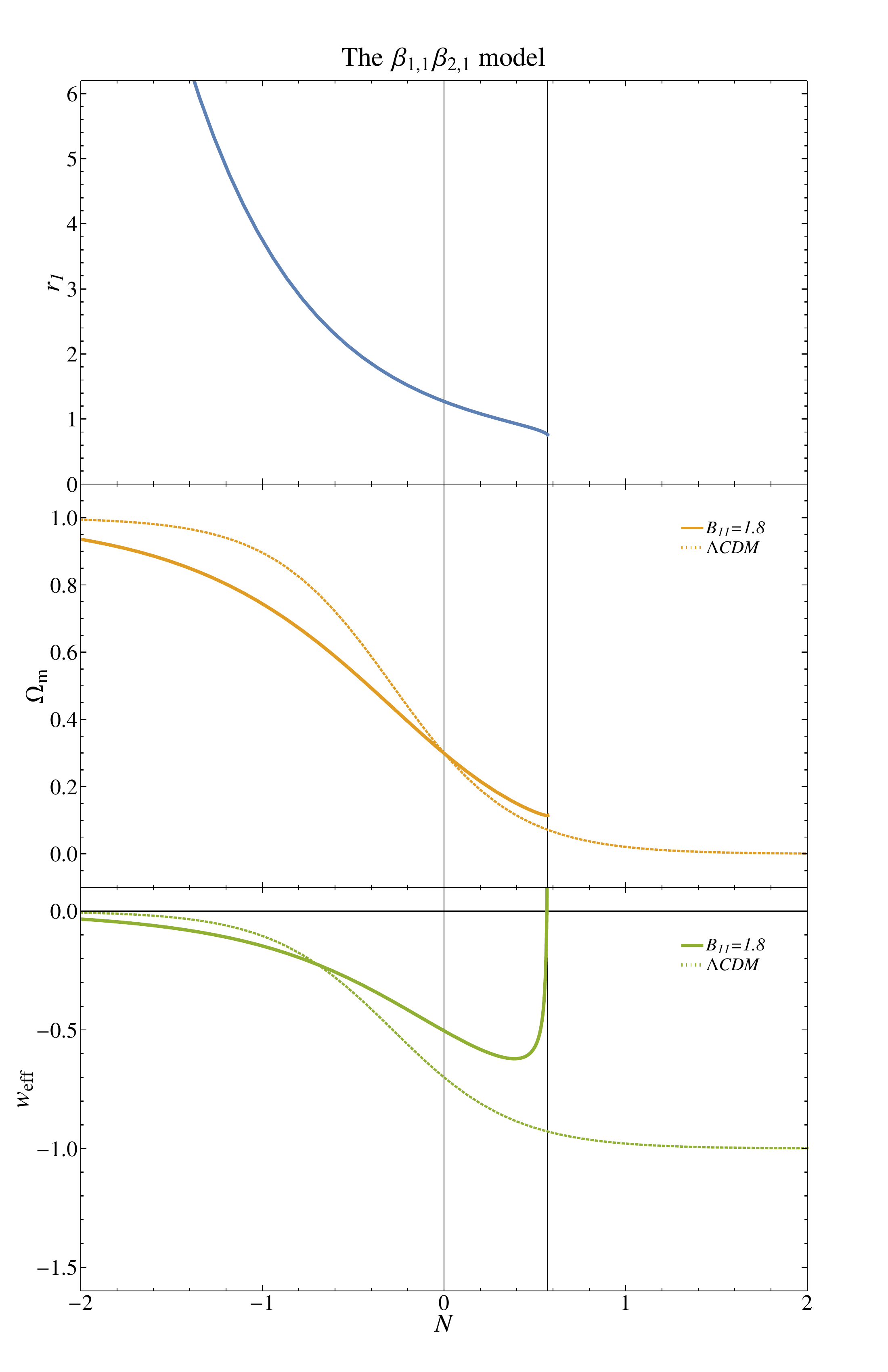}
						\caption{\label{fig:T2-11a-timeintegration}Evolution of $r_{1}$, $\Omega_{\text{m}}$, and $w_{\text{eff}}$ as functions of $N=\ln a$ in the $\beta_{1,1}\beta_{2,1}$ model of path trigravity with $B_{11}=1.8$ representing case (a).
						In order to compare to standard cosmology, the evolution of $\Omega_{\text{m}}$ and $w_{\text{eff}}$ are plotted also for $\Lambda$CDM with $\Omega_{\text{m},0}=0.3$.
						The left vertical line represents $N=0$ (today).
						Left panel: Evolution for the finite branch $[0,r_{1}^{\text{sing}}]$, where the right vertical line, at $N=0.28$ (in the future), represents the time at which $r_{1}$ takes its final value $r_{1}^{\text{sing}}$.
						Right panel: Evolution for the infinite branch $[r_{1}^{\text{sing}},\infty]$, where the right vertical line, at $N=0.57$, represents the time at which $r_{1}$ takes its final value $r_{1}^{\text{sing}}$.
						}
					\end{figure}

					These statements, however, do not rule this model out, and we should therefore analyze the model further as it has a new phenomenology; we do this
					by integrating the differential \cref{eq:T2-11-derivative} numerically for this finite branch with $B_{11}=1.8$.
					To integrate \cref{eq:T2-11-derivative}, we choose as the initial condition $r_{1,0}\equiv r_{1}(N=0)=0.52$.
					This value is chosen such that we obtain a present-time value of the matter density parameter of $\Omega_{\text{m},0}\approx0.3$.
					This should be considered only as a representative example; an extensive and careful statistical analysis is needed in order to see whether this model is consistent with cosmological observations.
					The time evolution of the quantities $r_{1}$, $\Omega_{\text{m}}$, and $w_{\text{eff}}$ are presented in \cref{fig:T2-11a-timeintegration} (left panel). For comparison, we have plotted also the time evolution of $\Omega_{\text{m}}$ and $w_{\text{eff}}$ for standard $\Lambda$CDM cosmology.
					The results show that, at late times, the effective equation of state is negative and with larger absolute values than the ones in $\Lambda$CDM for this path trigravity model.
					This however may not be a problem, as the average value of $w_{\text{eff}}$ at low redshifts, a quantity that is usually measured from observations, could be similar to the one in $\Lambda$CDM, and therefore, a more detailed statistical analysis is required to test the model observationally.
					The model however predicts that the evolution continues for only about $0.28$ $e$-foldings in the future.
					At that time, $r_1$ reaches its final value given by $r_1^\text{sing}$ at which $r_i^\prime\rightarrow\infty$ for both $i=1,2$.
					From $w_\text{eff}\rightarrow-\infty$ at the singular point we can deduce that $\mathcal H^\prime\rightarrow\infty$.
					To summarize, the model approaches a singular point after some finite time, a which the size of the Universe is finite ($\sim e^{0.28}$ times its size today), and the matter density parameter takes a finite value $\Omega_\text{m}>0$, but $\mathcal H^\prime\rightarrow\infty$.
					This does not rule out the model and one should compare the model's predictions to observational data.
					
					Let us now turn to the infinite branch.
					The matter density parameter behaves as in the finite branch with the same finite final value.
					However, the effective equation of state first decreases, starting with $w_\text{eff}^\text{init}=0$, during matter domination to a minimum value and then increases at late times towards $+\infty$.
					
					In order to further investigate this branch, we integrate the differential \cref{eq:T2-11-derivative}.
					As the initial condition we choose $r_{1,0}=r_{1}(N=0)=1.27$ in order to achieve a present-time value of $\Omega_\text{m,0}\approx0.3$.
					Again, this should be considered as a representative example.
					The time evolution of $r_1^\prime$, $\Omega_\text{m}$, and $w_\text{eff}$ are presented in the right panel of \cref{fig:T2-11a-timeintegration} where we have also included the time evolution of $\Omega_\text{m}$ and $w_\text{eff}$ for standard $\Lambda$CDM cosmology for comparison.
					The results show that the effective equation of state is larger than the one in $\Lambda$CDM at late times, but this might not be a problem as explained before.
					This model predicts that the evolution of the Universe continues only for about $0.57$ $e$-foldings in the future.
					The scale factor ratio takes its final value $r_1^\text{sing}$ and the matter density parameter approaches a constant but non-vanishing value $\Omega_\text{m}^\text{fin}$. While the size of the Universe is finite at this final point ($\sim e^{0.58}$ times its size today), the model predicts a decelerated expansion because $\mathcal H^\prime\rightarrow-\infty$.
					This branch is not ruled out by this analysis, and a detailed comparison to observational data is necessary in order to further constrain the model or to finally rule it out.
					
\begin{figure}
\centering{}\includegraphics[width=0.4\linewidth]{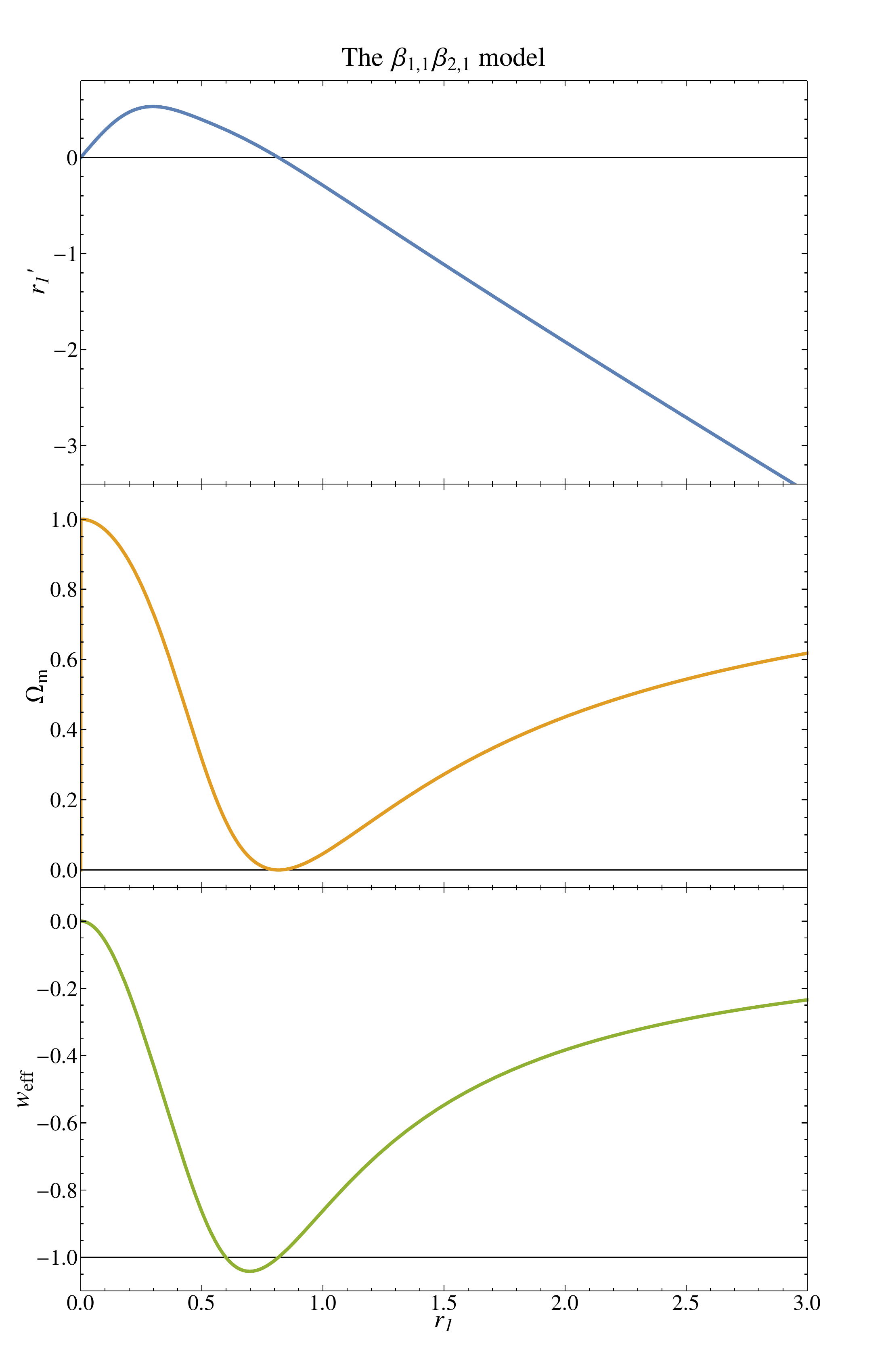}
\includegraphics[width=0.4\linewidth]{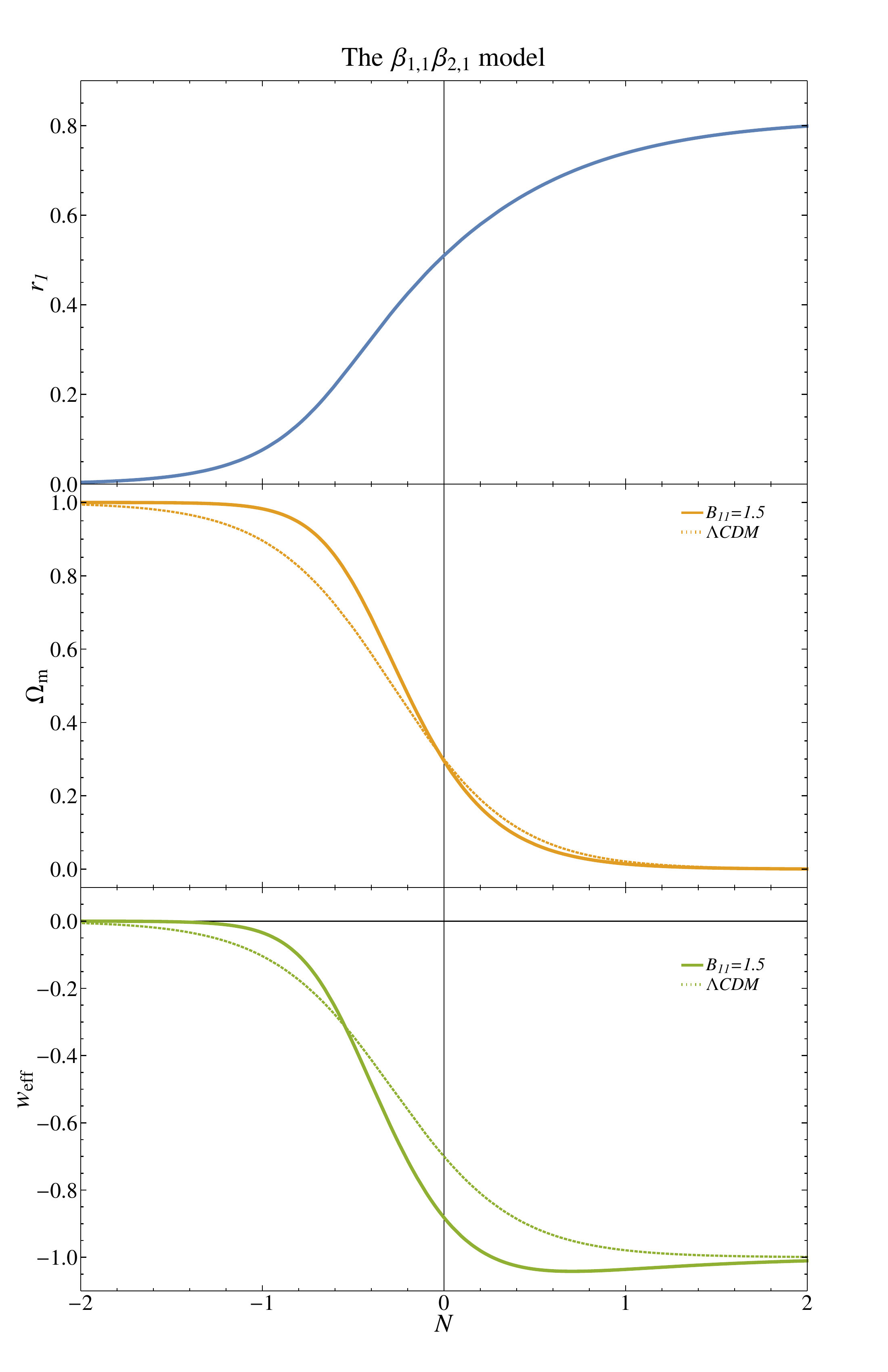} \protect\protect\caption{\label{fig:T2-11b}The $\beta_{1,1}\beta_{2,1}$ model of path trigravity
with the ratio of the interaction parameters $B_{11}=1.5$ representing
case (b). Left panel: Evolution of $r_{1}'$, $\Omega_{\text{m}}$,
and $w_{\text{eff}}$ as functions of $r_{1}$. Right panel: Evolution
of $r_{1}$, $\Omega_{\text{m}}$, and $w_{\text{eff}}$ as functions
of $N\equiv\ln a$ for the finite branch of the model. The vertical line represents $N=0$ (today). In
order to compare to standard cosmology, the evolution of $\Omega_{\text{m}}$
and $w_{\text{eff}}$ is plotted also for $\Lambda$CDM with $\Omega_{\text{m},0}=0.3$.}
\end{figure}

$\bullet$ \textbf{Case (b):} As shown in the left panel of \cref{fig:T2-11b}, this
case admits a finite and an infinite branch separated by the fixed point $r_1^\text{fix}=\frac{2}{3}\sqrt{\frac{3}{2}}$. 
In both branches the matter density parameter decreases from $\Omega_{\text{m}}^{\text{init}}=1$
to $\Omega_{\text{m}}^{\text{fin}}=0$ as in standard cosmology. The
effective equation of state starts off with $w_{\text{eff}}^{\text{init}}=0$
during matter domination and becomes $w_{\text{eff}}^{\text{fin}}=-1$ at late times, again for both branches.
Although the infinite branch has a standard phenomenology, on the finite branch, there is a period at late times when the effective equation of state is phantom, i.e., $w_\mathrm{eff}<-1$.
This branch has a non-standard phenomenology and needs further analysis in order
to check its viability. Therefore, we solve the differential \cref{eq:T2-11-derivative}
numerically subject to the initial condition $r_{1}(N=0)=0.51$, where
$N=0$ corresponds to today, and show the result in the right panel of \cref{fig:T2-11b}. This initial value has been chosen in
such a way that the model gives a present-time matter density parameter
of $\Omega_{\text{m},0}\approx0.3$. This then results in a present-time
effective equation of state of $w_{\text{eff},0}\approx-0.88$. Again,
as the measured value of $w_{\text{eff}}$ from observations is usually
an average over about one $e$-folding, this model could very well
be viable, as far as the background cosmology is concerned.
The standard $\Lambda$CDM evolution of $\Omega_{\text{m}}$ and $w_{\text{eff}}$
is also plotted for comparison.
We can therefore conclude that the
model has an interesting new phenomenology, allowing for a phantom
equation of state at late times, which cannot be ruled out at this
stage and needs further investigation. In order to find out whether
the model can successfully describe the late-time evolution of the
Universe, one needs to perform a statistical analysis and compare
the model to observational data; we leave this for future work.

\begin{figure}
\begin{centering}
\includegraphics[width=0.4\linewidth]{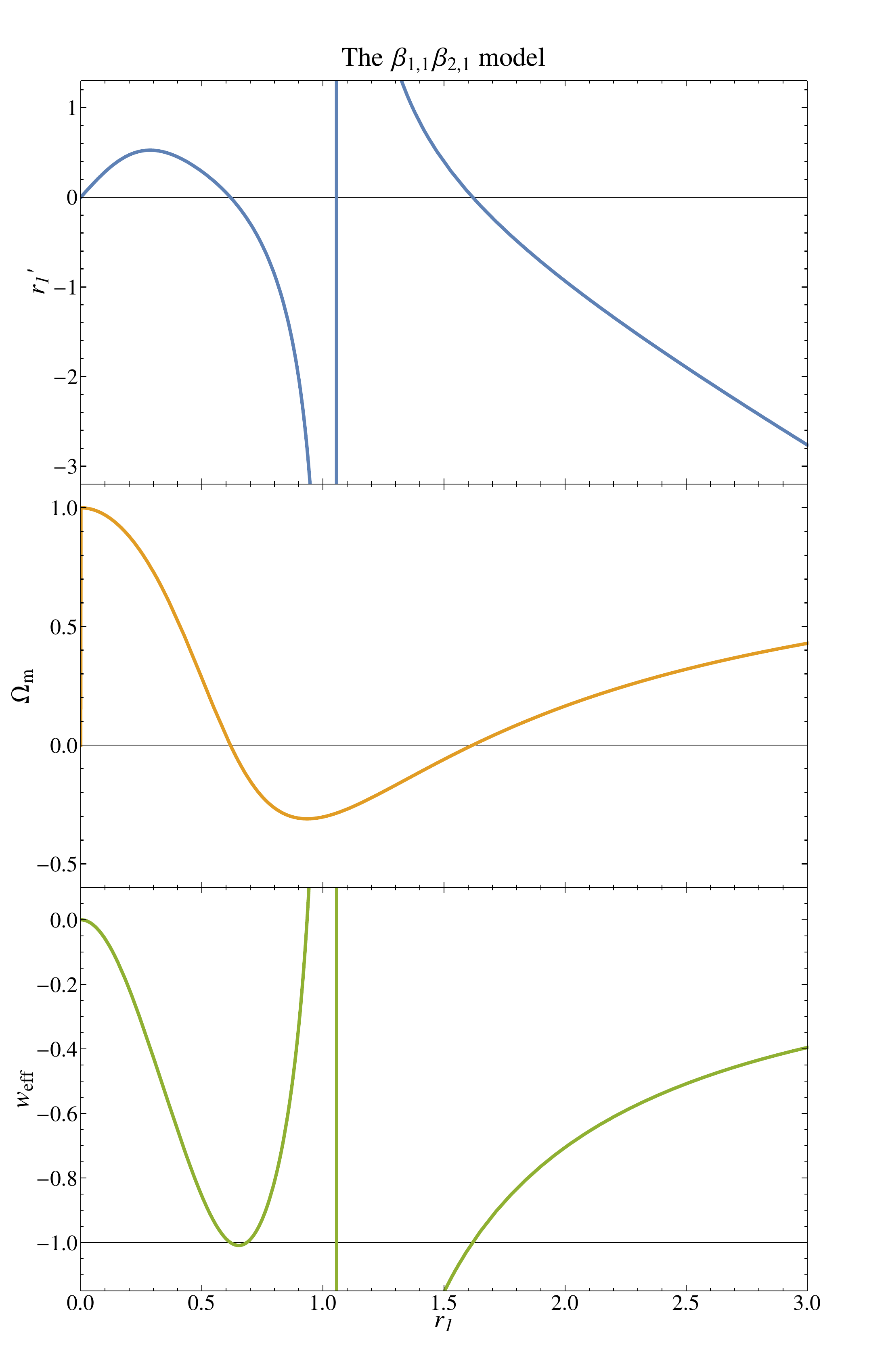}\includegraphics[width=0.4\linewidth]{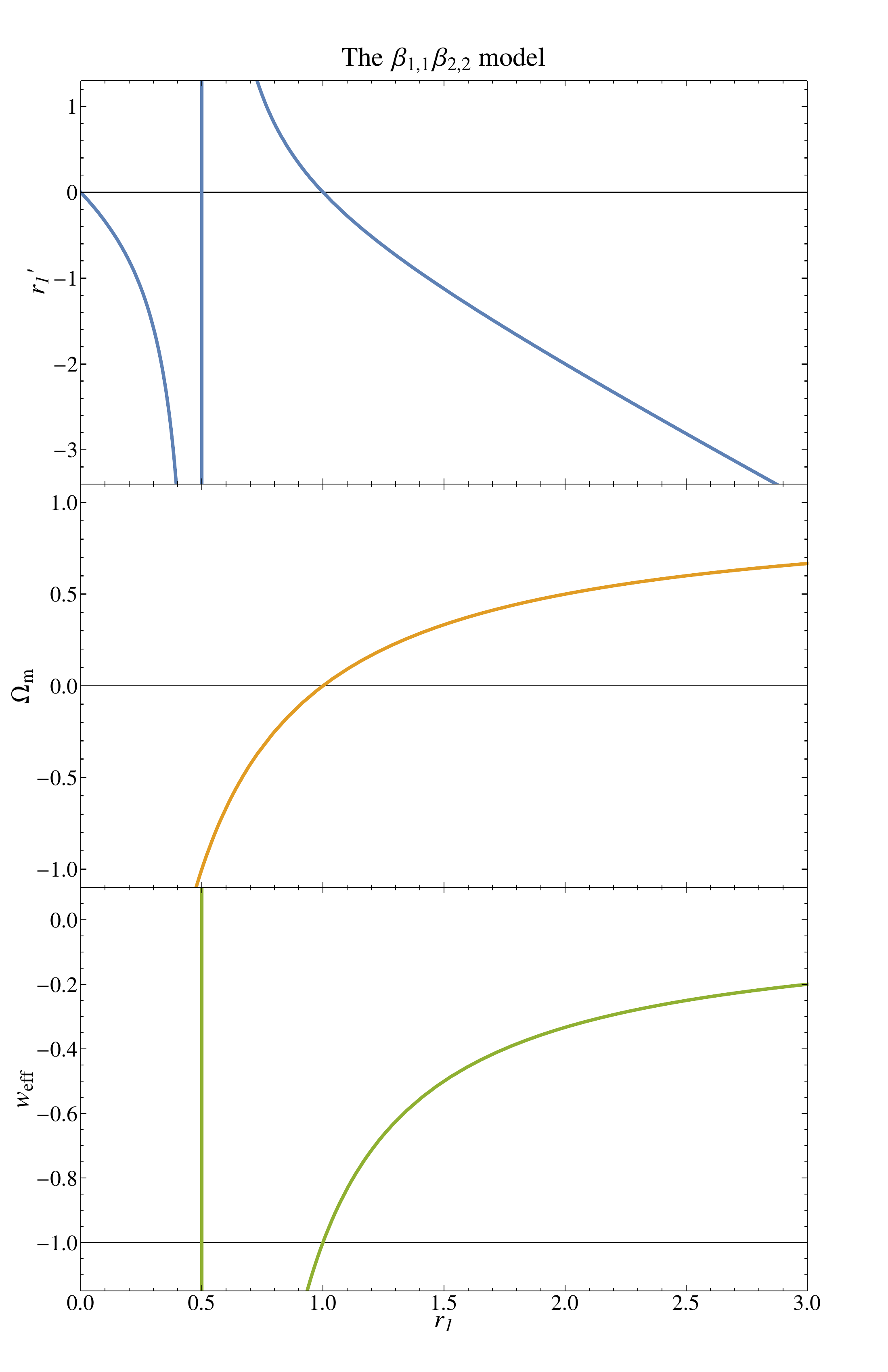} 
\par\end{centering}

\protect\protect\caption{\label{fig:T2-11c}\label{fig:T2-12}Left panel: The evolution of $r_{1}^{\prime}$, $\Omega_{\text{m}}$, and $w_\text{eff}$ as functions of $r_{1}$ for the $\beta_{1,1}\beta_{2,1}$ model of
path trigravity with $B_{11}=1$ representing case (c).
Right panel: The same for the $\beta_{1,1}\beta_{2,2}$ model of
path trigravity with the interaction parameter ratio of $B_{12}=1$.}
\end{figure}

$\bullet$ \textbf{Case (c):} The model admits a finite branch $[0,r_{1}^{\text{fix}-}]$, an infinite branch $[r_{1}^{\text{fix}+},\infty]$, and two intermediate finite branches $[r_{1}^{\text{fix}-},r_{1}^{\text{sing}}]$ and $[r_{1}^{\text{sing}},r_{1}^{\text{fix}+}]$ separated by a singular point $r_{1}^{\text{sing}}$ as can
be seen in the left panel of \cref{fig:T2-11c}. Neither of the intermediate branches is viable
as the matter density parameter is negative for the entire evolution.
On the finite and infinite branches, $\Omega_{\text{m}}$ and $w_{\text{eff}}$ follow
the standard behavior (as for the finite branches of bigravity) such
that these branches are viable at the background level.

\subsubsection{The $\beta_{1,1}\beta_{2,2}$ model}

This model is described by the following equations: 
\begin{flalign}
r_{2}^{\pm}=\pm\sqrt{1-\frac{1}{3B_{12}r_1^3}},\ \ r_{2}^{\prime}=\frac{1-6B_{12}r_1^3(r_2^2-1)}{6B_{12}r_{1}^{4}r_{2}}r_{1}^{\prime}.
\end{flalign}
Here, $r_{2}^{\pm}$ are the two roots of \cref{eq:T2-scalefactorrelation};
they both lead to the same results. The derivative of $r_{1}$ is
given by 
\begin{flalign}
r_{1}^{\prime} & =-3\frac{1-B_{12}r_1}{1-2B_{12}r_1}r_1,
\end{flalign}
from which we can read off two fixed and one singular point: 
\begin{flalign}
r_{1}^{\text{fix,I}}&=0,\\
r_{1}^{\text{fix,II}} & =B_{12}^{-1},\\
r_{1}^{\text{sing}} & =(2B_{12})^{-1}.
\end{flalign}
The matter density parameter and the effective equation of state are given by 
\begin{flalign}
\Omega_{\text{m}} & =1-B_{12}^{-1}r_{1},\\
w_{\text{eff}}&=-\frac{(r_1^\prime+3r_1)r_1}{1+3B_{12}r_1^3r_2^2}.
\end{flalign}

As presented in the right panel of \cref{fig:T2-12} for a representative value of $B_{12}$ (i.e., $B_{12}=1$), there are two finite branches, $[0,r_{1}^{\text{sing}}]$ and $[r_{1}^{\text{sing}},r_{1}^{\text{fix,II}}]$, and an infinite branch $[r_{1}^{\text{fix,II}},\infty]$.
Both finite branches are ruled out because $\Omega_\text{m}<0$ for the entire evolution of the Universe. 
The infinite branch however admits a standard phenomenology with $\Omega_\text{m}$ evolving from $1$ to $0$ and $w_\text{eff}$ evolving from $0$ to $-1$.

\subsubsection{The $\beta_{1,1}\beta_{2,3}$ model}
\label{sec:T2-M13}

For this model, \cref{eq:T2-scalefactorrelation} has three roots:
\begin{flalign}
	&r_{2}^\text{I}=\left(\frac{\sqrt{1-4B_{13}^{2}r_1^6}-1}{2B_{13}r_1^3}\right)^{\frac{1}{3}}+\left(\frac{\sqrt{1-4B_{13}^{2}r_1^6}-1}{2B_{13}r_1^3}\right)^{-\frac{1}{3}},\label{eq:T2-M13-scalefactorrelation}\\
	&r_{2}^\text{II}=-\frac{1-i\sqrt{3}}{2}\left(\frac{\sqrt{1-4B_{13}^2 r_1^6}-1}{2 B_{13}r_1^3}\right)^{\frac{1}{3}}-\frac{1+i\sqrt{3}}{2}\left(\frac{\sqrt{1-4B_{13}^2 r_1^6}-1}{2 B_{13}r_1^3}\right)^{-\frac{1}{3}},\\
	&r_{2}^\text{III}=-\frac{1+i\sqrt{3}}{2}\left(\frac{\sqrt{1-4B_{13}^2 r_1^6}-1}{2 B_{13}r_1^3}\right)^{\frac{1}{3}}-\frac{1-i\sqrt{3}}{2}\left(\frac{\sqrt{1-4B_{13}^2 r_1^6}-1}{2 B_{13}r_1^3}\right)^{-\frac{1}{3}}.
\end{flalign}
The derivatives of the two scale factor ratios are uniquely related via 
\begin{flalign}
r_{2}^{\prime} & =\frac{1-2B_{13}r_{1}^{3}r_{2}(r_2^2-3)}{3B_{13}r_{1}^{4}(r_{2}^{2}-1)}r_{1}^{\prime},\label{eq:T2-13-derivativerelation}
\end{flalign}
while the differential \cref{eq:T2-derivative} simplifies to 
\begin{flalign}
r_{1}^{\prime}=\frac{3r_{1}(1-r_{2}^{2})(1-3r_{1}^{2}+B_{13}r_{1}^{3}r_{2}^{3})}{1+4B_{13}r_1^3 r_2^3+3r_1^2(1-r_2^2)}.\label{eq:T2-13-derivative}
\end{flalign}
The matter density parameter and the effective equation of state for the model are 
\begin{flalign}
	\Omega_{\text{m}} & =1-\frac{3r_{1}^{2}}{1+B_{13}r_{1}^{3}r_{2}^{3}},\\
	w_{\text{eff}} & =-\frac{(r_{1}^{\prime}+3r_{1})r_{1}}{1+B_{13}r_{1}^{3}r_{2}^{3}}.\label{eq:T2-M13-weff}
\end{flalign}

\begin{figure}
\begin{centering}
\includegraphics[width=0.4\linewidth]{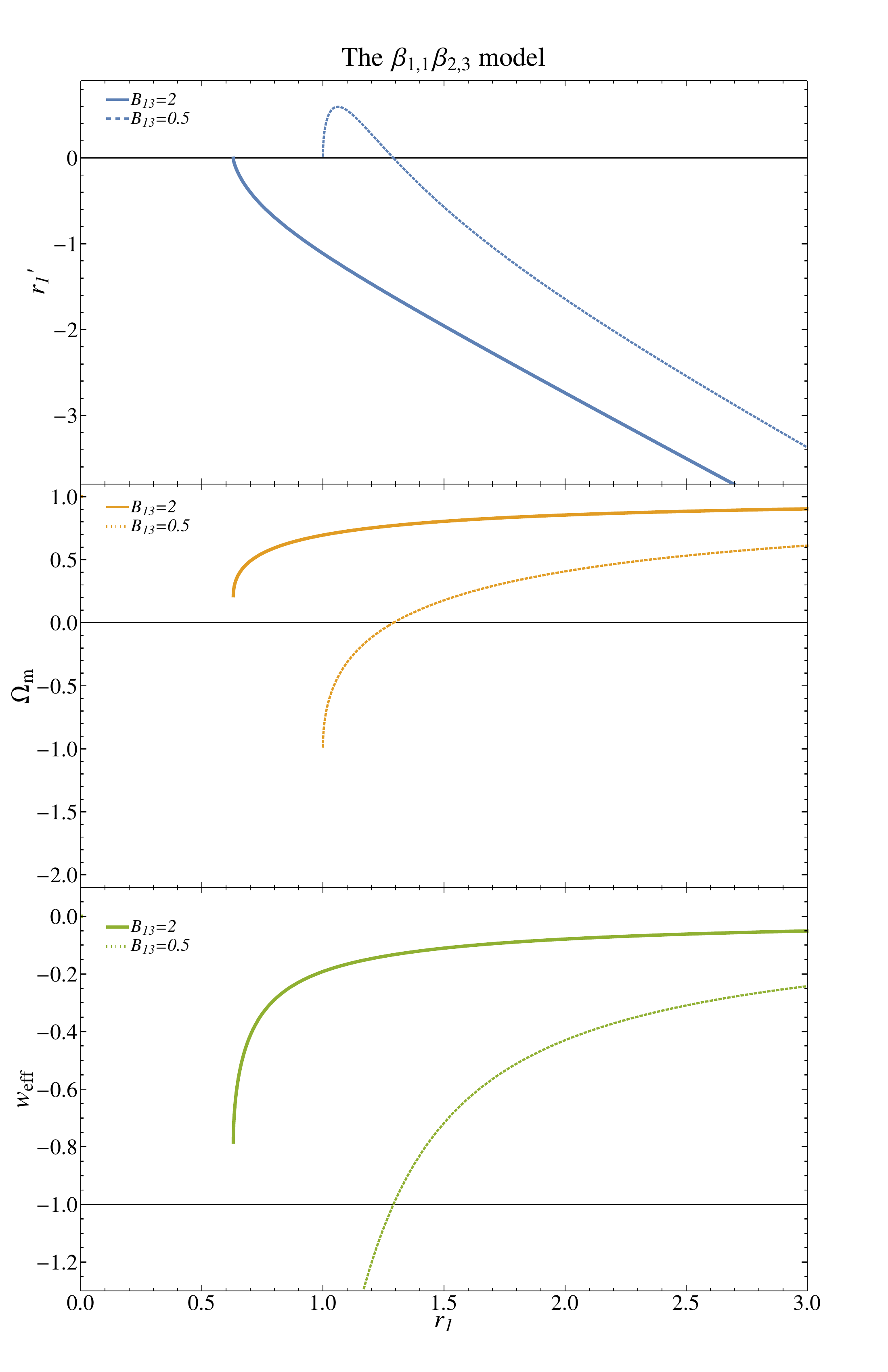}\includegraphics[width=0.4\linewidth]{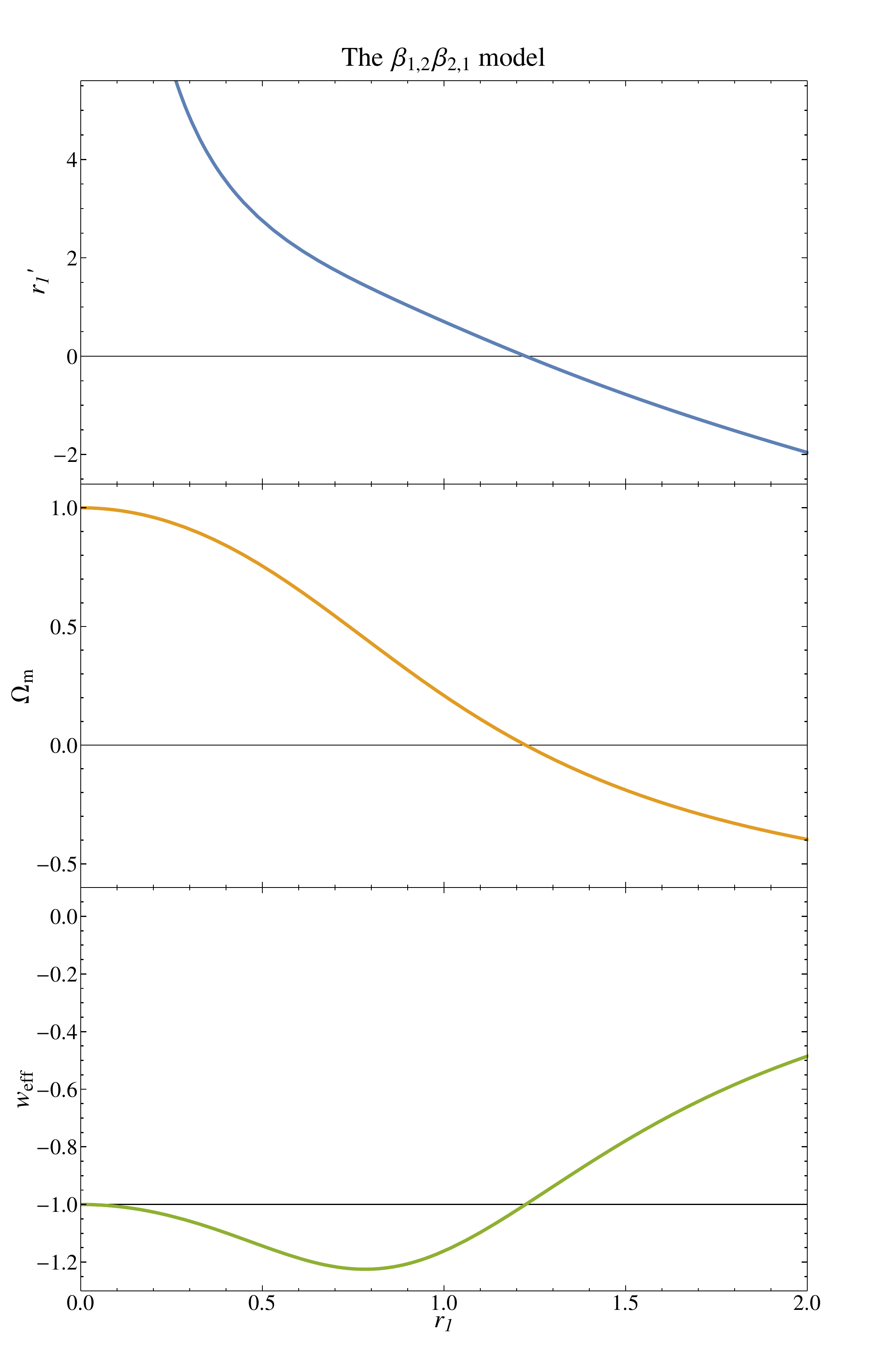} 
\par\end{centering}

\protect\protect\caption{\label{fig:T2-13} \label{fig:T2-21a}Left panel: Evolution of $r_{1}^{\prime}$, $\Omega_{\text{m}}$, and $w_\text{eff}$ as functions of $r_{1}$ for the $\beta_{1,1}\beta_{2,3}$
model of path trigravity with the interaction parameter ratios of $B_{13}=2$ (solid lines) representing case (Ia)
and $B_{13}=0.5$ (dotted lines) representing case (Ib).
Right panel: The same for the $\beta_{1,2}\beta_{2,1}$ model of path trigravity
with the interaction parameter ratio of $B_{21}=1$ representing case (a).}
\end{figure}

The root $r_2^\text{II}$ leads to a positive effective equation of state for any value of $B_{13}$ and we conclude that this root does not lead to viable results without presenting the phase space.
The root $r_2^\text{III}$ produces a negative matter density parameter, $\Omega_\text{m}<0$, for any value of $B_{13}$ and thus the phenomenology is not viable.
Only the root $r_2^\text{I}$ needs a more detailed analysis and we therefore focus on \cref{eq:T2-M13-scalefactorrelation} relating $r_2$ and $r_1$.
We have to distinguish the two cases (Ia) $B_{13}\gtrsim1.12$ with one fixed point $r_1^{\text{fix,I}}$ and (Ib) $B_{13}\lesssim1.12$ with two fixed points $r_1^{\text{fix,I}}$ and $r_1^{\text{fix,II}}$.
We do not present the analytical expressions of the fixed points here as they are quite lengthy.
Using \cref{eq:T2-M13-scalefactorrelation}, $r_{1}^{\prime}$ is real only in the interval $[r_{1}^{\mathrm{fix,I}},\infty]$.

$\bullet$ \textbf{Case (Ia):} As presented in \cref{fig:T2-13} (left panel) for the representative value of $B_{13}=2$, there is only an infinite branch $[r_{1}^{\text{fix,I}},\infty]$.
For $r_{1}<r_{1}^{\text{fix,I}}$, $r_{1}^{\prime}$ takes complex values and thus the model's phase space is limited to the interval $[r_{1}^{\text{fix,I}},\infty]$.
Although the matter density parameter and the effective equation of state initially behave as in the standard phenomenology, i.e., $\Omega_\text{m}^\text{init}=1$ and $w_\text{eff}^\text{init}=0$, they approach values $\Omega_\text{m}^\text{fin}>0$ and $w_\text{eff}^\text{fin}>-1$ in the future.
This is again a scaling solution.

A more careful investigation of the model, however, reveals some subtleties at and close to $r_{1}^{\text{fix,I}}$ which must be taken into account when interpreting its cosmological implications. If the point is really a fixed point, then the matter density parameter will become a constant in the future and since the continuity \cref{eq:continuity} implies $\rho_\text{m}\propto a^{-3}$, the Hubble parameter must evolve as $\mathcal H^2\propto a^{-3}$.
This implies $w_\text{eff}=0$, which contradicts \cref{eq:T2-M13-weff} as can be seen in \cref{fig:T2-13} (left panel) where $w_\text{eff}$ is negative in the future. The reason for this contradiction can however be understood by looking at how $r_1$, $r_2$, and their derivatives evolve with time.
At the point given by $r_1^\prime=0$ we have $r_2=1$, such that we divide by $0$ in \cref{eq:T2-13-derivativerelation}, and therefore \cref{eq:T2-derivative-rel} is not valid at that point.
In fact we have $r_2^\prime\ne0$ as can be found by taking the derivative of \cref{eq:T2-M13-scalefactorrelation}.
In addition, we find that $\mathrm{d}r_2/\mathrm{d}r_1$ and $\mathrm{d}r_2^\prime/\mathrm{d}r_1$ are both singular at this point.
This all means that the point with $r_1^\prime=0$ is not really a fixed point of the system because $r_2^\prime\ne0$.
Since the derivatives of some quantities are singular at this point, we call it a \textit{singular fixed point}.
With the analysis developed and used in this paper it is not possible to make predictions for singular fixed points, and their analysis is beyond the scope of this work.
We therefore leave a careful treatment of models with singular fixed points for future work.

Note however that our analysis does not rule this case out, if the singular fixed point can be pushed to a time far in the future. The question of what happens to the Universe when it approaches this singularity is an interesting one that needs to be explored. Situations with such singular fixed points occur in path trigravity $1+1$-parameter models only when $\beta_{2,3}\ne0$, as can be seen by looking at the denominator of \cref{eq:T2-derivative-rel}.
In that case, the denominator will have a term $\propto(1-r_2^2)$.
Whenever we get to a point with $r_2^2=1$ our analysis does not work because \cref{eq:T2-derivative-rel} is not valid.
One possible way to deal with such a singularity is to not use \cref{eq:T2-scalefactorrelation,eq:T2-derivative-rel} to rewrite $r_2$ in terms of $r_1$, but to treat $r_1$ and $r_2$ as two independent dynamical variables subject to the Friedmann \cref{eq:T2-g-friedmann,eq:T2-f1-friedmann,eq:T2-f2-friedmann}, and then to analyze the $2$-dimensional phase space numerically.
Since this requires a type of analysis that is different from our approach in this paper, we leave it for future work.

$\bullet$ \textbf{Case (Ib):} As presented in \cref{fig:T2-13} (left panel) for the representative value of $B_{13}=0.5$ there is one finite and one infinite branch.
The infinite branch $[r_{1}^{\text{fix,II}},\infty]$ produces the standard phenomenology with $\Omega_{\text{m}}$ evolving from $1$ to $0$ and $w_{\text{eff}}$ evolving from $0$ to $-1$.
The finite branch $[r_{1}^{\text{fix,I}},r_{1}^{\text{fix,II}}]$ is not viable because $\Omega_{\text{m}}<0$ always.
For $r_{1}<r_{1}^{\text{fix,I}}$, $r_{1}^{\prime}$ takes complex values and thus the phase space is limited to the interval $[r_{1}^{\text{fix,I}},\infty]$.

\subsubsection{The $\beta_{1,2}\beta_{2,1}$ model}

The phase space for this model is described by 
\begin{flalign}
r_{1}^{\prime}=\frac{3(1+3r_{2}^{2})(1-r_{1}^{2}+B_{21}r_{1}^{2}r_{2})}{2r_1(1-2B_{21}r_{2}+3r_{2}^{2})},
\end{flalign}
with $r_{2}$ and its time derivative given in terms of $r_{1}$ and
$r_{1}^{\prime}$ as 
\begin{flalign}
r_{2}^{\pm}=\frac{\pm\sqrt{9+12B_{21}^{2}r_{1}^{4}}-3}{6B_{21}r_{1}^{2}},\ \ r_{2}^{\prime}=\frac{2r_2-6r_2^3}{r_1+3r_{1}r_{2}^{2}}r_{1}^{\prime},
\end{flalign}
where $r_{2}^{\pm}$ are again the two roots of \cref{eq:T2-scalefactorrelation}.
The root $r_{2}^{-}$ leads to sulutions with $\Omega_\text{m}>1$ and $w_\text{eff}>0$ always, and we therefore focus only on $r_{2}^{+}$ when rewriting $r_{2}$ in terms of $r_{1}$ in the following expressions.
The model possesses one fixed point which is given by 
\begin{flalign}
r_{1}^{\text{fix}} =\sqrt{\frac{3}{3-B_{21}^2}}.
\end{flalign}
We have to distinguish the three cases (a) $B_{21}<\sqrt{3}$, (b) $B_{21}=\sqrt{3}$, and (c) $B_{21}>\sqrt{3}$.
The matter density parameter and the effective equation of state are
\begin{flalign}
\Omega_{\text{m}} & =1-\frac{r_{1}^{2}}{1+B_{21}r_{1}^{2}r_{2}},\\
w_{\text{eff}} & =-\frac{(2r_{1}^{\prime}+3r_{1})r_{1}}{3+3B_{21}r_{1}^{2}r_{2}}.
\end{flalign}

\begin{figure}
\begin{centering}
\includegraphics[width=0.4\linewidth]{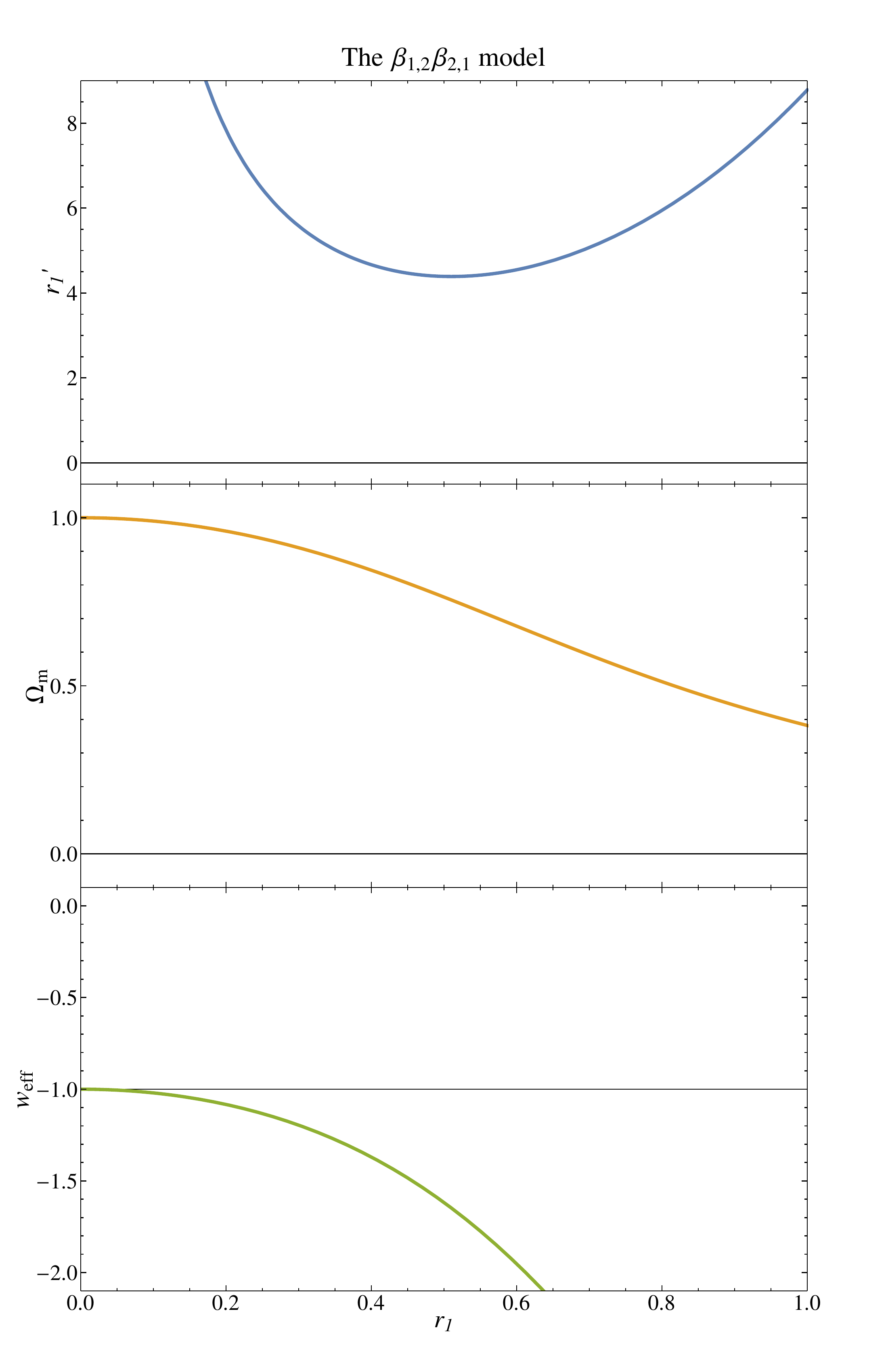}\includegraphics[width=0.4\linewidth]{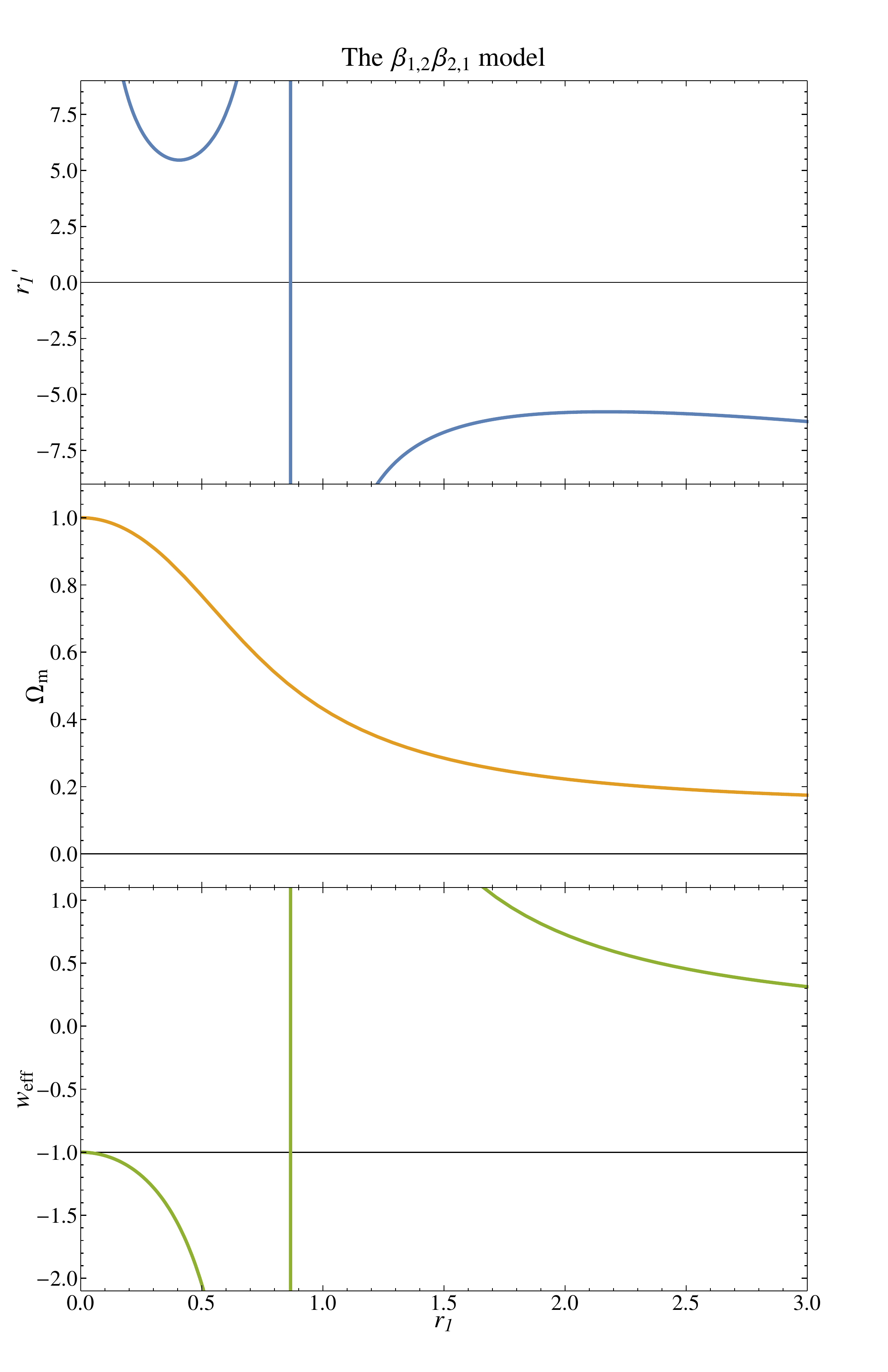} 
\par\end{centering}

\protect\protect\caption{\label{fig:T2-21b} \label{fig:T2-21c}Evolution of $r_{1}^{\prime}$, $\Omega_{\text{m}}$, and $w_\text{eff}$ as functions of $r_{1}$ for the $\beta_{1,2}\beta_{2,1}$
model of path trigravity with the interaction parameter ratios of $B_{21}=\sqrt{3}$ (left panel) representing case (b)
and $B_{21}=2$ (right panel) representing case (c).}
\end{figure}

$\bullet$ \textbf{Case (a):}
As presented in \cref{fig:T2-21a} (right panel) for the representative value $B_{21}=1$, the model contains a finite and an infinite branch.
The infinite branch $[r_{1}^{\text{fix}},\infty]$ is not viable as $\Omega_{\text{m}}<0$ always.
On the finite branch $[0,r_{1}^{\text{fix}}]$, the scale factor ratio $r_{1}$ increases starting at the singular point $r_{1}=0$.
The matter density parameter decreases from $1$ to $0$ as in standard cosmology, but the effective equation of state
is phantom for the entire evolution of the Universe, even during matter domination.
We therefore conclude that this case is not viable.

$\bullet$ \textbf{Case (b):}
As presented in \cref{fig:T2-21b} (left panel) for the value $B_{21}=\sqrt{3}$, the model contains only one finite branch $[0,\infty]$.
The scale factor ratio $r_{1}$ increases starting at the singular point $r_{1}=0$.
The matter density parameter decreases from $1$ to $0$ as in standard cosmology, but the effective equation of state
is phantom at all times.
Thus, this case does not have a viable cosmology.

$\bullet$ \textbf{Case (c):}
As presented in \cref{fig:T2-21c} (right panel) for the representative value $B_{21}=2$, the model contains a finite and an infinite branch, separated by a singular point.
While the effective equation of state is phantom on the finite branch during the entire evolution of the Universe, it is positive on the infinite branch.
Therefore, this case does not have a viable phenomenology.

\subsubsection{The $\beta_{1,2}\beta_{2,2}$ model}

The relation between $r_{2}$ and $r_1$ is given by 
\begin{flalign}
r_{2}^{\pm}=\pm\sqrt{1-\frac{1}{B_{22}r_1^2}},\ \ r_{2}^{\prime}=\frac{1-r_{2}^2}{r_{1}r_2}r_{1}^{\prime},
\end{flalign}
where both roots $r_{2}^{\pm}$ of \cref{eq:T2-scalefactorrelation} lead to the same results.
The evolution equation of $r_{1}$ is
\begin{flalign}
r_{1}^{\prime} & =-\frac{3}{2}r_1.
\end{flalign}
The only fixed point of the model can be read off as $r_{1}^{\text{fix}}=0$ and therefore there are no different cases in this model that we need to distinguish between.
The matter density parameter simplifies to  $\Omega_{\text{m}}=1-B_{22}^{-1}$ and thus is a constant that does not depend on $r_1$.
The effective equation of state is also a constant: $w_{\text{eff}}=0$.

Since the matter density parameter and the effective equation of state are both constants and do not depend on $r_1$, we can immediately conclude that this model is not viable. We therefore do not present its phase space.

\subsubsection{The $\beta_{1,2}\beta_{2,3}$ model}

The derivative of the scale factor ratio, \cref{eq:T2-derivative},
simplifies to 
\begin{flalign}
r_{1}^{\prime}=3\frac{(1-r_{2}^{2})(3-3r_{1}^{2}+B_{23}r_{1}^{2}r_{2}^{3})}{2r_{1}(3-3r_{2}^{2}+2B_{23}r_{2}^{3})}.
\end{flalign}
For this model, \cref{eq:T2-scalefactorrelation} leads to three different possible relations between $r_{1}$ and $r_{2}$ that are not redundant at the level of the Friedmann equations:
\begin{flalign}
r_{2}^\text{I}&=\left(\frac{\sqrt{9-4B_{23}^2 r_1^4}-3}{2B_{23}r_1^2}\right)^{-1/3}+\left(\frac{\sqrt{9-4B_{23}^2r_1^4}-3}{2B_{23}r_1^2}\right)^{1/3},\label{eq:T2-M23-scalefactorrelation}\\
r_{2}^\text{II}&=-\frac{1+i\sqrt{3}}{2}\left(\frac{\sqrt{9-4B_{23}^2 r_1^4}-3}{2B_{23}r_1^2}\right)^{-1/3}-\frac{1-i\sqrt{3}}{2}\left(\frac{\sqrt{9-4B_{23}^2r_1^4}-3}{2B_{23}r_1^2}\right)^{1/3},\\
r_{2}^\text{III}&=-\frac{1-i\sqrt{3}}{2}\left(\frac{\sqrt{9-4B_{23}^2 r_1^4}-3}{2B_{23}r_1^2}\right)^{-1/3}-\frac{1+i\sqrt{3}}{2}\left(\frac{\sqrt{9-4B_{23}^2r_1^4}-3}{2B_{23}r_1^2}\right)^{1/3}.
\end{flalign}
The quantity $r_2$ and its derivative are uniquely related as
\begin{flalign}
r_{2}^{\prime} & =\frac{2r_{2}(3-r_2^2)}{3r_{1}(r_{2}^{2}-1)}.
\end{flalign}
The matter density parameter and the effective equation of motion
are given by 
\begin{flalign}
\Omega_{\text{m}} & =1-\frac{3r_{1}^{2}}{3+B_{23}r_{1}^{2}r_{2}^{3}},\\
w_{\text{eff}} & =-\frac{(2r_{1}^{\prime}+3r_{1})r_{1}}{3+B_{23}r_{1}^{2}r_{2}^{3}}.
\end{flalign}

\begin{figure}
\begin{centering}
\includegraphics[width=0.4\linewidth]{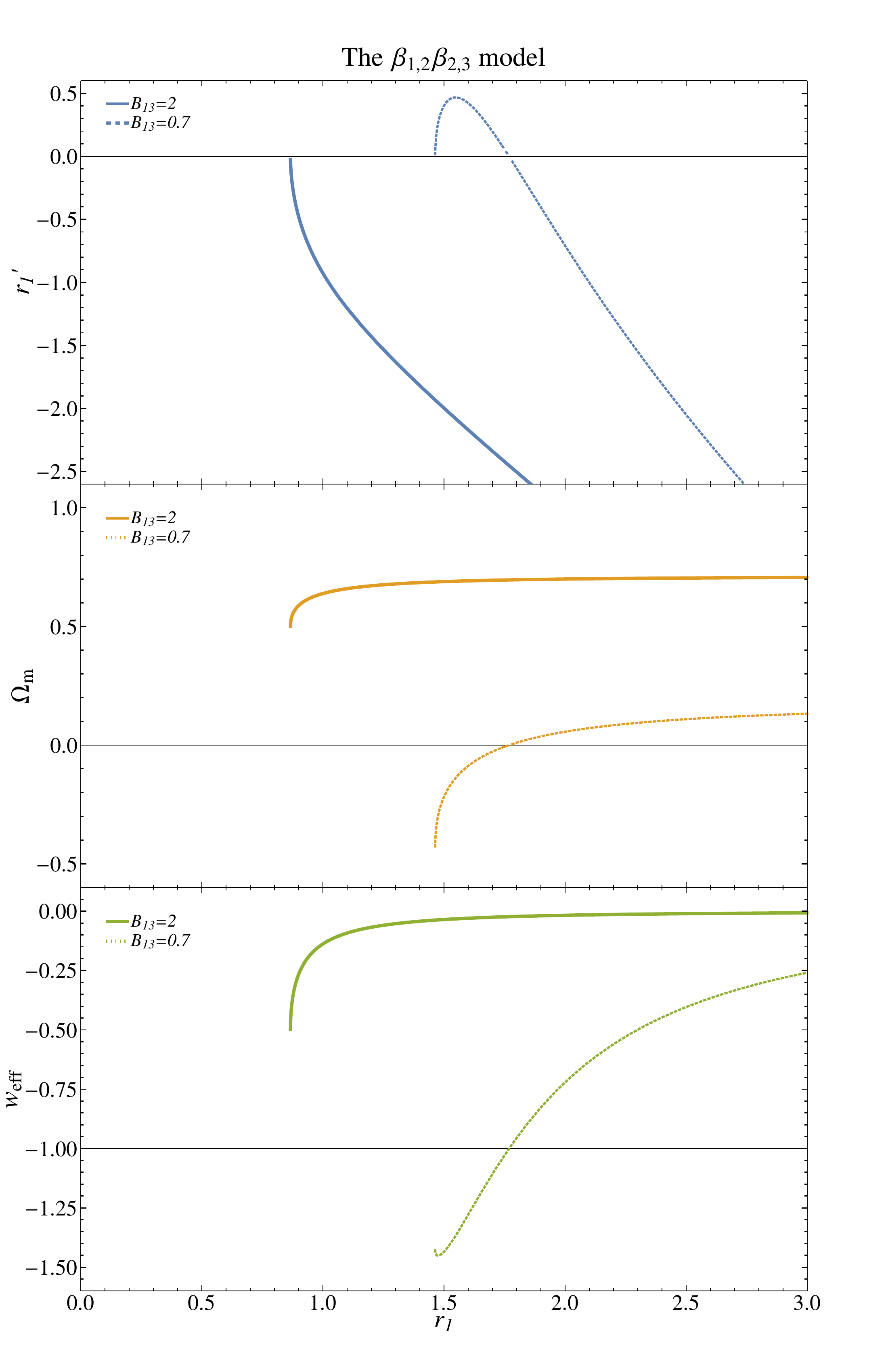} 
\par\end{centering}

\protect\protect\caption{\label{fig:T2-23}Evolution of $r_{1}^{\prime}$,
$\Omega_{\text{m}}$, and $w_{\text{eff}}$ as functions of $r_{1}$
for the $\beta_{1,2}\beta_{2,3}$ model of path trigravity with the
interaction parameter ratios of $B_{23}=2$ (solid lines) representing
case (Ia) and $B_{23}=0.7$ (dotted lines) representing case (Ib).}
\end{figure}

The root $r_2^\text{II}$ leads to $\Omega_\text{m}>1$ for any values of $B_{23}$ and $r_1$, and has therefore no viable phenomenology.
If we choose $r_2^\text{III}$ to relate the two scale factor ratios, the matter density is negative always, $\Omega_\text{m}<0$, leading to an unviable phenomenology.
These two cases are ruled and and we do not present their phase spaces.
We therefore focus only on $r_2^\text{I}$ relating $r_{2}$ and $r_{1}$.
We should distinguish between two different cases: (Ia) $B_{23}\gtrsim0.9$ with one fixed point $r_{1}^{\mathrm{fix,I}}$, and (Ib) $B_{23}\lesssim0.9$ with two fixed points $r_{1}^{\mathrm{fix,I}}$ and $r_{2}^{\mathrm{fix,II}}$.
The analytic expressions of the fixed points are quite lengthy and we do not present them here.
Moreover, for values $B_{23}\le0.5$, $r_1^\prime$ takes complex values and we are therefore limited to values of the interaction parameter ratio of $B_{23}>0.5$ in case (Ib); using \cref{eq:T2-M23-scalefactorrelation}, $r_{1}^{\prime}$ is real only in the interval $[0,r_{1}^{\mathrm{fix,I}}]$.

$\bullet$ \textbf{Case (Ia):} As presented in \cref{fig:T2-23} for the representative value of $B_{23}=2$, there is only an infinite branch $[r_{1}^{\text{fix,I}},\infty]$. For $r_{1}<r_{1}^{\text{fix,I}}$, $r_{1}^{\prime}$ takes complex values and thus the model's phase space is limited to the interval $[r_{1}^{\text{fix,I}},\infty]$.
This model yields a scaling solution with $\Omega_{\text{m}}^{\text{fin}}>0$ and $w_{\text{eff}}^{\text{fin}}>-1$.
This case is another example for a singular fixed point, discussed in \cref{sec:T2-M13}.
Since we cannot apply the method used in this paper to such cases, we leave a detailed analysis of this singular fixed point for future work.

$\bullet$ \textbf{Case (Ib):} As presented in \cref{fig:T2-23} for the representative value of $B_{23}=0.7$, there is one finite and one infinite branch. 
The infinite branch $[r_{1}^{\text{fix,II}},\infty]$ produces the standard phenomenology with $\Omega_{\text{m}}$ evolving from $1$ to $0$ and $w_{\text{eff}}$ evolving from $0$ to $-1$. 
The finite branch $[r_{1}^{\text{fix,I}},r_{1}^{\text{fix,II}}]$ is not viable because $\Omega_{\text{m}}<0$ and $w_{\text{eff}}<-1$ always.
For $r_{1}<r_{1}^{\text{fix,I}}$, $r_{1}^{\prime}$ takes complex values and thus the phase space is limited to the interval $[r_{1}^{\text{fix,I}},\infty]$.

\subsubsection{The $\beta_{1,3}\beta_{2,1}$ model}

For this model, the derivative of the scale factor ratio, \cref{eq:T2-derivative},
simplifies to 
\begin{flalign}	
r_{1}^{\prime}=\frac{r_1(3-r_1^2+3B_{31}r_1r_2)(1+3r_2^2)}{-1-4B_{31}r_1 r_2+r_1^2(1+3r_2^2)},
\end{flalign}
where $r_{2}$ and its derivative are given by 
\begin{flalign}
r_{2}^{\pm}&=\frac{\pm\sqrt{9+6r_1+r_1^2+12B_{31}r_1^2}-3-r_1}{6B_{31}r_{1}},\\
r_{2}^{\prime}&=-\frac{(3r_2-2B_{31}r_1+6B_{31}r_1r_2^2)r_2}{B_{31}r_1^2(1+3r_2^2)}r_{1}^{\prime},
\end{flalign}
with $r_{2}^{\pm}$ being the two roots of \cref{eq:T2-scalefactorrelation}.
The root $r_{2}^{-}$ does not lead to consistent results because $\Omega_\text{m}<0$ always, and we will therefore use only $r_{2}^{+}$ in our studies of the cosmological solutions for this model.
The model has two fixed points, $r_{1}^{\text{fix,I}}=0$ and $r_{1}^{\text{fix,II}}$, which exist for all values of $B_{31}$.
The analytic expression for $r_{1}^\text{fix,II}$ is quite lengthy and we do not present it here.
The matter density parameter and the effective equation of state are 
\begin{flalign}
\Omega_{\text{m}} & =1-\frac{r_{1}^{2}}{3+3B_{31}r_{1}r_{2}},\\
w_{\text{eff}} & =-\frac{(r_{1}^{\prime}+r_{1})r_{1}}{3+3B_{31}r_{1}r_{2}}.
\end{flalign}

\begin{figure}
\centering{}\includegraphics[width=0.4\linewidth]{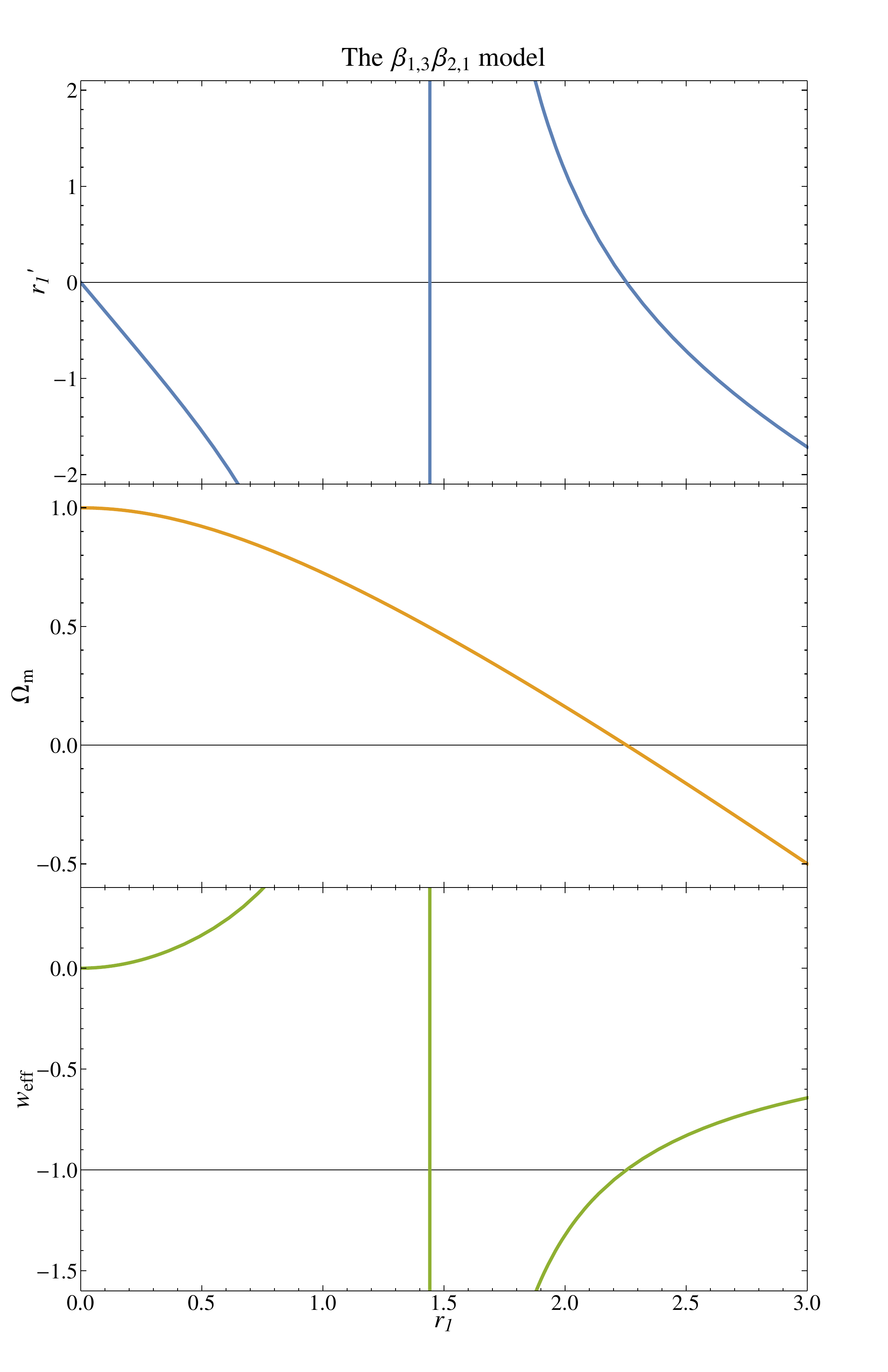}
\protect\protect\caption{\label{fig:T2-31}Evolution of $r_{1}^{\prime}$,
$\Omega_{\text{m}}$, and $w_{\text{eff}}$ as functions of $r_{1}$
for the $\beta_{1,3}\beta_{2,1}$ model of path trigravity with an
interaction parameter ratio of $B_{31}=1$.}
\end{figure}

As shown in \cref{fig:T2-31} for the representative
value $B_{31}=1$, there are three branches in this model. The infinite
branch $[r_{1}^{\text{fix,II}},\infty]$ is not viable as $\Omega_{\text{m}}<0$,
always. The finite branch $[0,r_{1}^{\text{sing}}]$ is not viable
either as $\Omega_{\text{m}}$ increases with time. In addition, $w_{\text{eff}}$
is always positive on this branch, which does not allow an accelerating
universe. On the finite branch $[r_{1}^{\text{sing}},r_{1}^{\text{fix,II}}]$,
the matter density parameter starts off with $\Omega_{\text{m}}^{\text{init}}<1$,
decreases in time, and vanishes in the infinite future, but the effective
equation of state is always phantom, rendering the model unviable.

\subsubsection{The $\beta_{1,3}\beta_{2,2}$ model}

The phase space of this model is described by 
\begin{flalign}
r_{1}^{\prime} & =3r_{1}^{-1}-r_{1}+3B_{32}r_{2}^{2},
\end{flalign}
where the two scale factor ratios $r_{1}$ and $r_{2}$, and their
time derivatives, are related via 
\begin{flalign}
r_{2}^{\pm}=\pm\sqrt{1-\frac{3+r_1}{3B_{32}r_1}},\ \ r_{2}^{\prime}=-\frac{1-2B_{32}r_{1}(1-r_{2}^{2})}{2B_{32}r_{1}^{2}r_{2}}r_{1}^{\prime}.\label{eq:T2-M32-scalefactorrelation}
\end{flalign}
Both roots $r_{2}^{\pm}$ of \cref{eq:T2-scalefactorrelation} lead
to the same results. The model has the fixed point 
\begin{flalign}
r_{1}^{\text{fix}}=3B_{32}-1,
\end{flalign}
such that we have to distinguish the cases $B_{32}>1/3$ with two fixed points $r_1^\text{fix}$ and $0$, and $B_{32}\le1/3$ with only one fixed point $0$.
The matter density parameter and the effective equation of state are 
\begin{flalign}
	\Omega_{\text{m}}&=1-\frac{r_{1}^{2}}{3+3B_{32}r_{1}r_{2}^{2}},\\
	w_\text{eff}&=-\frac{r_1^\prime+r_1}{3+3B_{32}r_1 r_2^2}.
\end{flalign}

We first rewrite $r_2$ in the expression of $\Omega_\text{m}$ using \cref{eq:T2-M32-scalefactorrelation} to get $\Omega_\text{m}=1-\frac{r_1}{3B_{32}-1}$.
We can infer that the values of the interaction parameter ratios $B_{32}=1/3$ and $B_{32}<1/3$ do not lead to viable phenomenologies.
For $B_{32}=1/3$ the matter density parameter is infinite and for $B_{32}<1/3$, it will be larger than $1$.
We therefore restrict our discussion to $B_{32}>1/3$.
The same procedure for the effective equation of state yields $w_\text{eff}=\frac{(1-B_{32})r_1}{(1-3B_{32})(2B_{32}-r_1)}$.
The value $B_{32}=1/3$ results in an infinite effective equation of state.
Additionally, the value $B_{32}=1$ is special because it leads to a constant effective equation of state $w_\text{eff}=0$.
Since we can already infer that $B_{32}\le1/3$ and $B_{32}=1$ do not yield viable phenomenologies, we do not present the corresponding phase spaces here.
We are left with two cases that need further investigation: (a) $B_{32}>1$ and (b) $1/3<B_{32}<1$.

\begin{figure}
\centering{}\includegraphics[width=0.4\linewidth]{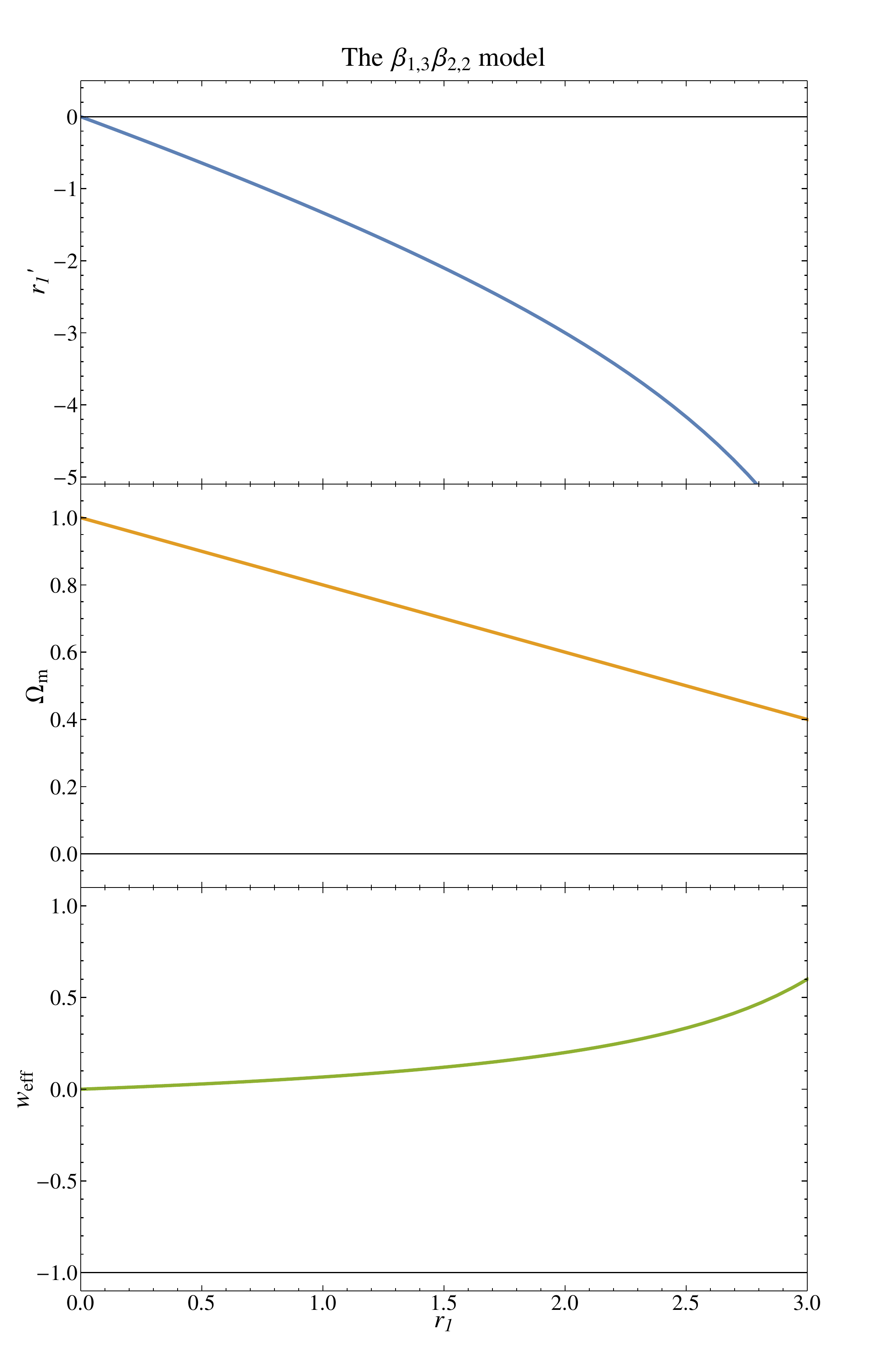}\includegraphics[width=0.4\linewidth]{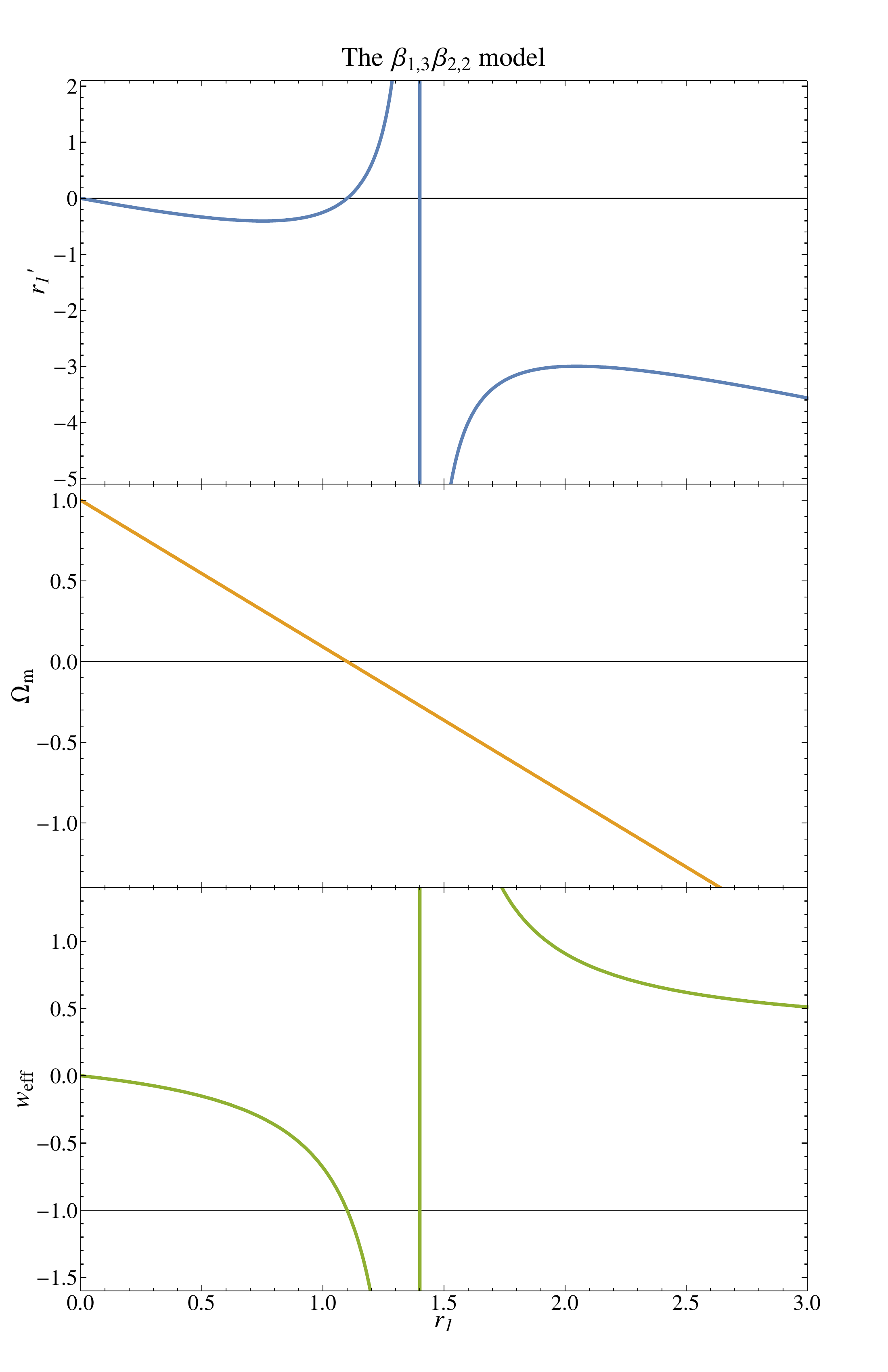}
\protect\protect\caption{\label{fig:T2-32}Evolution of $r_{1}^{\prime}$,
$\Omega_{\text{m}}$, and $w_{\text{eff}}$ as functions of $r_{1}$
for the $\beta_{1,3}\beta_{2,2}$ model of path trigravity with an
interaction parameter ratio of $B_{32}=2$ (left panel) representing case (a) and $B_{32}=0.7$ (right panel) representing case (b).}
\end{figure}

$\bullet$ \textbf{Case (a):} As \cref{fig:T2-32} (left panel) shows for the representative value of $B_{32}=2$, the model has only an infinite branch $[0,\infty]$.
The branch is not viable as $\Omega_{\text{m}}$ increases in time and $w_\text{eff}$ is positive during the entire evolution of the Universe.

$\bullet$ \textbf{Case (b):} As \cref{fig:T2-32} (right panel) shows for the representative value of $B_{32}=0.7$, the model has two finite and an infinite branch.
The finite branch $[0,r_1^\text{fix}]$ is not viable as $\Omega_{\text{m}}$ and $w_\text{eff}$ increase in time.
The finite branch $[r_1^\text{fix},r_1^\text{sing}]$ and the infinite branch $[r_1^\text{sing},\infty]$ are both unviable because the matter density parameter is always negative.
Therefore, this case does not have a viable cosmology.

\subsubsection{The $\beta_{1,3}\beta_{2,3}$ model}

For this model, the derivative of the scale factor ratio, \cref{eq:T2-derivative}, simplifies to 
\begin{flalign}
	r_{1}^{\prime}=\frac{3r_{1}(1-r_{2}^{2})(3-r_{1}^{2}+B_{33}r_{1}r_{2}^{3})}{-3+4r_{1}r_2^3+3r_{1}^{2}-3r_1^2r_2^2}.
\end{flalign}
There are three different solutions to \cref{eq:T2-scalefactorrelation}, given by
\begin{flalign}
	r_{2}^\text{I}=&\left(\frac{\sqrt{9+6r_1+r_1^2-4B_{33}^2 r_1^2}-3-r1}{2B_{33}r_1}\right)^{\frac{1}{3}}+\left(\frac{\sqrt{9+6r_1+r_1^2-4B_{33}^2 r_1^2}-3-r1}{2B_{33}r_1}\right)^{-\frac{1}{3}},\label{eq:T2-M33-scalefactorrelation}\\
	r_{2}^\text{II}=&-\frac{1-i\sqrt{3}}{2}\left(\frac{\sqrt{9+6r_1+r_1^2-4B_{33}^2 r_1^2}-3-r1}{2B_{33}r_1}\right)^{\frac{1}{3}}\nonumber\\
	&-\frac{1+i\sqrt{3}}{2}\left(\frac{\sqrt{9+6r_1+r_1^2-4B_{33}^2 r_1^2}-3-r1}{2B_{33}r_1}\right)^{-\frac{1}{3}},\\
	r_{2}^\text{III}=&-\frac{1+i\sqrt{3}}{2}\left(\frac{\sqrt{9+6r_1+r_1^2-4B_{33}^2 r_1^2}-3-r1}{2B_{33}r_1}\right)^{\frac{1}{3}}\nonumber\\
	&-\frac{1-i\sqrt{3}}{2}\left(\frac{\sqrt{9+6r_1+r_1^2-4B_{33}^2 r_1^2}-3-r1}{2B_{33}r_1}\right)^{-\frac{1}{3}}.
\end{flalign}
The derivatives are uniquely related via
\begin{flalign}
r_{2}^{\prime} & =\frac{3+2B_{33}r_{1}r_{2}(3-r_2^2)}{3B_{33}r_1^2(r_2^2-1)}r_{1}^{\prime}.
\end{flalign}
The matter density parameter and the effective equation of state are given by 
\begin{flalign}
\Omega_{\text{m}} & =1-\frac{r_{1}^{2}}{3+B_{33}r_{1}r_{2}^{3}},\\
w_{\text{eff}} & =-\frac{(r_{1}^{\prime}+r_{1})r_{1}}{3+B_{33}r_{1}r_{2}^{3}}.
\end{flalign}

Let us first focus on the roots $r_2^\text{II}$ and $r_2^\text{III}$ to relate the two scale factor ratios.
While the root $r_2^\text{II}$ leads to a matter density parameter with $\Omega_\text{m}>1$ always, the root $r_2^\text{III}$ leads to a phantom equation of state with $w_\text{eff}<-1$ always, for any values of $B_{33}$ and $r_1$.
We thus conclude the phenomenology of this model is not viable, if we choose $r_2^\text{II}$ or $r_2^\text{III}$ to relate the scale factor ratios.

Let us now turn to the root $r_2^\text{I}$ in order to relate $r_2$ and $r_1$.
The phase space of this model is quite complex and there exist numerous cases we have to distinguish, but our analysis shows that none of the cases and branches lead to a viable phenomenology as we discuss now without presenting the corresponding phase spaces.
For values $B_{33}\lesssim1.08$ the matter density parameter will be negative at all times, i.e., $\Omega_\text{m}\lesssim0$.
In the interval $1.08\lesssim B_{33}\lesssim1.13$ there is an intermediate finite branch where the matter density parameter starts at $0$, takes a maximum value of $\lesssim0.13$ and decreases to $0$ again, i.e., there is no matter-dominated era.
The other branches have $\Omega_\text{m}<0$.
For values $B_{33}\gtrsim1.13$, the matter density parameter is either negative or increasing in time.
In addition, the phase space contains a singular fixed point that requires further investigation.
To summarize, this model does not have a viable phenomenology.

\subsubsection{Summary}

We now summarize the phenomenology of the $1+1$-parameter models of
path trigravity. We give an overview of our results in \cref{tab:T2-summary},
where we briefly describe the behavior of $\Omega_{\text{m}}$ and
$w_{\text{eff}}$ for different models, their cases and branches.

A $\checkmark$ in the matter density parameter column means that
$\Omega_{\text{m}}$ starts off with an initial value $\Omega_{\text{m}}^{\text{init}}=1$
and decreases monotonically with time to the final value $\Omega_{\text{m}}^{\text{fin}}=0$,
i.e., as in standard $\Lambda$CDM cosmology. A $\checkmark$ in the
effective equation of state column means that $w_{\text{eff}}$ starts
off with the initial value $w_{\text{eff}}^{\text{init}}=0$ at early
times (matter-dominated epoch) and decreases to the final value $w_{\text{eff}}^{\text{fin}}=-1$
at late times, again as in $\Lambda$CDM. Otherwise, if $\Omega_{\text{m}}$
and/or $w_{\text{eff}}$ do not behave as in standard cosmology, we
briefly describe their behavior, and point out whether/why the phenomenology
of the model/branch is new or unviable.

As summarized in the table, we are left with four viable models:
\begin{itemize}
\item The $\beta_{1,1}\beta_{2,1}$-model infinite branches for the cases $B_{11}<3/2$ and $B_{11}=3/2$, as well as the finite branch of the case $B_{11}<3/2$, have fulfilled our viability criteria and have standard phenomenologies.
	The case $B_{11}=3/2$ has a viable finite branch with a phantom effective equation of state at late times and thus leads to a new phenomenology.
	Note that in the case $B_{11}=3/2$ the ratio of the interaction parameters is not a free parameter, and is fixed.
	In the case $B_{11}>3/2$ we find a finite and an infinite branch, both giving rise to new phenomenology.
	The final value of the matter density parameter is larger than $0$, such that in both branches the model does not approach a de Sitter point in the infinite future, resulting in the so-called scaling solutions.
	However, the effective equation of state will be singular at late times for both branches, but the initial conditions and $B_{11}$ can be chosen in such a way that the present time value of $w_\text{eff}$ is still consistent with observations.
	Additionally, this case predicts a singular point in the (near) future.
	One needs to systematically perform a statistical analysis and compare the model's predictions to data in order to be able to either finally rule this model out or make it a distinguishable alternative to $\Lambda$CDM.
\item The $\beta_{1,1}\beta_{2,2}$ model has an infinite branch with a standard phenomenology for any value of $B_{12}$.
\item The $\beta_{1,1}\beta_{2,3}$ model has two cases depending on the value of $B_{13}$.
	For the case $B_{13}\lesssim1.12$, the model has an infinite branch with standard phenomenology.
	For values $B_{13}\gtrsim1.12$ the phase space contains a singular fixed point.
	With the analysis performed in this paper it is not possible to make predictions about what will happen at such singular fixed points and thus we cannot rule out this model.
	In order to analyze the model further, one can for example treat both $r_1$ and $r_2$ as dynamical variables subject to the Friedmann equations and then analyze the full $2$-dimensional phase space.
\item For the $\beta_{1,2}\beta_{2,3}$ model, we need to distinguish between two cases depending on the value of $B_{23}$.
	For $B_{23}\lesssim0.9$ the phase space contains an infinite branch with standard phenomenology.
	In the case $B_{23}\gtrsim0.9$ the phase space contains a singular fixed point.
	In order to be able to rule out this case, one needs to perform a different analysis than the one performed in this paper. 
\end{itemize}

We have therefore found a number of models that produce the standard phenomenology.
That does not mean, however, that the phenomenologies of these models are completely indistinguishable from $\Lambda$CDM.
In order to find out whether these models are able to explain the late-time accelerated expansion of the Universe, one needs to perform a statistical analysis, comparing the model's predictions to observations.
Of course, the same needs to be done for the models with new phenomenology.

\begin{table}[H]
\begin{tabular}{lllp{5cm}p{5cm}l}
\toprule[0.5mm] 

& & & \multicolumn{2}{c}{{\bf Viability criterion}} & \tabularnewline
\cmidrule{4-5}
{\bf Model} & {\bf Case} & {\bf Branch} & $\Omega_{\text{m}}$ & $w_{\text{eff}}$ & {\bf Phenomenology}\tabularnewline
\midrule[0.5mm]

\multirow{10}{*}{$\beta_{1,1}\beta_{2,1}$} & \multirow{2}{*}{$B_{11}>3/2$} & Finite & $\Omega_\text{m}^\text{fin}>0$ & $w_\text{eff}^\text{fin}\rightarrow-\infty$ & New\tabularnewline
\cmidrule{3-6} 
& & Infinite & $\Omega_\text{m}^\text{fin}>0$ & $w_\text{eff}^\text{fin}\rightarrow\infty$ & New\tabularnewline
\cmidrule{2-6}
 & \multirow{2}{*}{$B_{11}=3/2$} & Finite & $\checkmark$ & $w_\text{eff}^\text{late}<-1$ & New\tabularnewline
 \cmidrule{3-6}
 & & Infinite & $\checkmark$ & $\checkmark$ & Standard\tabularnewline
\cmidrule{2-6} 
 & \multirow{5}{*}{$B_{11}<3/2$} & Finite & $\checkmark$  & $\checkmark$ & Standard\tabularnewline
 \cmidrule{3-6}
 & & $[r_{1}^{\text{fix}-},r_{1}^{\text{sing}}]$ & $\Omega_\text{m}<0$  & $w_\text{eff}^\text{init}\rightarrow\infty$ & Unviable\tabularnewline
 \cmidrule{3-6}
 & & $[r_{1}^{\text{sing}},r_{1}^{\text{fix}+}]$ & $\Omega_\text{m}<0$  & Phantom & Unviable\tabularnewline
 \cmidrule{3-6}
 & & Infinite & $\checkmark$  & $\checkmark$ & Standard\tabularnewline
\midrule[0.3mm]

\multirow{3}{*}{$\beta_{1,1}\beta_{2,2}$} & & $[0,r_1^\text{sing}]$ & $\Omega_\text{m}<0$ & $w_\text{eff}>0$ & Unviable\tabularnewline
\cmidrule{3-6}
& & $[r_1^\text{sing},r_1^\text{fix}]$ & $\Omega_\text{m}<0$  & Phantom & Unviable\tabularnewline
\cmidrule{3-6}
& & Infinite & $\checkmark$ & $\checkmark$ & Standard\tabularnewline
\midrule[0.3mm]

\multirow{4}{*}{$\beta_{1,1}\beta_{2,3}$} & $B_{13}\gtrsim1.12$ & Infinite & $\Omega_\text{m}^\text{fin}>0$ & $w_\text{eff}^\text{fin}>-1$ & New\tabularnewline
\cmidrule{2-6}
& \multirow{2}{*}{$B_{13}\lesssim1.12$} & Finite & $\Omega_\text{m}<0$ & Phantom & Unviable\tabularnewline
\cmidrule{3-6}
& & Infinite & $\checkmark$ & $\checkmark$ & Standard\tabularnewline
\midrule[0.3mm]

\multirow{6}{*}{$\beta_{1,2}\beta_{2,1}$} & \multirow{2}{*}{$B_{21}<\sqrt{3}$} & Finite &$\checkmark$ & Phantom & Unviable\tabularnewline
\cmidrule{3-6}
& & Infinite & $\Omega_\text{m}<0$ & $\checkmark$ & Unviable\tabularnewline
\cmidrule{2-6}
& $B_{21}=\sqrt{3}$ & Finite & $\checkmark$ & Phantom & Unviable\tabularnewline
\cmidrule{2-6}
& \multirow{2}{*}{$B_{21}>\sqrt{3}$} & Finite & $\Omega_\text{m}^\text{fin}>0$ & Phantom & Unviable\tabularnewline
\cmidrule{3-6}
& & Infinite & $\Omega_\text{m}^\text{init}<1$, increasing & $w_\text{eff}>0$ & Unviable\tabularnewline
\midrule[0.3mm]

$\beta_{1,2}\beta_{2,2}$ & & Finite & Constant & Constant & Unviable\tabularnewline
\midrule[0.3mm]

\multirow{3}{*}{$\beta_{1,2}\beta_{2,3}$} & $B_{23}\gtrsim0.9$ & Infinite & $\Omega_\text{m}^\text{fin}>0$ & $w_\text{eff}^\text{fin}>-1$ & New\tabularnewline
\cmidrule{2-6} 
& \multirow{2}{*}{$B_{23}\lesssim0.9$} & Finite & $\Omega_\text{m}<0$ & Phantom & Unviable\tabularnewline
\cmidrule{3-6} 
& & Ininite & $\checkmark$ & $\checkmark$ & Standard\tabularnewline
\midrule[0.3mm]

\multirow{3}{*}{$\beta_{1,3}\beta_{2,1}$} & & $[0,r_{1}^{\text{sing}}]$ & Increasing & $w_\text{eff}>0$ & Unviable\tabularnewline
\cmidrule{3-6} 
 & & $[r_{1}^{\text{sing}},r_{1}^{\text{fix}}]$ & $\Omega_{\text{m}}^{\text{init}}<1$ & Phantom & Unviable\tabularnewline
 \cmidrule{3-6} 
 & & Infinite & $\Omega_\text{m}<0$ & $\checkmark$ & Unviable\tabularnewline
\midrule[0.3mm]

\multirow{4}{*}{$\beta_{1,3}\beta_{2,2}$} & $B_{32}>1$ & Infinite & Increasing & $w_\text{eff}>0$ & Unviable\tabularnewline
\cmidrule{2-6} 
 & \multirow{4}{*}{$1/3<B_{32}<1$}& $[0,r_{1}^{\text{sing}}]$ & Increasing & Increasing & Unviable\tabularnewline
 \cmidrule{3-6} 
 & & $[r_{1}^{\text{sing}},r_{1}^{\text{fix}}]$ & $\Omega_\text{m}<0$ & Phantom & Unviable\tabularnewline
 \cmidrule{3-6} 
 & & Infinite & $\Omega_\text{m}<0$ & $w_\text{eff}>0$ & Unviable\tabularnewline
 \cmidrule{2-6} 
 & $B_{32}=1$ &  & &Constant: $w_\text{eff}=0$ & Unviable\tabularnewline
 \cmidrule{2-6} 
 & $B_{32}\le 1/3$ & & $\Omega_\text{m}>1$ or $\Omega_\text{m}=\infty$ & & Unviable\tabularnewline
\midrule[0.3mm]
$\beta_{1,3}\beta_{2,3}$ & & & & & Unviable\tabularnewline
\bottomrule[0.5mm]
\end{tabular}
\protect\caption{\label{tab:T2-summary}An overview of the cosmological viability of different $1+1$-parameter models in path trigravity.
We consider different branches for different cases in each model.}
\end{table}

\section{Novel phenomenology}
\label{sec:novelpheno}

Let us now take a closer look at the results of our analysis summarized
in \cref{tab:T1-summary,tab:T2-summary}. We have identified in
total one class of models in star trigravity and three in path trigravity
which a) are not immediately ruled out by observations, and b) possess
some new phenomenology (as defined in \cref{sec:cos-11params})
as far as the background expansion is concerned. These are the $\beta_{1,1}\beta_{2,3}$
star model in one particular branch (what we called intermediate finite
branch), as well as the path models $\beta_{1,1}\beta_{2,1}$ with
$B_{11}\ge1.5$, $\beta_{1,1}\beta_{2,3}$
with $B_{13}\gtrsim1.12$, and $\beta_{1,2}\beta_{2,3}$ with $B_{23}\gtrsim0.9$.

The first case, i.e., the $\beta_{1,1}\beta_{2,3}$ model of star
trigravity, has a matter density parameter $\Omega_{\text{m}}$ that
begins with a finite value in the past at $r_{1}=r_{1}^{\text{sing}}$
and ends up in a de Sitter state with $r_{1}=r_{1}^{\text{fix+}}$
and $\Omega_{\text{m}}=0,$ passing through a maximum value. At the
same time the effective equation of state $w_{\text{eff}}$ decreases
monotonically from infinity to $-1$. By adjusting the value of $B_{13}$
one can obtain a phenomenology that resembles $\Lambda$CDM, moving
the singularity far into the past. Therefore, although phenomenologically
new, this evolution is actually viable only in the limit in which
the model is indistinguishable from $\Lambda$CDM.

The $\beta_{1,1}\beta_{2,1}$ model of path trigravity has two cases giving rise to new phenomenology.
	For $B_{11}>1.5$ the phase space of the model contains one finite and one infinite branch, separated by a singular point.
	The past evolution is similar to the standard one on both branches with $\Omega_\text{m}^\text{init}=1$ and $w_\text{eff}^\text{init}=0$.
	Both branches contain a singularity that can be moved to the future.
	One can adjust the initial conditions and $B_{11}$ so as to get $\Omega_{\text{m},0}\approx0.3$ and $w_{\text{eff},0}\approx-0.9$ at present time, although rapidly varying with time.
	The evolution is therefore not standard and might be better constrained, or finally ruled out, by a full comparison with observational data.
	The other case with $B_{11}=1.5$ is particularly interesting because now the evolution on the finite branch has no singularities and no obvious problems.
	The ratio of the interaction parameters is not a free parameter in this case, i.e., the model has only one free parameter as in $\Lambda$CDM (either $\beta_{1,1}$ is a free parameter and $\beta_{2,1}$ is fixed or the other way around).
	This model predicts an effective equation of state smaller than $-1$, i.e., of phantom type, at late times.
	The asymptotic value is $w_\text{eff}^\text{fin}=-1$.
	Initial conditions can be adjusted so as to have $w_\text{eff}\approx-0.9$ today.
	It is possible to have a viable cosmology with a phantom crossing, i.e., evolution of $w_\text{eff}$ from above to below $-1$, contrary to the bimetric case where viable models never cross $w_\mathrm{eff}=-1$ \cite{Konnig:2013gxa}.
	Again a careful comparison with observations can rule out or confirm the validity of this case.
	
	The path models $\beta_{1,1}\beta_{2,3}$ with $B_{13}\gtrsim1.12$ and $\beta_{1,2}\beta_{2,3}$ with $B_{23}\gtrsim0.9$ have similar non-standard phenomenologies.
	The past evolution is standard for both models with $\Omega_\text{m}^\text{init}=1$ and $w_\text{eff}^\text{init}=0$, but the models evolve towards singular fixed points at which $\Omega_\text{m}>0$ and $w_\text{eff}>-1$.
	This would identify these solutions as scaling solutions.
	However, the singular fixed points are not real fixed points at which the evolution stops.
	In order to predict how the models will evolve when approaching the singular fixed points it might be necessary to analyze their $2$-dimensional phase space.

\section{Conclusions and outlook}

One of our goals in studying multimetric gravity is to find
non-trivial, consistent, simple, and viable alternatives to $\Lambda$CDM.
Such cosmologies should be clearly distinguishable from $\Lambda$CDM,
be free of ghosts and other instabilities, and possess a small number of free parameters.
They should also avoid obvious inconsistencies with observation which can arise in these theories, such as singularities in the
observable past, absence of late-time acceleration, and the nonexistence of a matter era. This
goal has not yet been reached with theories of massive and bimetric gravity. This failure has prompted us to
investigate in detail the cosmology of trimetric gravity in search of alternatives to $\Lambda$CDM. In particular, we
explored in some detail all possible forms of trimetric gravity with
two free interaction parameters (one for each pair of interacting metrics) and no cosmological constants---the minimal non-trivial models in this framework.
We have shown that the phase space of these
models in most cases is simple and $1$-dimensional, i.e., the equations for $r_{i}'(r)$ depend only on the ratio of
the two interaction parameters. For each
model, we discussed analytically and numerically the cosmic evolution
and determined whether it was compatible with the current understanding
of our Universe.

Our main result is that, in addition to many unviable cases, there are
a number of models in which the evolution is compatible with observations at the background level,
although most are practically indistinguishable from $\Lambda$CDM.
In fact, perhaps surprisingly, we find only
three cases that appear to be promising alternatives to
standard cosmology, all of which have the ``path'' configuration of interactions in which the two additional metrics couple to the physical spacetime metric but not to each other. The first
viable model is that in which only the couplings $\beta_{1,1}$ and $\beta_{2,1}$
are switched on; in particular, if the ratio of the coupling constants
is $1.5$, such that the model has only one free parameter, just like $\Lambda$CDM, we find a non-trivial evolution without obvious problems.
This case is interesting also because it crosses the phantom divide $w_{\mathrm{eff}}=-1$, contrary to what happens in bimetric models. We have additionally found two cases with
scaling solutions, i.e., solutions which do not asymptote to de Sitter at late times. These models contain singular fixed points in the future with possibly interesting implications for cosmology. It is necessary to perform a more detailed ($2$-dimensional) phase-space analysis near the singular points in order to understand how these solutions would evolve when passing through the fixed points.
	
All the viable trimetric models that we have found in this work, with either new or standard phenomenology, deserve a more detailed treatment in terms
of both comparison to observational data and analysis of
perturbations. Naturally, the question of whether these models contain
instabilities of any kind is of particular importance, as even models with standard phenomenologies at the background level, i.e., the ones that behave similarly to $\Lambda$CDM or bigravity models, may very well behave differently at the perturbative level. In particular, bimetric models with viable backgrounds have been shown to contain instabilities at the level of linear perturbations. With the extra freedom afforded by trimetric gravity, we are optimistic about finding viable and stable alternatives to the standard cosmology within the framework of massive, multimetric theories of gravity. All these questions will be explored in future work.

\begin{acknowledgments}
We are grateful to Frank Könnig for helpful discussions. Y.A. and
L.A. acknowledge support from DFG through the TRR33 project ``The
Dark Universe.'' The work of A.R.S. is supported in part by US Department
of Energy (HEP) Award DE-SC0013528 and by funds provided by the University
of Pennsylvania. 
\end{acknowledgments}

\bibliography{bibliography}

\begin{thebibliography}{93}%
\makeatletter
\providecommand \@ifxundefined [1]{%
 \@ifx{#1\undefined}
}%
\providecommand \@ifnum [1]{%
 \ifnum #1\expandafter \@firstoftwo
 \else \expandafter \@secondoftwo
 \fi
}%
\providecommand \@ifx [1]{%
 \ifx #1\expandafter \@firstoftwo
 \else \expandafter \@secondoftwo
 \fi
}%
\providecommand \natexlab [1]{#1}%
\providecommand \enquote  [1]{``#1''}%
\providecommand \bibnamefont  [1]{#1}%
\providecommand \bibfnamefont [1]{#1}%
\providecommand \citenamefont [1]{#1}%
\providecommand \href@noop [0]{\@secondoftwo}%
\providecommand \href [0]{\begingroup \@sanitize@url \@href}%
\providecommand \@href[1]{\@@startlink{#1}\@@href}%
\providecommand \@@href[1]{\endgroup#1\@@endlink}%
\providecommand \@sanitize@url [0]{\catcode `\\12\catcode `\$12\catcode
  `\&12\catcode `\#12\catcode `\^12\catcode `\_12\catcode `\%12\relax}%
\providecommand \@@startlink[1]{}%
\providecommand \@@endlink[0]{}%
\providecommand \url  [0]{\begingroup\@sanitize@url \@url }%
\providecommand \@url [1]{\endgroup\@href {#1}{\urlprefix }}%
\providecommand \urlprefix  [0]{URL }%
\providecommand \Eprint [0]{\href }%
\providecommand \doibase [0]{http://dx.doi.org/}%
\providecommand \selectlanguage [0]{\@gobble}%
\providecommand \bibinfo  [0]{\@secondoftwo}%
\providecommand \bibfield  [0]{\@secondoftwo}%
\providecommand \translation [1]{[#1]}%
\providecommand \BibitemOpen [0]{}%
\providecommand \bibitemStop [0]{}%
\providecommand \bibitemNoStop [0]{.\EOS\space}%
\providecommand \EOS [0]{\spacefactor3000\relax}%
\providecommand \BibitemShut  [1]{\csname bibitem#1\endcsname}%
\let\auto@bib@innerbib\@empty
%</preamble>
\bibitem [{\citenamefont {Gupta}(1954)}]{Gupta:1954zz}%
  \BibitemOpen
  \bibfield  {author} {\bibinfo {author} {\bibfnamefont {Suraj~N.}\
  \bibnamefont {Gupta}},\ }\bibfield  {title} {\enquote {\bibinfo {title}
  {{Gravitation and Electromagnetism}},}\ }\href {\doibase
  10.1103/PhysRev.96.1683} {\bibfield  {journal} {\bibinfo  {journal}
  {Phys.Rev.}\ }\textbf {\bibinfo {volume} {96}},\ \bibinfo {pages}
  {1683--1685} (\bibinfo {year} {1954})}\BibitemShut {NoStop}%
%\%CITATION = PHRVA,96,1683;\%\%
\bibitem [{\citenamefont {Weinberg}(1965)}]{Weinberg:1965rz}%
  \BibitemOpen
  \bibfield  {author} {\bibinfo {author} {\bibfnamefont {Steven}\ \bibnamefont
  {Weinberg}},\ }\bibfield  {title} {\enquote {\bibinfo {title} {{Photons and
  gravitons in perturbation theory: Derivation of Maxwell's and Einstein's
  equations}},}\ }\href {\doibase 10.1103/PhysRev.138.B988} {\bibfield
  {journal} {\bibinfo  {journal} {Phys.Rev.}\ }\textbf {\bibinfo {volume}
  {138}},\ \bibinfo {pages} {B988--B1002} (\bibinfo {year} {1965})}\BibitemShut
  {NoStop}%
%\%CITATION = PHRVA,138,B988;\%\%
\bibitem [{\citenamefont {Deser}(1970)}]{Deser:1969wk}%
  \BibitemOpen
  \bibfield  {author} {\bibinfo {author} {\bibfnamefont {Stanley}\ \bibnamefont
  {Deser}},\ }\bibfield  {title} {\enquote {\bibinfo {title} {{Selfinteraction
  and gauge invariance}},}\ }\href {\doibase 10.1007/BF00759198} {\bibfield
  {journal} {\bibinfo  {journal} {Gen.Rel.Grav.}\ }\textbf {\bibinfo {volume}
  {1}},\ \bibinfo {pages} {9--18} (\bibinfo {year} {1970})},\ \Eprint
  {http://arxiv.org/abs/gr-qc/0411023} {arXiv:gr-qc/0411023 [gr-qc]}
  \BibitemShut {NoStop}%
%\%CITATION = GR-QC/0411023;\%\%
\bibitem [{\citenamefont {Boulware}\ and\ \citenamefont
  {Deser}(1975)}]{Boulware:1974sr}%
  \BibitemOpen
  \bibfield  {author} {\bibinfo {author} {\bibfnamefont {David~G.}\
  \bibnamefont {Boulware}}\ and\ \bibinfo {author} {\bibfnamefont {Stanley}\
  \bibnamefont {Deser}},\ }\bibfield  {title} {\enquote {\bibinfo {title}
  {{Classical General Relativity Derived from Quantum Gravity}},}\ }\href
  {\doibase 10.1016/0003-4916(75)90302-4} {\bibfield  {journal} {\bibinfo
  {journal} {Annals Phys.}\ }\textbf {\bibinfo {volume} {89}},\ \bibinfo
  {pages} {193} (\bibinfo {year} {1975})}\BibitemShut {NoStop}%
%\%CITATION = APNYA,89,193;\%\%
\bibitem [{\citenamefont {Feynman}(1996)}]{Feynman:1996kb}%
  \BibitemOpen
  \bibfield  {author} {\bibinfo {author} {\bibfnamefont {R.P.}\ \bibnamefont
  {Feynman}},\ }\href@noop {} {\emph {\bibinfo {title} {{Feynman lectures on
  gravitation}}}},\ edited by\ \bibinfo {editor} {\bibfnamefont {F.B.}\
  \bibnamefont {Morinigo}}, \bibinfo {editor} {\bibfnamefont {W.G.}\
  \bibnamefont {Wagner}}, \ and\ \bibinfo {editor} {\bibfnamefont
  {B.}~\bibnamefont {Hatfield}}\ (\bibinfo  {publisher} {{Addison-Wesley}},\
  \bibinfo {year} {1996})\BibitemShut {NoStop}%
%\%CITATION = INSPIRE-427379;\%\%
\bibitem [{\citenamefont {Boulware}\ and\ \citenamefont
  {Deser}(1972)}]{Boulware:1973my}%
  \BibitemOpen
  \bibfield  {author} {\bibinfo {author} {\bibfnamefont {D.G.}\ \bibnamefont
  {Boulware}}\ and\ \bibinfo {author} {\bibfnamefont {Stanley}\ \bibnamefont
  {Deser}},\ }\bibfield  {title} {\enquote {\bibinfo {title} {{Can gravitation
  have a finite range?}}}\ }\href {\doibase 10.1103/PhysRevD.6.3368} {\bibfield
   {journal} {\bibinfo  {journal} {Phys.Rev.}\ }\textbf {\bibinfo {volume}
  {D6}},\ \bibinfo {pages} {3368--3382} (\bibinfo {year} {1972})}\BibitemShut
  {NoStop}%
%%CITATION = PHRVA,D6,3368;%%
\bibitem [{\citenamefont {Fierz}\ and\ \citenamefont
  {Pauli}(1939)}]{Fierz:1939ix}%
  \BibitemOpen
  \bibfield  {author} {\bibinfo {author} {\bibfnamefont {M.}~\bibnamefont
  {Fierz}}\ and\ \bibinfo {author} {\bibfnamefont {W.}~\bibnamefont {Pauli}},\
  }\bibfield  {title} {\enquote {\bibinfo {title} {{On relativistic wave
  equations for particles of arbitrary spin in an electromagnetic field}},}\
  }\href {\doibase 10.1098/rspa.1939.0140} {\bibfield  {journal} {\bibinfo
  {journal} {Proc.Roy.Soc.Lond.}\ }\textbf {\bibinfo {volume} {A173}},\
  \bibinfo {pages} {211--232} (\bibinfo {year} {1939})}\BibitemShut {NoStop}%
%%CITATION = PRSLA,A173,211;%%
\bibitem [{\citenamefont {Arkani-Hamed}\ \emph {et~al.}(2003)\citenamefont
  {Arkani-Hamed}, \citenamefont {Georgi},\ and\ \citenamefont
  {Schwartz}}]{ArkaniHamed:2002sp}%
  \BibitemOpen
  \bibfield  {author} {\bibinfo {author} {\bibfnamefont {Nima}\ \bibnamefont
  {Arkani-Hamed}}, \bibinfo {author} {\bibfnamefont {Howard}\ \bibnamefont
  {Georgi}}, \ and\ \bibinfo {author} {\bibfnamefont {Matthew~D.}\ \bibnamefont
  {Schwartz}},\ }\bibfield  {title} {\enquote {\bibinfo {title} {{Effective
  field theory for massive gravitons and gravity in theory space}},}\ }\href
  {\doibase 10.1016/S0003-4916(03)00068-X} {\bibfield  {journal} {\bibinfo
  {journal} {Annals Phys.}\ }\textbf {\bibinfo {volume} {305}},\ \bibinfo
  {pages} {96--118} (\bibinfo {year} {2003})},\ \Eprint
  {http://arxiv.org/abs/hep-th/0210184} {arXiv:hep-th/0210184 [hep-th]}
  \BibitemShut {NoStop}%
%%CITATION = HEP-TH/0210184;%%
\bibitem [{\citenamefont {Creminelli}\ \emph {et~al.}(2005)\citenamefont
  {Creminelli}, \citenamefont {Nicolis}, \citenamefont {Papucci},\ and\
  \citenamefont {Trincherini}}]{Creminelli:2005qk}%
  \BibitemOpen
  \bibfield  {author} {\bibinfo {author} {\bibfnamefont {Paolo}\ \bibnamefont
  {Creminelli}}, \bibinfo {author} {\bibfnamefont {Alberto}\ \bibnamefont
  {Nicolis}}, \bibinfo {author} {\bibfnamefont {Michele}\ \bibnamefont
  {Papucci}}, \ and\ \bibinfo {author} {\bibfnamefont {Enrico}\ \bibnamefont
  {Trincherini}},\ }\bibfield  {title} {\enquote {\bibinfo {title} {{Ghosts in
  massive gravity}},}\ }\href {\doibase 10.1088/1126-6708/2005/09/003}
  {\bibfield  {journal} {\bibinfo  {journal} {JHEP}\ }\textbf {\bibinfo
  {volume} {0509}},\ \bibinfo {pages} {003} (\bibinfo {year} {2005})},\ \Eprint
  {http://arxiv.org/abs/hep-th/0505147} {arXiv:hep-th/0505147 [hep-th]}
  \BibitemShut {NoStop}%
%%CITATION = HEP-TH/0505147;%%
\bibitem [{\citenamefont {de~Rham}\ and\ \citenamefont
  {Gabadadze}(2010)}]{deRham:2010ik}%
  \BibitemOpen
  \bibfield  {author} {\bibinfo {author} {\bibfnamefont {Claudia}\ \bibnamefont
  {de~Rham}}\ and\ \bibinfo {author} {\bibfnamefont {Gregory}\ \bibnamefont
  {Gabadadze}},\ }\bibfield  {title} {\enquote {\bibinfo {title}
  {{Generalization of the Fierz-Pauli Action}},}\ }\href {\doibase
  10.1103/PhysRevD.82.044020} {\bibfield  {journal} {\bibinfo  {journal}
  {Phys.Rev.}\ }\textbf {\bibinfo {volume} {D82}},\ \bibinfo {pages} {044020}
  (\bibinfo {year} {2010})},\ \Eprint {http://arxiv.org/abs/1007.0443}
  {arXiv:1007.0443 [hep-th]} \BibitemShut {NoStop}%
%%CITATION = ARXIV:1007.0443;%%
\bibitem [{\citenamefont {de~Rham}\ \emph
  {et~al.}(2011{\natexlab{a}})\citenamefont {de~Rham}, \citenamefont
  {Gabadadze},\ and\ \citenamefont {Tolley}}]{deRham:2010kj}%
  \BibitemOpen
  \bibfield  {author} {\bibinfo {author} {\bibfnamefont {Claudia}\ \bibnamefont
  {de~Rham}}, \bibinfo {author} {\bibfnamefont {Gregory}\ \bibnamefont
  {Gabadadze}}, \ and\ \bibinfo {author} {\bibfnamefont {Andrew~J.}\
  \bibnamefont {Tolley}},\ }\bibfield  {title} {\enquote {\bibinfo {title}
  {{Resummation of Massive Gravity}},}\ }\href {\doibase
  10.1103/PhysRevLett.106.231101} {\bibfield  {journal} {\bibinfo  {journal}
  {Phys.Rev.Lett.}\ }\textbf {\bibinfo {volume} {106}},\ \bibinfo {pages}
  {231101} (\bibinfo {year} {2011}{\natexlab{a}})},\ \Eprint
  {http://arxiv.org/abs/1011.1232} {arXiv:1011.1232 [hep-th]} \BibitemShut
  {NoStop}%
%%CITATION = ARXIV:1011.1232;%%
\bibitem [{\citenamefont {Hassan}\ and\ \citenamefont
  {Rosen}(2011)}]{Hassan:2011vm}%
  \BibitemOpen
  \bibfield  {author} {\bibinfo {author} {\bibfnamefont {S.F.}\ \bibnamefont
  {Hassan}}\ and\ \bibinfo {author} {\bibfnamefont {Rachel~A.}\ \bibnamefont
  {Rosen}},\ }\bibfield  {title} {\enquote {\bibinfo {title} {{On Non-Linear
  Actions for Massive Gravity}},}\ }\href {\doibase 10.1007/JHEP07(2011)009}
  {\bibfield  {journal} {\bibinfo  {journal} {JHEP}\ }\textbf {\bibinfo
  {volume} {1107}},\ \bibinfo {pages} {009} (\bibinfo {year} {2011})},\ \Eprint
  {http://arxiv.org/abs/1103.6055} {arXiv:1103.6055 [hep-th]} \BibitemShut
  {NoStop}%
%%CITATION = ARXIV:1103.6055;%%
\bibitem [{\citenamefont {Hassan}\ and\ \citenamefont
  {Rosen}(2012{\natexlab{a}})}]{Hassan:2011hr}%
  \BibitemOpen
  \bibfield  {author} {\bibinfo {author} {\bibfnamefont {S.F.}\ \bibnamefont
  {Hassan}}\ and\ \bibinfo {author} {\bibfnamefont {Rachel~A.}\ \bibnamefont
  {Rosen}},\ }\bibfield  {title} {\enquote {\bibinfo {title} {{Resolving the
  Ghost Problem in non-Linear Massive Gravity}},}\ }\href {\doibase
  10.1103/PhysRevLett.108.041101} {\bibfield  {journal} {\bibinfo  {journal}
  {Phys.Rev.Lett.}\ }\textbf {\bibinfo {volume} {108}},\ \bibinfo {pages}
  {041101} (\bibinfo {year} {2012}{\natexlab{a}})},\ \Eprint
  {http://arxiv.org/abs/1106.3344} {arXiv:1106.3344 [hep-th]} \BibitemShut
  {NoStop}%
%%CITATION = ARXIV:1106.3344;%%
\bibitem [{\citenamefont {de~Rham}\ \emph {et~al.}(2012)\citenamefont
  {de~Rham}, \citenamefont {Gabadadze},\ and\ \citenamefont
  {Tolley}}]{deRham:2011rn}%
  \BibitemOpen
  \bibfield  {author} {\bibinfo {author} {\bibfnamefont {Claudia}\ \bibnamefont
  {de~Rham}}, \bibinfo {author} {\bibfnamefont {Gregory}\ \bibnamefont
  {Gabadadze}}, \ and\ \bibinfo {author} {\bibfnamefont {Andrew~J.}\
  \bibnamefont {Tolley}},\ }\bibfield  {title} {\enquote {\bibinfo {title}
  {{Ghost free Massive Gravity in the St{\"u}ckelberg language}},}\ }\href
  {\doibase 10.1016/j.physletb.2012.03.081} {\bibfield  {journal} {\bibinfo
  {journal} {Phys.Lett.}\ }\textbf {\bibinfo {volume} {B711}},\ \bibinfo
  {pages} {190--195} (\bibinfo {year} {2012})},\ \Eprint
  {http://arxiv.org/abs/1107.3820} {arXiv:1107.3820 [hep-th]} \BibitemShut
  {NoStop}%
%%CITATION = ARXIV:1107.3820;%%
\bibitem [{\citenamefont {de~Rham}\ \emph
  {et~al.}(2011{\natexlab{b}})\citenamefont {de~Rham}, \citenamefont
  {Gabadadze},\ and\ \citenamefont {Tolley}}]{deRham:2011qq}%
  \BibitemOpen
  \bibfield  {author} {\bibinfo {author} {\bibfnamefont {Claudia}\ \bibnamefont
  {de~Rham}}, \bibinfo {author} {\bibfnamefont {Gregory}\ \bibnamefont
  {Gabadadze}}, \ and\ \bibinfo {author} {\bibfnamefont {Andrew~J.}\
  \bibnamefont {Tolley}},\ }\bibfield  {title} {\enquote {\bibinfo {title}
  {{Helicity Decomposition of Ghost-free Massive Gravity}},}\ }\href {\doibase
  10.1007/JHEP11(2011)093} {\bibfield  {journal} {\bibinfo  {journal} {JHEP}\
  }\textbf {\bibinfo {volume} {1111}},\ \bibinfo {pages} {093} (\bibinfo {year}
  {2011}{\natexlab{b}})},\ \Eprint {http://arxiv.org/abs/1108.4521}
  {arXiv:1108.4521 [hep-th]} \BibitemShut {NoStop}%
%%CITATION = ARXIV:1108.4521;%%
\bibitem [{\citenamefont {Hassan}\ \emph
  {et~al.}(2012{\natexlab{a}})\citenamefont {Hassan}, \citenamefont {Rosen},\
  and\ \citenamefont {Schmidt-May}}]{Hassan:2011tf}%
  \BibitemOpen
  \bibfield  {author} {\bibinfo {author} {\bibfnamefont {S.F.}\ \bibnamefont
  {Hassan}}, \bibinfo {author} {\bibfnamefont {Rachel~A.}\ \bibnamefont
  {Rosen}}, \ and\ \bibinfo {author} {\bibfnamefont {Angnis}\ \bibnamefont
  {Schmidt-May}},\ }\bibfield  {title} {\enquote {\bibinfo {title} {{Ghost-free
  Massive Gravity with a General Reference Metric}},}\ }\href {\doibase
  10.1007/JHEP02(2012)026} {\bibfield  {journal} {\bibinfo  {journal} {JHEP}\
  }\textbf {\bibinfo {volume} {1202}},\ \bibinfo {pages} {026} (\bibinfo {year}
  {2012}{\natexlab{a}})},\ \Eprint {http://arxiv.org/abs/1109.3230}
  {arXiv:1109.3230 [hep-th]} \BibitemShut {NoStop}%
%%CITATION = ARXIV:1109.3230;%%
\bibitem [{\citenamefont {Hassan}\ and\ \citenamefont
  {Rosen}(2012{\natexlab{b}})}]{Hassan:2011ea}%
  \BibitemOpen
  \bibfield  {author} {\bibinfo {author} {\bibfnamefont {S.F.}\ \bibnamefont
  {Hassan}}\ and\ \bibinfo {author} {\bibfnamefont {Rachel~A.}\ \bibnamefont
  {Rosen}},\ }\bibfield  {title} {\enquote {\bibinfo {title} {{Confirmation of
  the Secondary Constraint and Absence of Ghost in Massive Gravity and Bimetric
  Gravity}},}\ }\href {\doibase 10.1007/JHEP04(2012)123} {\bibfield  {journal}
  {\bibinfo  {journal} {JHEP}\ }\textbf {\bibinfo {volume} {1204}},\ \bibinfo
  {pages} {123} (\bibinfo {year} {2012}{\natexlab{b}})},\ \Eprint
  {http://arxiv.org/abs/1111.2070} {arXiv:1111.2070 [hep-th]} \BibitemShut
  {NoStop}%
%%CITATION = ARXIV:1111.2070;%%
\bibitem [{\citenamefont {Hassan}\ \emph
  {et~al.}(2012{\natexlab{b}})\citenamefont {Hassan}, \citenamefont
  {Schmidt-May},\ and\ \citenamefont {von Strauss}}]{Hassan:2012qv}%
  \BibitemOpen
  \bibfield  {author} {\bibinfo {author} {\bibfnamefont {S.F.}\ \bibnamefont
  {Hassan}}, \bibinfo {author} {\bibfnamefont {Angnis}\ \bibnamefont
  {Schmidt-May}}, \ and\ \bibinfo {author} {\bibfnamefont {Mikael}\
  \bibnamefont {von Strauss}},\ }\bibfield  {title} {\enquote {\bibinfo {title}
  {{Proof of Consistency of Nonlinear Massive Gravity in the St{\"u}ckelberg
  Formulation}},}\ }\href {\doibase 10.1016/j.physletb.2012.07.018} {\bibfield
  {journal} {\bibinfo  {journal} {Phys.Lett.}\ }\textbf {\bibinfo {volume}
  {B715}},\ \bibinfo {pages} {335--339} (\bibinfo {year}
  {2012}{\natexlab{b}})},\ \Eprint {http://arxiv.org/abs/1203.5283}
  {arXiv:1203.5283 [hep-th]} \BibitemShut {NoStop}%
%%CITATION = ARXIV:1203.5283;%%
\bibitem [{\citenamefont {Hinterbichler}\ and\ \citenamefont
  {Rosen}(2012)}]{Hinterbichler:2012cn}%
  \BibitemOpen
  \bibfield  {author} {\bibinfo {author} {\bibfnamefont {Kurt}\ \bibnamefont
  {Hinterbichler}}\ and\ \bibinfo {author} {\bibfnamefont {Rachel~A.}\
  \bibnamefont {Rosen}},\ }\bibfield  {title} {\enquote {\bibinfo {title}
  {{Interacting Spin-2 Fields}},}\ }\href {\doibase 10.1007/JHEP07(2012)047}
  {\bibfield  {journal} {\bibinfo  {journal} {JHEP}\ }\textbf {\bibinfo
  {volume} {1207}},\ \bibinfo {pages} {047} (\bibinfo {year} {2012})},\ \Eprint
  {http://arxiv.org/abs/1203.5783} {arXiv:1203.5783 [hep-th]} \BibitemShut
  {NoStop}%
%%CITATION = ARXIV:1203.5783;%%
\bibitem [{\citenamefont {Hassan}\ and\ \citenamefont
  {Rosen}(2012{\natexlab{c}})}]{Hassan:2011zd}%
  \BibitemOpen
  \bibfield  {author} {\bibinfo {author} {\bibfnamefont {S.F.}\ \bibnamefont
  {Hassan}}\ and\ \bibinfo {author} {\bibfnamefont {Rachel~A.}\ \bibnamefont
  {Rosen}},\ }\bibfield  {title} {\enquote {\bibinfo {title} {{Bimetric Gravity
  from Ghost-free Massive Gravity}},}\ }\href {\doibase
  10.1007/JHEP02(2012)126} {\bibfield  {journal} {\bibinfo  {journal} {JHEP}\
  }\textbf {\bibinfo {volume} {1202}},\ \bibinfo {pages} {126} (\bibinfo {year}
  {2012}{\natexlab{c}})},\ \Eprint {http://arxiv.org/abs/1109.3515}
  {arXiv:1109.3515 [hep-th]} \BibitemShut {NoStop}%
%%CITATION = ARXIV:1109.3515;%%
\bibitem [{\citenamefont {de~Rham}(2014)}]{deRham:2014zqa}%
  \BibitemOpen
  \bibfield  {author} {\bibinfo {author} {\bibfnamefont {Claudia}\ \bibnamefont
  {de~Rham}},\ }\bibfield  {title} {\enquote {\bibinfo {title} {{Massive
  Gravity}},}\ }\href {\doibase 10.12942/lrr-2014-7} {\bibfield  {journal}
  {\bibinfo  {journal} {Living Rev.Rel.}\ }\textbf {\bibinfo {volume} {17}},\
  \bibinfo {pages} {7} (\bibinfo {year} {2014})},\ \Eprint
  {http://arxiv.org/abs/1401.4173} {arXiv:1401.4173 [hep-th]} \BibitemShut
  {NoStop}%
%%CITATION = ARXIV:1401.4173;%%
\bibitem [{\citenamefont {Hinterbichler}(2012)}]{Hinterbichler:2011tt}%
  \BibitemOpen
  \bibfield  {author} {\bibinfo {author} {\bibfnamefont {Kurt}\ \bibnamefont
  {Hinterbichler}},\ }\bibfield  {title} {\enquote {\bibinfo {title}
  {{Theoretical Aspects of Massive Gravity}},}\ }\href {\doibase
  10.1103/RevModPhys.84.671} {\bibfield  {journal} {\bibinfo  {journal}
  {Rev.Mod.Phys.}\ }\textbf {\bibinfo {volume} {84}},\ \bibinfo {pages}
  {671--710} (\bibinfo {year} {2012})},\ \Eprint
  {http://arxiv.org/abs/1105.3735} {arXiv:1105.3735 [hep-th]} \BibitemShut
  {NoStop}%
%%CITATION = ARXIV:1105.3735;%%
\bibitem [{\citenamefont {Schmidt-May}\ and\ \citenamefont {von
  Strauss}(2016)}]{Schmidt-May:2015vnx}%
  \BibitemOpen
  \bibfield  {author} {\bibinfo {author} {\bibfnamefont {Angnis}\ \bibnamefont
  {Schmidt-May}}\ and\ \bibinfo {author} {\bibfnamefont {Mikael}\ \bibnamefont
  {von Strauss}},\ }\bibfield  {title} {\enquote {\bibinfo {title} {{Recent
  developments in bimetric theory}},}\ }\href {\doibase
  10.1088/1751-8113/49/18/183001} {\bibfield  {journal} {\bibinfo  {journal}
  {J. Phys.}\ }\textbf {\bibinfo {volume} {A49}},\ \bibinfo {pages} {183001}
  (\bibinfo {year} {2016})},\ \Eprint {http://arxiv.org/abs/1512.00021}
  {arXiv:1512.00021 [hep-th]} \BibitemShut {NoStop}%
%%CITATION = ARXIV:1512.00021;%%
\bibitem [{\citenamefont {Solomon}(2015)}]{Solomon:2015hja}%
  \BibitemOpen
  \bibfield  {author} {\bibinfo {author} {\bibfnamefont {Adam~R.}\ \bibnamefont
  {Solomon}},\ }\emph {\bibinfo {title} {{Cosmology Beyond Einstein}}},\ \href
  {http://inspirehep.net/record/1390097/files/arXiv:1508.06859.pdf} {Ph.D.
  thesis},\ \bibinfo  {school} {Cambridge U.} (\bibinfo {year} {2015}),\
  \Eprint {http://arxiv.org/abs/1508.06859} {arXiv:1508.06859 [gr-qc]}
  \BibitemShut {NoStop}%
%%CITATION = ARXIV:1508.06859;%%
\bibitem [{\citenamefont {Nomura}\ and\ \citenamefont
  {Soda}(2012)}]{Nomura:2012xr}%
  \BibitemOpen
  \bibfield  {author} {\bibinfo {author} {\bibfnamefont {Kouichi}\ \bibnamefont
  {Nomura}}\ and\ \bibinfo {author} {\bibfnamefont {Jiro}\ \bibnamefont
  {Soda}},\ }\bibfield  {title} {\enquote {\bibinfo {title} {{When is
  Multimetric Gravity Ghost-free?}}}\ }\href {\doibase
  10.1103/PhysRevD.86.084052} {\bibfield  {journal} {\bibinfo  {journal}
  {Phys.Rev.}\ }\textbf {\bibinfo {volume} {D86}},\ \bibinfo {pages} {084052}
  (\bibinfo {year} {2012})},\ \Eprint {http://arxiv.org/abs/1207.3637}
  {arXiv:1207.3637 [hep-th]} \BibitemShut {NoStop}%
%%CITATION = ARXIV:1207.3637;%%
\bibitem [{\citenamefont {Scargill}\ \emph {et~al.}(2014)\citenamefont
  {Scargill}, \citenamefont {Noller},\ and\ \citenamefont
  {Ferreira}}]{Scargill:2014wya}%
  \BibitemOpen
  \bibfield  {author} {\bibinfo {author} {\bibfnamefont {James H.~C.}\
  \bibnamefont {Scargill}}, \bibinfo {author} {\bibfnamefont {Johannes}\
  \bibnamefont {Noller}}, \ and\ \bibinfo {author} {\bibfnamefont {Pedro~G.}\
  \bibnamefont {Ferreira}},\ }\bibfield  {title} {\enquote {\bibinfo {title}
  {{Cycles of interactions in multi-gravity theories}},}\ }\href {\doibase
  10.1007/JHEP12(2014)160} {\bibfield  {journal} {\bibinfo  {journal} {JHEP}\
  }\textbf {\bibinfo {volume} {12}},\ \bibinfo {pages} {160} (\bibinfo {year}
  {2014})},\ \Eprint {http://arxiv.org/abs/1410.7774} {arXiv:1410.7774
  [hep-th]} \BibitemShut {NoStop}%
%%CITATION = ARXIV:1410.7774;%%
\bibitem [{\citenamefont {de~Rham}\ and\ \citenamefont
  {Tolley}(2015)}]{deRham:2015cha}%
  \BibitemOpen
  \bibfield  {author} {\bibinfo {author} {\bibfnamefont {Claudia}\ \bibnamefont
  {de~Rham}}\ and\ \bibinfo {author} {\bibfnamefont {Andrew~J.}\ \bibnamefont
  {Tolley}},\ }\bibfield  {title} {\enquote {\bibinfo {title} {{Vielbein to the
  rescue? Breaking the symmetric vielbein condition in massive gravity and
  multigravity}},}\ }\href {\doibase 10.1103/PhysRevD.92.024024} {\bibfield
  {journal} {\bibinfo  {journal} {Phys. Rev.}\ }\textbf {\bibinfo {volume}
  {D92}},\ \bibinfo {pages} {024024} (\bibinfo {year} {2015})},\ \Eprint
  {http://arxiv.org/abs/1505.01450} {arXiv:1505.01450 [hep-th]} \BibitemShut
  {NoStop}%
%%CITATION = ARXIV:1505.01450;%%
\bibitem [{\citenamefont {de~Rham}\ \emph
  {et~al.}(2015{\natexlab{a}})\citenamefont {de~Rham}, \citenamefont {Matas},\
  and\ \citenamefont {Tolley}}]{deRham:2015rxa}%
  \BibitemOpen
  \bibfield  {author} {\bibinfo {author} {\bibfnamefont {Claudia}\ \bibnamefont
  {de~Rham}}, \bibinfo {author} {\bibfnamefont {Andrew}\ \bibnamefont {Matas}},
  \ and\ \bibinfo {author} {\bibfnamefont {Andrew~J.}\ \bibnamefont {Tolley}},\
  }\bibfield  {title} {\enquote {\bibinfo {title} {{New Kinetic Terms for
  Massive Gravity and Multi-gravity: A No-Go in Vielbein Form}},}\ }\href
  {\doibase 10.1088/0264-9381/32/21/215027} {\bibfield  {journal} {\bibinfo
  {journal} {Class. Quant. Grav.}\ }\textbf {\bibinfo {volume} {32}},\ \bibinfo
  {pages} {215027} (\bibinfo {year} {2015}{\natexlab{a}})},\ \Eprint
  {http://arxiv.org/abs/1505.00831} {arXiv:1505.00831 [hep-th]} \BibitemShut
  {NoStop}%
%%CITATION = ARXIV:1505.00831;%%
\bibitem [{\citenamefont {Hassan}\ \emph {et~al.}(2013)\citenamefont {Hassan},
  \citenamefont {Schmidt-May},\ and\ \citenamefont {von
  Strauss}}]{Hassan:2012wr}%
  \BibitemOpen
  \bibfield  {author} {\bibinfo {author} {\bibfnamefont {S.F.}\ \bibnamefont
  {Hassan}}, \bibinfo {author} {\bibfnamefont {Angnis}\ \bibnamefont
  {Schmidt-May}}, \ and\ \bibinfo {author} {\bibfnamefont {Mikael}\
  \bibnamefont {von Strauss}},\ }\bibfield  {title} {\enquote {\bibinfo {title}
  {{On Consistent Theories of Massive Spin-2 Fields Coupled to Gravity}},}\
  }\href {\doibase 10.1007/JHEP05(2013)086} {\bibfield  {journal} {\bibinfo
  {journal} {JHEP}\ }\textbf {\bibinfo {volume} {1305}},\ \bibinfo {pages}
  {086} (\bibinfo {year} {2013})},\ \Eprint {http://arxiv.org/abs/1208.1515}
  {arXiv:1208.1515 [hep-th]} \BibitemShut {NoStop}%
%%CITATION = ARXIV:1208.1515;%%
\bibitem [{\citenamefont {Bull}\ \emph {et~al.}(2016)\citenamefont {Bull} \emph
  {et~al.}}]{Bull:2015stt}%
  \BibitemOpen
  \bibfield  {author} {\bibinfo {author} {\bibfnamefont {Philip}\ \bibnamefont
  {Bull}} \emph {et~al.},\ }\bibfield  {title} {\enquote {\bibinfo {title}
  {{Beyond $\Lambda$CDM: Problems, solutions, and the road ahead}},}\ }\href
  {\doibase 10.1016/j.dark.2016.02.001} {\bibfield  {journal} {\bibinfo
  {journal} {Phys. Dark Univ.}\ }\textbf {\bibinfo {volume} {12}},\ \bibinfo
  {pages} {56--99} (\bibinfo {year} {2016})},\ \Eprint
  {http://arxiv.org/abs/1512.05356} {arXiv:1512.05356 [astro-ph.CO]}
  \BibitemShut {NoStop}%
%%CITATION = ARXIV:1512.05356;%%
\bibitem [{\citenamefont {Riess}\ \emph {et~al.}(1998)\citenamefont {Riess}
  \emph {et~al.}}]{Riess:1998cb}%
  \BibitemOpen
  \bibfield  {author} {\bibinfo {author} {\bibfnamefont {Adam~G.}\ \bibnamefont
  {Riess}} \emph {et~al.} (\bibinfo {collaboration} {Supernova Search Team}),\
  }\bibfield  {title} {\enquote {\bibinfo {title} {{Observational evidence from
  supernovae for an accelerating universe and a cosmological constant}},}\
  }\href {\doibase 10.1086/300499} {\bibfield  {journal} {\bibinfo  {journal}
  {Astron.J.}\ }\textbf {\bibinfo {volume} {116}},\ \bibinfo {pages}
  {1009--1038} (\bibinfo {year} {1998})},\ \Eprint
  {http://arxiv.org/abs/astro-ph/9805201} {arXiv:astro-ph/9805201 [astro-ph]}
  \BibitemShut {NoStop}%
%%CITATION = ASTRO-PH/9805201;%%
\bibitem [{\citenamefont {Perlmutter}\ \emph {et~al.}(1999)\citenamefont
  {Perlmutter} \emph {et~al.}}]{Perlmutter:1998np}%
  \BibitemOpen
  \bibfield  {author} {\bibinfo {author} {\bibfnamefont {S.}~\bibnamefont
  {Perlmutter}} \emph {et~al.} (\bibinfo {collaboration} {Supernova Cosmology
  Project}),\ }\bibfield  {title} {\enquote {\bibinfo {title} {{Measurements of
  Omega and Lambda from 42 high redshift supernovae}},}\ }\href {\doibase
  10.1086/307221} {\bibfield  {journal} {\bibinfo  {journal} {Astrophys.J.}\
  }\textbf {\bibinfo {volume} {517}},\ \bibinfo {pages} {565--586} (\bibinfo
  {year} {1999})},\ \Eprint {http://arxiv.org/abs/astro-ph/9812133}
  {arXiv:astro-ph/9812133 [astro-ph]} \BibitemShut {NoStop}%
%%CITATION = ASTRO-PH/9812133;%%
\bibitem [{\citenamefont {Clifton}\ \emph {et~al.}(2012)\citenamefont
  {Clifton}, \citenamefont {Ferreira}, \citenamefont {Padilla},\ and\
  \citenamefont {Skordis}}]{Clifton:2011jh}%
  \BibitemOpen
  \bibfield  {author} {\bibinfo {author} {\bibfnamefont {Timothy}\ \bibnamefont
  {Clifton}}, \bibinfo {author} {\bibfnamefont {Pedro~G.}\ \bibnamefont
  {Ferreira}}, \bibinfo {author} {\bibfnamefont {Antonio}\ \bibnamefont
  {Padilla}}, \ and\ \bibinfo {author} {\bibfnamefont {Constantinos}\
  \bibnamefont {Skordis}},\ }\bibfield  {title} {\enquote {\bibinfo {title}
  {{Modified Gravity and Cosmology}},}\ }\href {\doibase
  10.1016/j.physrep.2012.01.001} {\bibfield  {journal} {\bibinfo  {journal}
  {Phys. Rept.}\ }\textbf {\bibinfo {volume} {513}},\ \bibinfo {pages} {1--189}
  (\bibinfo {year} {2012})},\ \Eprint {http://arxiv.org/abs/1106.2476}
  {arXiv:1106.2476 [astro-ph.CO]} \BibitemShut {NoStop}%
%%CITATION = ARXIV:1106.2476;%%
\bibitem [{\citenamefont {D'Amico}\ \emph {et~al.}(2011)\citenamefont
  {D'Amico}, \citenamefont {de~Rham}, \citenamefont {Dubovsky}, \citenamefont
  {Gabadadze}, \citenamefont {Pirtskhalava} \emph {et~al.}}]{D'Amico:2011jj}%
  \BibitemOpen
  \bibfield  {author} {\bibinfo {author} {\bibfnamefont {G.}~\bibnamefont
  {D'Amico}}, \bibinfo {author} {\bibfnamefont {C.}~\bibnamefont {de~Rham}},
  \bibinfo {author} {\bibfnamefont {S.}~\bibnamefont {Dubovsky}}, \bibinfo
  {author} {\bibfnamefont {G.}~\bibnamefont {Gabadadze}}, \bibinfo {author}
  {\bibfnamefont {D.}~\bibnamefont {Pirtskhalava}},  \emph {et~al.},\
  }\bibfield  {title} {\enquote {\bibinfo {title} {{Massive Cosmologies}},}\
  }\href {\doibase 10.1103/PhysRevD.84.124046} {\bibfield  {journal} {\bibinfo
  {journal} {Phys.Rev.}\ }\textbf {\bibinfo {volume} {D84}},\ \bibinfo {pages}
  {124046} (\bibinfo {year} {2011})},\ \Eprint {http://arxiv.org/abs/1108.5231}
  {arXiv:1108.5231 [hep-th]} \BibitemShut {NoStop}%
%%CITATION = ARXIV:1108.5231;%%
\bibitem [{\citenamefont {G{\"u}mr{\"u}k{\c c}{\"u}o{\u g}lu}\ \emph
  {et~al.}(2011)\citenamefont {G{\"u}mr{\"u}k{\c c}{\"u}o{\u g}lu},
  \citenamefont {Lin},\ and\ \citenamefont {Mukohyama}}]{Gumrukcuoglu:2011ew}%
  \BibitemOpen
  \bibfield  {author} {\bibinfo {author} {\bibfnamefont {A.~Emir}\ \bibnamefont
  {G{\"u}mr{\"u}k{\c c}{\"u}o{\u g}lu}}, \bibinfo {author} {\bibfnamefont
  {Chunshan}\ \bibnamefont {Lin}}, \ and\ \bibinfo {author} {\bibfnamefont
  {Shinji}\ \bibnamefont {Mukohyama}},\ }\bibfield  {title} {\enquote {\bibinfo
  {title} {{Open FRW universes and self-acceleration from nonlinear massive
  gravity}},}\ }\href {\doibase 10.1088/1475-7516/2011/11/030} {\bibfield
  {journal} {\bibinfo  {journal} {JCAP}\ }\textbf {\bibinfo {volume} {1111}},\
  \bibinfo {pages} {030} (\bibinfo {year} {2011})},\ \Eprint
  {http://arxiv.org/abs/1109.3845} {arXiv:1109.3845 [hep-th]} \BibitemShut
  {NoStop}%
%%CITATION = ARXIV:1109.3845;%%
\bibitem [{\citenamefont {G{\"u}mr{\"u}k{\c c}{\"u}o{\u g}lu}\ \emph
  {et~al.}(2012)\citenamefont {G{\"u}mr{\"u}k{\c c}{\"u}o{\u g}lu},
  \citenamefont {Lin},\ and\ \citenamefont {Mukohyama}}]{Gumrukcuoglu:2011zh}%
  \BibitemOpen
  \bibfield  {author} {\bibinfo {author} {\bibfnamefont {A.~Emir}\ \bibnamefont
  {G{\"u}mr{\"u}k{\c c}{\"u}o{\u g}lu}}, \bibinfo {author} {\bibfnamefont
  {Chunshan}\ \bibnamefont {Lin}}, \ and\ \bibinfo {author} {\bibfnamefont
  {Shinji}\ \bibnamefont {Mukohyama}},\ }\bibfield  {title} {\enquote {\bibinfo
  {title} {{Cosmological perturbations of self-accelerating universe in
  nonlinear massive gravity}},}\ }\href {\doibase
  10.1088/1475-7516/2012/03/006} {\bibfield  {journal} {\bibinfo  {journal}
  {JCAP}\ }\textbf {\bibinfo {volume} {1203}},\ \bibinfo {pages} {006}
  (\bibinfo {year} {2012})},\ \Eprint {http://arxiv.org/abs/1111.4107}
  {arXiv:1111.4107 [hep-th]} \BibitemShut {NoStop}%
%%CITATION = ARXIV:1111.4107;%%
\bibitem [{\citenamefont {Vakili}\ and\ \citenamefont
  {Khosravi}(2012)}]{Vakili:2012tm}%
  \BibitemOpen
  \bibfield  {author} {\bibinfo {author} {\bibfnamefont {Babak}\ \bibnamefont
  {Vakili}}\ and\ \bibinfo {author} {\bibfnamefont {Nima}\ \bibnamefont
  {Khosravi}},\ }\bibfield  {title} {\enquote {\bibinfo {title} {{Classical and
  quantum massive cosmology for the open FRW universe}},}\ }\href {\doibase
  10.1103/PhysRevD.85.083529} {\bibfield  {journal} {\bibinfo  {journal}
  {Phys.Rev.}\ }\textbf {\bibinfo {volume} {D85}},\ \bibinfo {pages} {083529}
  (\bibinfo {year} {2012})},\ \Eprint {http://arxiv.org/abs/1204.1456}
  {arXiv:1204.1456 [gr-qc]} \BibitemShut {NoStop}%
%%CITATION = ARXIV:1204.1456;%%
\bibitem [{\citenamefont {De~Felice}\ \emph {et~al.}(2012)\citenamefont
  {De~Felice}, \citenamefont {G{\"u}mr{\"u}k{\c c}{\"u}o{\u g}lu},\ and\
  \citenamefont {Mukohyama}}]{DeFelice:2012mx}%
  \BibitemOpen
  \bibfield  {author} {\bibinfo {author} {\bibfnamefont {Antonio}\ \bibnamefont
  {De~Felice}}, \bibinfo {author} {\bibfnamefont {A.~Emir}\ \bibnamefont
  {G{\"u}mr{\"u}k{\c c}{\"u}o{\u g}lu}}, \ and\ \bibinfo {author}
  {\bibfnamefont {Shinji}\ \bibnamefont {Mukohyama}},\ }\bibfield  {title}
  {\enquote {\bibinfo {title} {{Massive gravity: nonlinear instability of the
  homogeneous and isotropic universe}},}\ }\href {\doibase
  10.1103/PhysRevLett.109.171101} {\bibfield  {journal} {\bibinfo  {journal}
  {Phys.Rev.Lett.}\ }\textbf {\bibinfo {volume} {109}},\ \bibinfo {pages}
  {171101} (\bibinfo {year} {2012})},\ \Eprint {http://arxiv.org/abs/1206.2080}
  {arXiv:1206.2080 [hep-th]} \BibitemShut {NoStop}%
%%CITATION = ARXIV:1206.2080;%%
\bibitem [{\citenamefont {Fasiello}\ and\ \citenamefont
  {Tolley}(2012)}]{Fasiello:2012rw}%
  \BibitemOpen
  \bibfield  {author} {\bibinfo {author} {\bibfnamefont {Matteo}\ \bibnamefont
  {Fasiello}}\ and\ \bibinfo {author} {\bibfnamefont {Andrew~J.}\ \bibnamefont
  {Tolley}},\ }\bibfield  {title} {\enquote {\bibinfo {title} {{Cosmological
  perturbations in Massive Gravity and the Higuchi bound}},}\ }\href {\doibase
  10.1088/1475-7516/2012/11/035} {\bibfield  {journal} {\bibinfo  {journal}
  {JCAP}\ }\textbf {\bibinfo {volume} {1211}},\ \bibinfo {pages} {035}
  (\bibinfo {year} {2012})},\ \Eprint {http://arxiv.org/abs/1206.3852}
  {arXiv:1206.3852 [hep-th]} \BibitemShut {NoStop}%
%%CITATION = ARXIV:1206.3852;%%
\bibitem [{\citenamefont {De~Felice}\ \emph {et~al.}(2013)\citenamefont
  {De~Felice}, \citenamefont {G{\"u}mr{\"u}k{\c c}{\"u}o{\u g}lu},
  \citenamefont {Lin},\ and\ \citenamefont {Mukohyama}}]{DeFelice:2013awa}%
  \BibitemOpen
  \bibfield  {author} {\bibinfo {author} {\bibfnamefont {Antonio}\ \bibnamefont
  {De~Felice}}, \bibinfo {author} {\bibfnamefont {A.~Emir}\ \bibnamefont
  {G{\"u}mr{\"u}k{\c c}{\"u}o{\u g}lu}}, \bibinfo {author} {\bibfnamefont
  {Chunshan}\ \bibnamefont {Lin}}, \ and\ \bibinfo {author} {\bibfnamefont
  {Shinji}\ \bibnamefont {Mukohyama}},\ }\bibfield  {title} {\enquote {\bibinfo
  {title} {{Nonlinear stability of cosmological solutions in massive
  gravity}},}\ }\href {\doibase 10.1088/1475-7516/2013/05/035} {\bibfield
  {journal} {\bibinfo  {journal} {JCAP}\ }\textbf {\bibinfo {volume} {1305}},\
  \bibinfo {pages} {035} (\bibinfo {year} {2013})},\ \Eprint
  {http://arxiv.org/abs/1303.4154} {arXiv:1303.4154 [hep-th]} \BibitemShut
  {NoStop}%
%%CITATION = ARXIV:1303.4154;%%
\bibitem [{\citenamefont {Volkov}(2012)}]{Volkov:2011an}%
  \BibitemOpen
  \bibfield  {author} {\bibinfo {author} {\bibfnamefont {Mikhail~S.}\
  \bibnamefont {Volkov}},\ }\bibfield  {title} {\enquote {\bibinfo {title}
  {{Cosmological solutions with massive gravitons in the bigravity theory}},}\
  }\href {\doibase 10.1007/JHEP01(2012)035} {\bibfield  {journal} {\bibinfo
  {journal} {JHEP}\ }\textbf {\bibinfo {volume} {1201}},\ \bibinfo {pages}
  {035} (\bibinfo {year} {2012})},\ \Eprint {http://arxiv.org/abs/1110.6153}
  {arXiv:1110.6153 [hep-th]} \BibitemShut {NoStop}%
%%CITATION = ARXIV:1110.6153;%%
\bibitem [{\citenamefont {Comelli}\ \emph
  {et~al.}(2012{\natexlab{a}})\citenamefont {Comelli}, \citenamefont
  {Crisostomi}, \citenamefont {Nesti},\ and\ \citenamefont
  {Pilo}}]{Comelli:2011zm}%
  \BibitemOpen
  \bibfield  {author} {\bibinfo {author} {\bibfnamefont {D.}~\bibnamefont
  {Comelli}}, \bibinfo {author} {\bibfnamefont {M.}~\bibnamefont {Crisostomi}},
  \bibinfo {author} {\bibfnamefont {F.}~\bibnamefont {Nesti}}, \ and\ \bibinfo
  {author} {\bibfnamefont {L.}~\bibnamefont {Pilo}},\ }\bibfield  {title}
  {\enquote {\bibinfo {title} {{FRW Cosmology in Ghost Free Massive
  Gravity}},}\ }\href {\doibase 10.1007/JHEP06(2012)020,
  10.1007/JHEP03(2012)067} {\bibfield  {journal} {\bibinfo  {journal} {JHEP}\
  }\textbf {\bibinfo {volume} {1203}},\ \bibinfo {pages} {067} (\bibinfo {year}
  {2012}{\natexlab{a}})},\ \Eprint {http://arxiv.org/abs/1111.1983}
  {arXiv:1111.1983 [hep-th]} \BibitemShut {NoStop}%
%%CITATION = ARXIV:1111.1983;%%
\bibitem [{\citenamefont {von Strauss}\ \emph {et~al.}(2012)\citenamefont {von
  Strauss}, \citenamefont {Schmidt-May}, \citenamefont {Enander}, \citenamefont
  {M{\"o}rtsell},\ and\ \citenamefont {Hassan}}]{vonStrauss:2011mq}%
  \BibitemOpen
  \bibfield  {author} {\bibinfo {author} {\bibfnamefont {Mikael}\ \bibnamefont
  {von Strauss}}, \bibinfo {author} {\bibfnamefont {Angnis}\ \bibnamefont
  {Schmidt-May}}, \bibinfo {author} {\bibfnamefont {Jonas}\ \bibnamefont
  {Enander}}, \bibinfo {author} {\bibfnamefont {Edvard}\ \bibnamefont
  {M{\"o}rtsell}}, \ and\ \bibinfo {author} {\bibfnamefont {S.F.}\ \bibnamefont
  {Hassan}},\ }\bibfield  {title} {\enquote {\bibinfo {title} {{Cosmological
  Solutions in Bimetric Gravity and their Observational Tests}},}\ }\href
  {\doibase 10.1088/1475-7516/2012/03/042} {\bibfield  {journal} {\bibinfo
  {journal} {JCAP}\ }\textbf {\bibinfo {volume} {1203}},\ \bibinfo {pages}
  {042} (\bibinfo {year} {2012})},\ \Eprint {http://arxiv.org/abs/1111.1655}
  {arXiv:1111.1655 [gr-qc]} \BibitemShut {NoStop}%
%%CITATION = ARXIV:1111.1655;%%
\bibitem [{\citenamefont {Akrami}\ \emph
  {et~al.}(2013{\natexlab{a}})\citenamefont {Akrami}, \citenamefont
  {Koivisto},\ and\ \citenamefont {Sandstad}}]{Akrami:2012vf}%
  \BibitemOpen
  \bibfield  {author} {\bibinfo {author} {\bibfnamefont {Yashar}\ \bibnamefont
  {Akrami}}, \bibinfo {author} {\bibfnamefont {Tomi~S.}\ \bibnamefont
  {Koivisto}}, \ and\ \bibinfo {author} {\bibfnamefont {Marit}\ \bibnamefont
  {Sandstad}},\ }\bibfield  {title} {\enquote {\bibinfo {title} {{Accelerated
  expansion from ghost-free bigravity: a statistical analysis with improved
  generality}},}\ }\href {\doibase 10.1007/JHEP03(2013)099} {\bibfield
  {journal} {\bibinfo  {journal} {JHEP}\ }\textbf {\bibinfo {volume} {1303}},\
  \bibinfo {pages} {099} (\bibinfo {year} {2013}{\natexlab{a}})},\ \Eprint
  {http://arxiv.org/abs/1209.0457} {arXiv:1209.0457 [astro-ph.CO]} \BibitemShut
  {NoStop}%
%%CITATION = ARXIV:1209.0457;%%
\bibitem [{\citenamefont {Akrami}\ \emph
  {et~al.}(2013{\natexlab{b}})\citenamefont {Akrami}, \citenamefont
  {Koivisto},\ and\ \citenamefont {Sandstad}}]{Akrami:2013pna}%
  \BibitemOpen
  \bibfield  {author} {\bibinfo {author} {\bibfnamefont {Yashar}\ \bibnamefont
  {Akrami}}, \bibinfo {author} {\bibfnamefont {Tomi~S.}\ \bibnamefont
  {Koivisto}}, \ and\ \bibinfo {author} {\bibfnamefont {Marit}\ \bibnamefont
  {Sandstad}},\ }\bibfield  {title} {\enquote {\bibinfo {title} {{Cosmological
  constraints on ghost-free bigravity: background dynamics and late-time
  acceleration}},}\ }\href@noop {} {\  (\bibinfo {year}
  {2013}{\natexlab{b}})},\ \Eprint {http://arxiv.org/abs/1302.5268}
  {arXiv:1302.5268 [astro-ph.CO]} \BibitemShut {NoStop}%
%%CITATION = ARXIV:1302.5268;%%
\bibitem [{\citenamefont {K{\"o}nnig}\ \emph
  {et~al.}(2014{\natexlab{a}})\citenamefont {K{\"o}nnig}, \citenamefont
  {Patil},\ and\ \citenamefont {Amendola}}]{Konnig:2013gxa}%
  \BibitemOpen
  \bibfield  {author} {\bibinfo {author} {\bibfnamefont {Frank}\ \bibnamefont
  {K{\"o}nnig}}, \bibinfo {author} {\bibfnamefont {Aashay}\ \bibnamefont
  {Patil}}, \ and\ \bibinfo {author} {\bibfnamefont {Luca}\ \bibnamefont
  {Amendola}},\ }\bibfield  {title} {\enquote {\bibinfo {title} {{Viable
  cosmological solutions in massive bimetric gravity}},}\ }\href {\doibase
  10.1088/1475-7516/2014/03/029} {\bibfield  {journal} {\bibinfo  {journal}
  {JCAP}\ }\textbf {\bibinfo {volume} {1403}},\ \bibinfo {pages} {029}
  (\bibinfo {year} {2014}{\natexlab{a}})},\ \Eprint
  {http://arxiv.org/abs/1312.3208} {arXiv:1312.3208 [astro-ph.CO]} \BibitemShut
  {NoStop}%
%%CITATION = ARXIV:1312.3208;%%
\bibitem [{\citenamefont {Enander}\ \emph
  {et~al.}(2015{\natexlab{a}})\citenamefont {Enander}, \citenamefont {Solomon},
  \citenamefont {Akrami},\ and\ \citenamefont {Mortsell}}]{Enander:2014xga}%
  \BibitemOpen
  \bibfield  {author} {\bibinfo {author} {\bibfnamefont {Jonas}\ \bibnamefont
  {Enander}}, \bibinfo {author} {\bibfnamefont {Adam~R.}\ \bibnamefont
  {Solomon}}, \bibinfo {author} {\bibfnamefont {Yashar}\ \bibnamefont
  {Akrami}}, \ and\ \bibinfo {author} {\bibfnamefont {Edvard}\ \bibnamefont
  {Mortsell}},\ }\bibfield  {title} {\enquote {\bibinfo {title} {{Cosmic
  expansion histories in massive bigravity with symmetric matter coupling}},}\
  }\href {\doibase 10.1088/1475-7516/2015/01/006} {\bibfield  {journal}
  {\bibinfo  {journal} {JCAP}\ }\textbf {\bibinfo {volume} {01}},\ \bibinfo
  {pages} {006} (\bibinfo {year} {2015}{\natexlab{a}})},\ \Eprint
  {http://arxiv.org/abs/1409.2860} {arXiv:1409.2860 [astro-ph.CO]} \BibitemShut
  {NoStop}%
%%CITATION = ARXIV:1409.2860;%%
\bibitem [{\citenamefont {Comelli}\ \emph
  {et~al.}(2012{\natexlab{b}})\citenamefont {Comelli}, \citenamefont
  {Crisostomi},\ and\ \citenamefont {Pilo}}]{Comelli:2012db}%
  \BibitemOpen
  \bibfield  {author} {\bibinfo {author} {\bibfnamefont {D.}~\bibnamefont
  {Comelli}}, \bibinfo {author} {\bibfnamefont {M.}~\bibnamefont {Crisostomi}},
  \ and\ \bibinfo {author} {\bibfnamefont {L.}~\bibnamefont {Pilo}},\
  }\bibfield  {title} {\enquote {\bibinfo {title} {{Perturbations in Massive
  Gravity Cosmology}},}\ }\href {\doibase 10.1007/JHEP06(2012)085} {\bibfield
  {journal} {\bibinfo  {journal} {JHEP}\ }\textbf {\bibinfo {volume} {1206}},\
  \bibinfo {pages} {085} (\bibinfo {year} {2012}{\natexlab{b}})},\ \Eprint
  {http://arxiv.org/abs/1202.1986} {arXiv:1202.1986 [hep-th]} \BibitemShut
  {NoStop}%
%%CITATION = ARXIV:1202.1986;%%
\bibitem [{\citenamefont {Khosravi}\ \emph
  {et~al.}(2012{\natexlab{a}})\citenamefont {Khosravi}, \citenamefont
  {Sepangi},\ and\ \citenamefont {Shahidi}}]{Khosravi:2012rk}%
  \BibitemOpen
  \bibfield  {author} {\bibinfo {author} {\bibfnamefont {Nima}\ \bibnamefont
  {Khosravi}}, \bibinfo {author} {\bibfnamefont {Hamid~Reza}\ \bibnamefont
  {Sepangi}}, \ and\ \bibinfo {author} {\bibfnamefont {Shahab}\ \bibnamefont
  {Shahidi}},\ }\bibfield  {title} {\enquote {\bibinfo {title} {{Massive
  cosmological scalar perturbations}},}\ }\href {\doibase
  10.1103/PhysRevD.86.043517} {\bibfield  {journal} {\bibinfo  {journal}
  {Phys.Rev.}\ }\textbf {\bibinfo {volume} {D86}},\ \bibinfo {pages} {043517}
  (\bibinfo {year} {2012}{\natexlab{a}})},\ \Eprint
  {http://arxiv.org/abs/1202.2767} {arXiv:1202.2767 [gr-qc]} \BibitemShut
  {NoStop}%
%%CITATION = ARXIV:1202.2767;%%
\bibitem [{\citenamefont {Berg}\ \emph {et~al.}(2012)\citenamefont {Berg},
  \citenamefont {Buchberger}, \citenamefont {Enander}, \citenamefont
  {M{\"o}rtsell},\ and\ \citenamefont {Sj{\"o}rs}}]{Berg:2012kn}%
  \BibitemOpen
  \bibfield  {author} {\bibinfo {author} {\bibfnamefont {Marcus}\ \bibnamefont
  {Berg}}, \bibinfo {author} {\bibfnamefont {Igor}\ \bibnamefont {Buchberger}},
  \bibinfo {author} {\bibfnamefont {Jonas}\ \bibnamefont {Enander}}, \bibinfo
  {author} {\bibfnamefont {Edvard}\ \bibnamefont {M{\"o}rtsell}}, \ and\
  \bibinfo {author} {\bibfnamefont {Stefan}\ \bibnamefont {Sj{\"o}rs}},\
  }\bibfield  {title} {\enquote {\bibinfo {title} {{Growth Histories in
  Bimetric Massive Gravity}},}\ }\href {\doibase 10.1088/1475-7516/2012/12/021}
  {\bibfield  {journal} {\bibinfo  {journal} {JCAP}\ }\textbf {\bibinfo
  {volume} {1212}},\ \bibinfo {pages} {021} (\bibinfo {year} {2012})},\ \Eprint
  {http://arxiv.org/abs/1206.3496} {arXiv:1206.3496 [gr-qc]} \BibitemShut
  {NoStop}%
%%CITATION = ARXIV:1206.3496;%%
\bibitem [{\citenamefont {K{\"o}nnig}\ and\ \citenamefont
  {Amendola}(2014)}]{Konnig:2014dna}%
  \BibitemOpen
  \bibfield  {author} {\bibinfo {author} {\bibfnamefont {Frank}\ \bibnamefont
  {K{\"o}nnig}}\ and\ \bibinfo {author} {\bibfnamefont {Luca}\ \bibnamefont
  {Amendola}},\ }\bibfield  {title} {\enquote {\bibinfo {title} {{Instability
  in a minimal bimetric gravity model}},}\ }\href {\doibase
  10.1103/PhysRevD.90.044030} {\bibfield  {journal} {\bibinfo  {journal}
  {Phys.Rev.}\ }\textbf {\bibinfo {volume} {D90}},\ \bibinfo {pages} {044030}
  (\bibinfo {year} {2014})},\ \Eprint {http://arxiv.org/abs/1402.1988}
  {arXiv:1402.1988 [astro-ph.CO]} \BibitemShut {NoStop}%
%%CITATION = ARXIV:1402.1988;%%
\bibitem [{\citenamefont {Solomon}\ \emph {et~al.}(2014)\citenamefont
  {Solomon}, \citenamefont {Akrami},\ and\ \citenamefont
  {Koivisto}}]{Solomon:2014dua}%
  \BibitemOpen
  \bibfield  {author} {\bibinfo {author} {\bibfnamefont {Adam~R.}\ \bibnamefont
  {Solomon}}, \bibinfo {author} {\bibfnamefont {Yashar}\ \bibnamefont
  {Akrami}}, \ and\ \bibinfo {author} {\bibfnamefont {Tomi~S.}\ \bibnamefont
  {Koivisto}},\ }\bibfield  {title} {\enquote {\bibinfo {title} {{Linear growth
  of structure in massive bigravity}},}\ }\href {\doibase
  10.1088/1475-7516/2014/10/066} {\bibfield  {journal} {\bibinfo  {journal}
  {JCAP}\ }\textbf {\bibinfo {volume} {1410}},\ \bibinfo {pages} {066}
  (\bibinfo {year} {2014})},\ \Eprint {http://arxiv.org/abs/1404.4061}
  {arXiv:1404.4061 [astro-ph.CO]} \BibitemShut {NoStop}%
%%CITATION = ARXIV:1404.4061;%%
\bibitem [{\citenamefont {K{\"o}nnig}\ \emph
  {et~al.}(2014{\natexlab{b}})\citenamefont {K{\"o}nnig}, \citenamefont
  {Akrami}, \citenamefont {Amendola}, \citenamefont {Motta},\ and\
  \citenamefont {Solomon}}]{Konnig:2014xva}%
  \BibitemOpen
  \bibfield  {author} {\bibinfo {author} {\bibfnamefont {Frank}\ \bibnamefont
  {K{\"o}nnig}}, \bibinfo {author} {\bibfnamefont {Yashar}\ \bibnamefont
  {Akrami}}, \bibinfo {author} {\bibfnamefont {Luca}\ \bibnamefont {Amendola}},
  \bibinfo {author} {\bibfnamefont {Mariele}\ \bibnamefont {Motta}}, \ and\
  \bibinfo {author} {\bibfnamefont {Adam~R.}\ \bibnamefont {Solomon}},\
  }\bibfield  {title} {\enquote {\bibinfo {title} {{Stable and unstable
  cosmological models in bimetric massive gravity}},}\ }\href {\doibase
  10.1103/PhysRevD.90.124014} {\bibfield  {journal} {\bibinfo  {journal}
  {Phys.Rev.}\ }\textbf {\bibinfo {volume} {D90}},\ \bibinfo {pages} {124014}
  (\bibinfo {year} {2014}{\natexlab{b}})},\ \Eprint
  {http://arxiv.org/abs/1407.4331} {arXiv:1407.4331 [astro-ph.CO]} \BibitemShut
  {NoStop}%
%%CITATION = ARXIV:1407.4331;%%
\bibitem [{\citenamefont {Lagos}\ and\ \citenamefont
  {Ferreira}(2014)}]{Lagos:2014lca}%
  \BibitemOpen
  \bibfield  {author} {\bibinfo {author} {\bibfnamefont {Macarena}\
  \bibnamefont {Lagos}}\ and\ \bibinfo {author} {\bibfnamefont {Pedro~G.}\
  \bibnamefont {Ferreira}},\ }\bibfield  {title} {\enquote {\bibinfo {title}
  {{Cosmological perturbations in massive bigravity}},}\ }\href {\doibase
  10.1088/1475-7516/2014/12/026} {\bibfield  {journal} {\bibinfo  {journal}
  {JCAP}\ }\textbf {\bibinfo {volume} {1412}},\ \bibinfo {pages} {026}
  (\bibinfo {year} {2014})},\ \Eprint {http://arxiv.org/abs/1410.0207}
  {arXiv:1410.0207 [gr-qc]} \BibitemShut {NoStop}%
%%CITATION = ARXIV:1410.0207;%%
\bibitem [{\citenamefont {Cusin}\ \emph {et~al.}(2015)\citenamefont {Cusin},
  \citenamefont {Durrer}, \citenamefont {Guarato},\ and\ \citenamefont
  {Motta}}]{Cusin:2014psa}%
  \BibitemOpen
  \bibfield  {author} {\bibinfo {author} {\bibfnamefont {Giulia}\ \bibnamefont
  {Cusin}}, \bibinfo {author} {\bibfnamefont {Ruth}\ \bibnamefont {Durrer}},
  \bibinfo {author} {\bibfnamefont {Pietro}\ \bibnamefont {Guarato}}, \ and\
  \bibinfo {author} {\bibfnamefont {Mariele}\ \bibnamefont {Motta}},\
  }\bibfield  {title} {\enquote {\bibinfo {title} {{Gravitational waves in
  bigravity cosmology}},}\ }\href {\doibase 10.1088/1475-7516/2015/05/030}
  {\bibfield  {journal} {\bibinfo  {journal} {JCAP}\ }\textbf {\bibinfo
  {volume} {1505}},\ \bibinfo {pages} {030} (\bibinfo {year} {2015})},\ \Eprint
  {http://arxiv.org/abs/1412.5979} {arXiv:1412.5979 [astro-ph.CO]} \BibitemShut
  {NoStop}%
%%CITATION = ARXIV:1412.5979;%%
\bibitem [{\citenamefont {Yamashita}\ and\ \citenamefont
  {Tanaka}(2014)}]{Yamashita:2014cra}%
  \BibitemOpen
  \bibfield  {author} {\bibinfo {author} {\bibfnamefont {Yasuho}\ \bibnamefont
  {Yamashita}}\ and\ \bibinfo {author} {\bibfnamefont {Takahiro}\ \bibnamefont
  {Tanaka}},\ }\bibfield  {title} {\enquote {\bibinfo {title} {{Mapping the
  ghost free bigravity into braneworld setup}},}\ }\href {\doibase
  10.1088/1475-7516/2014/06/004} {\bibfield  {journal} {\bibinfo  {journal}
  {JCAP}\ }\textbf {\bibinfo {volume} {1406}},\ \bibinfo {pages} {004}
  (\bibinfo {year} {2014})},\ \Eprint {http://arxiv.org/abs/1401.4336}
  {arXiv:1401.4336 [hep-th]} \BibitemShut {NoStop}%
%%CITATION = ARXIV:1401.4336;%%
\bibitem [{\citenamefont {De~Felice}\ \emph {et~al.}(2014)\citenamefont
  {De~Felice}, \citenamefont {G{\"u}mr{\"u}k{\c c}{\"u}o{\u g}lu},
  \citenamefont {Mukohyama}, \citenamefont {Tanahashi},\ and\ \citenamefont
  {Tanaka}}]{DeFelice:2014nja}%
  \BibitemOpen
  \bibfield  {author} {\bibinfo {author} {\bibfnamefont {Antonio}\ \bibnamefont
  {De~Felice}}, \bibinfo {author} {\bibfnamefont {A.~Emir}\ \bibnamefont
  {G{\"u}mr{\"u}k{\c c}{\"u}o{\u g}lu}}, \bibinfo {author} {\bibfnamefont
  {Shinji}\ \bibnamefont {Mukohyama}}, \bibinfo {author} {\bibfnamefont
  {Norihiro}\ \bibnamefont {Tanahashi}}, \ and\ \bibinfo {author}
  {\bibfnamefont {Takahiro}\ \bibnamefont {Tanaka}},\ }\bibfield  {title}
  {\enquote {\bibinfo {title} {{Viable cosmology in bimetric theory}},}\ }\href
  {\doibase 10.1088/1475-7516/2014/06/037} {\bibfield  {journal} {\bibinfo
  {journal} {JCAP}\ }\textbf {\bibinfo {volume} {1406}},\ \bibinfo {pages}
  {037} (\bibinfo {year} {2014})},\ \Eprint {http://arxiv.org/abs/1404.0008}
  {arXiv:1404.0008 [hep-th]} \BibitemShut {NoStop}%
%%CITATION = ARXIV:1404.0008;%%
\bibitem [{\citenamefont {Fasiello}\ and\ \citenamefont
  {Tolley}(2013)}]{Fasiello:2013woa}%
  \BibitemOpen
  \bibfield  {author} {\bibinfo {author} {\bibfnamefont {Matteo}\ \bibnamefont
  {Fasiello}}\ and\ \bibinfo {author} {\bibfnamefont {Andrew~J.}\ \bibnamefont
  {Tolley}},\ }\bibfield  {title} {\enquote {\bibinfo {title} {{Cosmological
  Stability Bound in Massive Gravity and Bigravity}},}\ }\href {\doibase
  10.1088/1475-7516/2013/12/002} {\bibfield  {journal} {\bibinfo  {journal}
  {JCAP}\ }\textbf {\bibinfo {volume} {1312}},\ \bibinfo {pages} {002}
  (\bibinfo {year} {2013})},\ \Eprint {http://arxiv.org/abs/1308.1647}
  {arXiv:1308.1647 [hep-th]} \BibitemShut {NoStop}%
%%CITATION = ARXIV:1308.1647;%%
\bibitem [{\citenamefont {Enander}\ \emph
  {et~al.}(2015{\natexlab{b}})\citenamefont {Enander}, \citenamefont {Akrami},
  \citenamefont {M{\"o}rtsell}, \citenamefont {Renneby},\ and\ \citenamefont
  {Solomon}}]{Enander:2015vja}%
  \BibitemOpen
  \bibfield  {author} {\bibinfo {author} {\bibfnamefont {Jonas}\ \bibnamefont
  {Enander}}, \bibinfo {author} {\bibfnamefont {Yashar}\ \bibnamefont
  {Akrami}}, \bibinfo {author} {\bibfnamefont {Edvard}\ \bibnamefont
  {M{\"o}rtsell}}, \bibinfo {author} {\bibfnamefont {Malin}\ \bibnamefont
  {Renneby}}, \ and\ \bibinfo {author} {\bibfnamefont {Adam~R.}\ \bibnamefont
  {Solomon}},\ }\bibfield  {title} {\enquote {\bibinfo {title} {{Integrated
  Sachs-Wolfe effect in massive bigravity}},}\ }\href {\doibase
  10.1103/PhysRevD.91.084046} {\bibfield  {journal} {\bibinfo  {journal}
  {Phys.Rev.}\ }\textbf {\bibinfo {volume} {D91}},\ \bibinfo {pages} {084046}
  (\bibinfo {year} {2015}{\natexlab{b}})},\ \Eprint
  {http://arxiv.org/abs/1501.02140} {arXiv:1501.02140 [astro-ph.CO]}
  \BibitemShut {NoStop}%
%%CITATION = ARXIV:1501.02140;%%
\bibitem [{\citenamefont {Amendola}\ \emph {et~al.}(2015)\citenamefont
  {Amendola}, \citenamefont {K{\"o}nnig}, \citenamefont {Martinelli},
  \citenamefont {Pettorino},\ and\ \citenamefont
  {Zumalacarregui}}]{Amendola:2015tua}%
  \BibitemOpen
  \bibfield  {author} {\bibinfo {author} {\bibfnamefont {Luca}\ \bibnamefont
  {Amendola}}, \bibinfo {author} {\bibfnamefont {Frank}\ \bibnamefont
  {K{\"o}nnig}}, \bibinfo {author} {\bibfnamefont {Matteo}\ \bibnamefont
  {Martinelli}}, \bibinfo {author} {\bibfnamefont {Valeria}\ \bibnamefont
  {Pettorino}}, \ and\ \bibinfo {author} {\bibfnamefont {Miguel}\ \bibnamefont
  {Zumalacarregui}},\ }\bibfield  {title} {\enquote {\bibinfo {title} {{Surfing
  gravitational waves: can bigravity survive growing tensor modes?}}}\ }\href
  {\doibase 10.1088/1475-7516/2015/05/052} {\bibfield  {journal} {\bibinfo
  {journal} {JCAP}\ }\textbf {\bibinfo {volume} {1505}},\ \bibinfo {pages}
  {052} (\bibinfo {year} {2015})},\ \Eprint {http://arxiv.org/abs/1503.02490}
  {arXiv:1503.02490 [astro-ph.CO]} \BibitemShut {NoStop}%
%%CITATION = ARXIV:1503.02490;%%
\bibitem [{\citenamefont {Johnson}\ and\ \citenamefont
  {Terrana}(2015)}]{Johnson:2015tfa}%
  \BibitemOpen
  \bibfield  {author} {\bibinfo {author} {\bibfnamefont {Matthew}\ \bibnamefont
  {Johnson}}\ and\ \bibinfo {author} {\bibfnamefont {Alexandra}\ \bibnamefont
  {Terrana}},\ }\bibfield  {title} {\enquote {\bibinfo {title} {{Tensor Modes
  in Bigravity: Primordial to Present}},}\ }\href {\doibase
  10.1103/PhysRevD.92.044001} {\bibfield  {journal} {\bibinfo  {journal} {Phys.
  Rev.}\ }\textbf {\bibinfo {volume} {D92}},\ \bibinfo {pages} {044001}
  (\bibinfo {year} {2015})},\ \Eprint {http://arxiv.org/abs/1503.05560}
  {arXiv:1503.05560 [astro-ph.CO]} \BibitemShut {NoStop}%
%%CITATION = ARXIV:1503.05560;%%
\bibitem [{\citenamefont {K{\"o}nnig}(2015)}]{Konnig:2015lfa}%
  \BibitemOpen
  \bibfield  {author} {\bibinfo {author} {\bibfnamefont {Frank}\ \bibnamefont
  {K{\"o}nnig}},\ }\bibfield  {title} {\enquote {\bibinfo {title} {{Higuchi
  Ghosts and Gradient Instabilities in Bimetric Gravity}},}\ }\href {\doibase
  10.1103/PhysRevD.91.104019} {\bibfield  {journal} {\bibinfo  {journal}
  {Phys.Rev.}\ }\textbf {\bibinfo {volume} {D91}},\ \bibinfo {pages} {104019}
  (\bibinfo {year} {2015})},\ \Eprint {http://arxiv.org/abs/1503.07436}
  {arXiv:1503.07436 [astro-ph.CO]} \BibitemShut {NoStop}%
%%CITATION = ARXIV:1503.07436;%%
\bibitem [{\citenamefont {Akrami}\ \emph {et~al.}(2015)\citenamefont {Akrami},
  \citenamefont {Hassan}, \citenamefont {K{\" o}nnig}, \citenamefont
  {Schmidt-May},\ and\ \citenamefont {Solomon}}]{Akrami:2015qga}%
  \BibitemOpen
  \bibfield  {author} {\bibinfo {author} {\bibfnamefont {Yashar}\ \bibnamefont
  {Akrami}}, \bibinfo {author} {\bibfnamefont {S.~F.}\ \bibnamefont {Hassan}},
  \bibinfo {author} {\bibfnamefont {Frank}\ \bibnamefont {K{\" o}nnig}},
  \bibinfo {author} {\bibfnamefont {Angnis}\ \bibnamefont {Schmidt-May}}, \
  and\ \bibinfo {author} {\bibfnamefont {Adam~R.}\ \bibnamefont {Solomon}},\
  }\bibfield  {title} {\enquote {\bibinfo {title} {{Bimetric gravity is
  cosmologically viable}},}\ }\href {\doibase 10.1016/j.physletb.2015.06.062}
  {\bibfield  {journal} {\bibinfo  {journal} {Phys. Lett.}\ }\textbf {\bibinfo
  {volume} {B748}},\ \bibinfo {pages} {37--44} (\bibinfo {year} {2015})},\
  \Eprint {http://arxiv.org/abs/1503.07521} {arXiv:1503.07521 [gr-qc]}
  \BibitemShut {NoStop}%
%%CITATION = ARXIV:1503.07521;%%
\bibitem [{\citenamefont {Mortsell}\ and\ \citenamefont
  {Enander}(2015)}]{Mortsell:2015exa}%
  \BibitemOpen
  \bibfield  {author} {\bibinfo {author} {\bibfnamefont {E.}~\bibnamefont
  {Mortsell}}\ and\ \bibinfo {author} {\bibfnamefont {J.}~\bibnamefont
  {Enander}},\ }\bibfield  {title} {\enquote {\bibinfo {title} {{Scalar
  instabilities in bimetric gravity: The Vainshtein mechanism and structure
  formation}},}\ }\href {\doibase 10.1088/1475-7516/2015/10/044} {\bibfield
  {journal} {\bibinfo  {journal} {JCAP}\ }\textbf {\bibinfo {volume} {1510}},\
  \bibinfo {pages} {044} (\bibinfo {year} {2015})},\ \Eprint
  {http://arxiv.org/abs/1506.04977} {arXiv:1506.04977 [astro-ph.CO]}
  \BibitemShut {NoStop}%
%%CITATION = ARXIV:1506.04977;%%
\bibitem [{\citenamefont {Vainshtein}(1972)}]{Vainshtein:1972sx}%
  \BibitemOpen
  \bibfield  {author} {\bibinfo {author} {\bibfnamefont {A.I.}\ \bibnamefont
  {Vainshtein}},\ }\bibfield  {title} {\enquote {\bibinfo {title} {{To the
  problem of nonvanishing gravitation mass}},}\ }\href {\doibase
  10.1016/0370-2693(72)90147-5} {\bibfield  {journal} {\bibinfo  {journal}
  {Phys.Lett.}\ }\textbf {\bibinfo {volume} {B39}},\ \bibinfo {pages}
  {393--394} (\bibinfo {year} {1972})}\BibitemShut {NoStop}%
%%CITATION = PHLTA,B39,393;%%
\bibitem [{\citenamefont {Babichev}\ and\ \citenamefont
  {Deffayet}(2013)}]{Babichev:2013usa}%
  \BibitemOpen
  \bibfield  {author} {\bibinfo {author} {\bibfnamefont {Eugeny}\ \bibnamefont
  {Babichev}}\ and\ \bibinfo {author} {\bibfnamefont {C{\'{e}}dric}\
  \bibnamefont {Deffayet}},\ }\bibfield  {title} {\enquote {\bibinfo {title}
  {{An introduction to the Vainshtein mechanism}},}\ }\href {\doibase
  10.1088/0264-9381/30/18/184001} {\bibfield  {journal} {\bibinfo  {journal}
  {Class.Quant.Grav.}\ }\textbf {\bibinfo {volume} {30}},\ \bibinfo {pages}
  {184001} (\bibinfo {year} {2013})},\ \Eprint {http://arxiv.org/abs/1304.7240}
  {arXiv:1304.7240 [gr-qc]} \BibitemShut {NoStop}%
%%CITATION = ARXIV:1304.7240;%%
\bibitem [{\citenamefont {Khosravi}\ \emph
  {et~al.}(2012{\natexlab{b}})\citenamefont {Khosravi}, \citenamefont
  {Rahmanpour}, \citenamefont {Sepangi},\ and\ \citenamefont
  {Shahidi}}]{Khosravi:2011zi}%
  \BibitemOpen
  \bibfield  {author} {\bibinfo {author} {\bibfnamefont {Nima}\ \bibnamefont
  {Khosravi}}, \bibinfo {author} {\bibfnamefont {Nafiseh}\ \bibnamefont
  {Rahmanpour}}, \bibinfo {author} {\bibfnamefont {Hamid~Reza}\ \bibnamefont
  {Sepangi}}, \ and\ \bibinfo {author} {\bibfnamefont {Shahab}\ \bibnamefont
  {Shahidi}},\ }\bibfield  {title} {\enquote {\bibinfo {title} {{Multi-Metric
  Gravity via Massive Gravity}},}\ }\href {\doibase 10.1103/PhysRevD.85.024049}
  {\bibfield  {journal} {\bibinfo  {journal} {Phys.Rev.}\ }\textbf {\bibinfo
  {volume} {D85}},\ \bibinfo {pages} {024049} (\bibinfo {year}
  {2012}{\natexlab{b}})},\ \Eprint {http://arxiv.org/abs/1111.5346}
  {arXiv:1111.5346 [hep-th]} \BibitemShut {NoStop}%
%%CITATION = ARXIV:1111.5346;%%
\bibitem [{\citenamefont {Tamanini}\ \emph {et~al.}(2014)\citenamefont
  {Tamanini}, \citenamefont {Saridakis},\ and\ \citenamefont
  {Koivisto}}]{Tamanini:2013xia}%
  \BibitemOpen
  \bibfield  {author} {\bibinfo {author} {\bibfnamefont {Nicola}\ \bibnamefont
  {Tamanini}}, \bibinfo {author} {\bibfnamefont {Emmanuel~N.}\ \bibnamefont
  {Saridakis}}, \ and\ \bibinfo {author} {\bibfnamefont {Tomi~S.}\ \bibnamefont
  {Koivisto}},\ }\bibfield  {title} {\enquote {\bibinfo {title} {{The Cosmology
  of Interacting Spin-2 Fields}},}\ }\href {\doibase
  10.1088/1475-7516/2014/02/015} {\bibfield  {journal} {\bibinfo  {journal}
  {JCAP}\ }\textbf {\bibinfo {volume} {1402}},\ \bibinfo {pages} {015}
  (\bibinfo {year} {2014})},\ \Eprint {http://arxiv.org/abs/1307.5984}
  {arXiv:1307.5984 [hep-th]} \BibitemShut {NoStop}%
%%CITATION = ARXIV:1307.5984;%%
\bibitem [{\citenamefont {Baldacchino}\ and\ \citenamefont
  {Schmidt-May}(2016)}]{Baldacchino:2016jsz}%
  \BibitemOpen
  \bibfield  {author} {\bibinfo {author} {\bibfnamefont {Oliver}\ \bibnamefont
  {Baldacchino}}\ and\ \bibinfo {author} {\bibfnamefont {Angnis}\ \bibnamefont
  {Schmidt-May}},\ }\bibfield  {title} {\enquote {\bibinfo {title} {{Structures
  in multiple spin-2 interactions}},}\ }\href@noop {} {\  (\bibinfo {year}
  {2016})},\ \Eprint {http://arxiv.org/abs/1604.04354} {arXiv:1604.04354
  [gr-qc]} \BibitemShut {NoStop}%
%%CITATION = ARXIV:1604.04354;%%
\bibitem [{\citenamefont {Akrami}\ \emph
  {et~al.}(2013{\natexlab{c}})\citenamefont {Akrami}, \citenamefont {Koivisto},
  \citenamefont {Mota},\ and\ \citenamefont {Sandstad}}]{Akrami:2013ffa}%
  \BibitemOpen
  \bibfield  {author} {\bibinfo {author} {\bibfnamefont {Yashar}\ \bibnamefont
  {Akrami}}, \bibinfo {author} {\bibfnamefont {Tomi~S.}\ \bibnamefont
  {Koivisto}}, \bibinfo {author} {\bibfnamefont {David~F.}\ \bibnamefont
  {Mota}}, \ and\ \bibinfo {author} {\bibfnamefont {Marit}\ \bibnamefont
  {Sandstad}},\ }\bibfield  {title} {\enquote {\bibinfo {title} {{Bimetric
  gravity doubly coupled to matter: theory and cosmological implications}},}\
  }\href {\doibase 10.1088/1475-7516/2013/10/046} {\bibfield  {journal}
  {\bibinfo  {journal} {JCAP}\ }\textbf {\bibinfo {volume} {1310}},\ \bibinfo
  {pages} {046} (\bibinfo {year} {2013}{\natexlab{c}})},\ \Eprint
  {http://arxiv.org/abs/1306.0004} {arXiv:1306.0004 [hep-th]} \BibitemShut
  {NoStop}%
%%CITATION = ARXIV:1306.0004;%%
\bibitem [{\citenamefont {Akrami}\ \emph {et~al.}(2014)\citenamefont {Akrami},
  \citenamefont {Koivisto},\ and\ \citenamefont {Solomon}}]{Akrami:2014lja}%
  \BibitemOpen
  \bibfield  {author} {\bibinfo {author} {\bibfnamefont {Yashar}\ \bibnamefont
  {Akrami}}, \bibinfo {author} {\bibfnamefont {Tomi~S.}\ \bibnamefont
  {Koivisto}}, \ and\ \bibinfo {author} {\bibfnamefont {Adam~R.}\ \bibnamefont
  {Solomon}},\ }\bibfield  {title} {\enquote {\bibinfo {title} {{The nature of
  spacetime in bigravity: two metrics or none?}}}\ }\href {\doibase
  10.1007/s10714-014-1838-4} {\bibfield  {journal} {\bibinfo  {journal}
  {Gen.Rel.Grav.}\ }\textbf {\bibinfo {volume} {47}},\ \bibinfo {pages} {1838}
  (\bibinfo {year} {2014})},\ \Eprint {http://arxiv.org/abs/1404.0006}
  {arXiv:1404.0006 [gr-qc]} \BibitemShut {NoStop}%
%%CITATION = ARXIV:1404.0006;%%
\bibitem [{\citenamefont {Yamashita}\ \emph {et~al.}(2014)\citenamefont
  {Yamashita}, \citenamefont {De~Felice},\ and\ \citenamefont
  {Tanaka}}]{Yamashita:2014fga}%
  \BibitemOpen
  \bibfield  {author} {\bibinfo {author} {\bibfnamefont {Yasuho}\ \bibnamefont
  {Yamashita}}, \bibinfo {author} {\bibfnamefont {Antonio}\ \bibnamefont
  {De~Felice}}, \ and\ \bibinfo {author} {\bibfnamefont {Takahiro}\
  \bibnamefont {Tanaka}},\ }\bibfield  {title} {\enquote {\bibinfo {title}
  {{Appearance of Boulware-Deser ghost in bigravity with doubly coupled
  matter}},}\ }\href {\doibase 10.1142/S0218271814430032} {\bibfield  {journal}
  {\bibinfo  {journal} {Int.J.Mod.Phys.}\ }\textbf {\bibinfo {volume} {D23}},\
  \bibinfo {pages} {3003} (\bibinfo {year} {2014})},\ \Eprint
  {http://arxiv.org/abs/1408.0487} {arXiv:1408.0487 [hep-th]} \BibitemShut
  {NoStop}%
%%CITATION = ARXIV:1408.0487;%%
\bibitem [{\citenamefont {de~Rham}\ \emph
  {et~al.}(2015{\natexlab{b}})\citenamefont {de~Rham}, \citenamefont
  {Heisenberg},\ and\ \citenamefont {Ribeiro}}]{deRham:2014naa}%
  \BibitemOpen
  \bibfield  {author} {\bibinfo {author} {\bibfnamefont {Claudia}\ \bibnamefont
  {de~Rham}}, \bibinfo {author} {\bibfnamefont {Lavinia}\ \bibnamefont
  {Heisenberg}}, \ and\ \bibinfo {author} {\bibfnamefont {Raquel~H.}\
  \bibnamefont {Ribeiro}},\ }\bibfield  {title} {\enquote {\bibinfo {title}
  {{On couplings to matter in massive (bi-)gravity}},}\ }\href {\doibase
  10.1088/0264-9381/32/3/035022} {\bibfield  {journal} {\bibinfo  {journal}
  {Class.Quant.Grav.}\ }\textbf {\bibinfo {volume} {32}},\ \bibinfo {pages}
  {035022} (\bibinfo {year} {2015}{\natexlab{b}})},\ \Eprint
  {http://arxiv.org/abs/1408.1678} {arXiv:1408.1678 [hep-th]} \BibitemShut
  {NoStop}%
%%CITATION = ARXIV:1408.1678;%%
\bibitem [{\citenamefont {Hassan}\ \emph {et~al.}(2014)\citenamefont {Hassan},
  \citenamefont {Kocic},\ and\ \citenamefont {Schmidt-May}}]{Hassan:2014gta}%
  \BibitemOpen
  \bibfield  {author} {\bibinfo {author} {\bibfnamefont {S.F.}\ \bibnamefont
  {Hassan}}, \bibinfo {author} {\bibfnamefont {Mikica}\ \bibnamefont {Kocic}},
  \ and\ \bibinfo {author} {\bibfnamefont {Angnis}\ \bibnamefont
  {Schmidt-May}},\ }\bibfield  {title} {\enquote {\bibinfo {title} {{Absence of
  ghost in a new bimetric-matter coupling}},}\ }\href@noop {} {\  (\bibinfo
  {year} {2014})},\ \Eprint {http://arxiv.org/abs/1409.1909} {arXiv:1409.1909
  [hep-th]} \BibitemShut {NoStop}%
%%CITATION = ARXIV:1409.1909;%%
\bibitem [{\citenamefont {Solomon}\ \emph {et~al.}(2015)\citenamefont
  {Solomon}, \citenamefont {Enander}, \citenamefont {Akrami}, \citenamefont
  {Koivisto}, \citenamefont {K{\"o}nnig} \emph {et~al.}}]{Solomon:2014iwa}%
  \BibitemOpen
  \bibfield  {author} {\bibinfo {author} {\bibfnamefont {Adam~R.}\ \bibnamefont
  {Solomon}}, \bibinfo {author} {\bibfnamefont {Jonas}\ \bibnamefont
  {Enander}}, \bibinfo {author} {\bibfnamefont {Yashar}\ \bibnamefont
  {Akrami}}, \bibinfo {author} {\bibfnamefont {Tomi~S.}\ \bibnamefont
  {Koivisto}}, \bibinfo {author} {\bibfnamefont {Frank}\ \bibnamefont
  {K{\"o}nnig}},  \emph {et~al.},\ }\bibfield  {title} {\enquote {\bibinfo
  {title} {{Cosmological viability of massive gravity with generalized matter
  coupling}},}\ }\href {\doibase 10.1088/1475-7516/2015/04/027} {\bibfield
  {journal} {\bibinfo  {journal} {JCAP}\ }\textbf {\bibinfo {volume} {1504}},\
  \bibinfo {pages} {027} (\bibinfo {year} {2015})},\ \Eprint
  {http://arxiv.org/abs/1409.8300} {arXiv:1409.8300 [astro-ph.CO]} \BibitemShut
  {NoStop}%
%%CITATION = ARXIV:1409.8300;%%
\bibitem [{\citenamefont {Schmidt-May}(2015)}]{Schmidt-May:2014xla}%
  \BibitemOpen
  \bibfield  {author} {\bibinfo {author} {\bibfnamefont {Angnis}\ \bibnamefont
  {Schmidt-May}},\ }\bibfield  {title} {\enquote {\bibinfo {title} {{Mass
  eigenstates in bimetric theory with matter coupling}},}\ }\href {\doibase
  10.1088/1475-7516/2015/01/039} {\bibfield  {journal} {\bibinfo  {journal}
  {JCAP}\ }\textbf {\bibinfo {volume} {1501}},\ \bibinfo {pages} {039}
  (\bibinfo {year} {2015})},\ \Eprint {http://arxiv.org/abs/1409.3146}
  {arXiv:1409.3146 [gr-qc]} \BibitemShut {NoStop}%
%%CITATION = ARXIV:1409.3146;%%
\bibitem [{\citenamefont {de~Rham}\ \emph {et~al.}(2014)\citenamefont
  {de~Rham}, \citenamefont {Heisenberg},\ and\ \citenamefont
  {Ribeiro}}]{deRham:2014fha}%
  \BibitemOpen
  \bibfield  {author} {\bibinfo {author} {\bibfnamefont {Claudia}\ \bibnamefont
  {de~Rham}}, \bibinfo {author} {\bibfnamefont {Lavinia}\ \bibnamefont
  {Heisenberg}}, \ and\ \bibinfo {author} {\bibfnamefont {Raquel~H.}\
  \bibnamefont {Ribeiro}},\ }\bibfield  {title} {\enquote {\bibinfo {title}
  {{Ghosts \& Matter Couplings in Massive (bi-\&multi-)Gravity}},}\ }\href
  {\doibase 10.1103/PhysRevD.90.124042} {\bibfield  {journal} {\bibinfo
  {journal} {Phys.Rev.}\ }\textbf {\bibinfo {volume} {D90}},\ \bibinfo {pages}
  {124042} (\bibinfo {year} {2014})},\ \Eprint {http://arxiv.org/abs/1409.3834}
  {arXiv:1409.3834 [hep-th]} \BibitemShut {NoStop}%
%%CITATION = ARXIV:1409.3834;%%
\bibitem [{\citenamefont {G{\"u}mr{\"u}k{\c c}{\"u}o{\u g}lu}\ \emph
  {et~al.}(2015{\natexlab{a}})\citenamefont {G{\"u}mr{\"u}k{\c c}{\"u}o{\u
  g}lu}, \citenamefont {Heisenberg},\ and\ \citenamefont
  {Mukohyama}}]{Gumrukcuoglu:2014xba}%
  \BibitemOpen
  \bibfield  {author} {\bibinfo {author} {\bibfnamefont {A.~Emir}\ \bibnamefont
  {G{\"u}mr{\"u}k{\c c}{\"u}o{\u g}lu}}, \bibinfo {author} {\bibfnamefont
  {Lavinia}\ \bibnamefont {Heisenberg}}, \ and\ \bibinfo {author}
  {\bibfnamefont {Shinji}\ \bibnamefont {Mukohyama}},\ }\bibfield  {title}
  {\enquote {\bibinfo {title} {{Cosmological perturbations in massive gravity
  with doubly coupled matter}},}\ }\href {\doibase
  10.1088/1475-7516/2015/02/022} {\bibfield  {journal} {\bibinfo  {journal}
  {JCAP}\ }\textbf {\bibinfo {volume} {1502}},\ \bibinfo {pages} {022}
  (\bibinfo {year} {2015}{\natexlab{a}})},\ \Eprint
  {http://arxiv.org/abs/1409.7260} {arXiv:1409.7260 [hep-th]} \BibitemShut
  {NoStop}%
%%CITATION = ARXIV:1409.7260;%%
\bibitem [{\citenamefont
  {Heisenberg}(2015{\natexlab{a}})}]{Heisenberg:2014rka}%
  \BibitemOpen
  \bibfield  {author} {\bibinfo {author} {\bibfnamefont {Lavinia}\ \bibnamefont
  {Heisenberg}},\ }\bibfield  {title} {\enquote {\bibinfo {title} {{Quantum
  corrections in massive bigravity and new effective composite metrics}},}\
  }\href {\doibase 10.1088/0264-9381/32/10/105011} {\bibfield  {journal}
  {\bibinfo  {journal} {Class.Quant.Grav.}\ }\textbf {\bibinfo {volume} {32}},\
  \bibinfo {pages} {105011} (\bibinfo {year} {2015}{\natexlab{a}})},\ \Eprint
  {http://arxiv.org/abs/1410.4239} {arXiv:1410.4239 [hep-th]} \BibitemShut
  {NoStop}%
%%CITATION = ARXIV:1410.4239;%%
\bibitem [{\citenamefont {G{\"u}mr{\"u}k{\c c}{\"u}o{\u g}lu}\ \emph
  {et~al.}(2015{\natexlab{b}})\citenamefont {G{\"u}mr{\"u}k{\c c}{\"u}o{\u
  g}lu}, \citenamefont {Heisenberg}, \citenamefont {Mukohyama},\ and\
  \citenamefont {Tanahashi}}]{Gumrukcuoglu:2015nua}%
  \BibitemOpen
  \bibfield  {author} {\bibinfo {author} {\bibfnamefont {A.~Emir}\ \bibnamefont
  {G{\"u}mr{\"u}k{\c c}{\"u}o{\u g}lu}}, \bibinfo {author} {\bibfnamefont
  {Lavinia}\ \bibnamefont {Heisenberg}}, \bibinfo {author} {\bibfnamefont
  {Shinji}\ \bibnamefont {Mukohyama}}, \ and\ \bibinfo {author} {\bibfnamefont
  {Norihiro}\ \bibnamefont {Tanahashi}},\ }\bibfield  {title} {\enquote
  {\bibinfo {title} {{Cosmology in bimetric theory with an effective composite
  coupling to matter}},}\ }\href {\doibase 10.1088/1475-7516/2015/04/008}
  {\bibfield  {journal} {\bibinfo  {journal} {JCAP}\ }\textbf {\bibinfo
  {volume} {1504}},\ \bibinfo {pages} {008} (\bibinfo {year}
  {2015}{\natexlab{b}})},\ \Eprint {http://arxiv.org/abs/1501.02790}
  {arXiv:1501.02790 [hep-th]} \BibitemShut {NoStop}%
%%CITATION = ARXIV:1501.02790;%%
\bibitem [{\citenamefont {Hinterbichler}\ and\ \citenamefont
  {Rosen}(2015)}]{Hinterbichler:2015yaa}%
  \BibitemOpen
  \bibfield  {author} {\bibinfo {author} {\bibfnamefont {Kurt}\ \bibnamefont
  {Hinterbichler}}\ and\ \bibinfo {author} {\bibfnamefont {Rachel~A.}\
  \bibnamefont {Rosen}},\ }\bibfield  {title} {\enquote {\bibinfo {title}
  {{Note on ghost-free matter couplings in massive gravity and
  multigravity}},}\ }\href {\doibase 10.1103/PhysRevD.92.024030} {\bibfield
  {journal} {\bibinfo  {journal} {Phys. Rev.}\ }\textbf {\bibinfo {volume}
  {D92}},\ \bibinfo {pages} {024030} (\bibinfo {year} {2015})},\ \Eprint
  {http://arxiv.org/abs/1503.06796} {arXiv:1503.06796 [hep-th]} \BibitemShut
  {NoStop}%
%%CITATION = ARXIV:1503.06796;%%
\bibitem [{\citenamefont
  {Heisenberg}(2015{\natexlab{b}})}]{Heisenberg:2015iqa}%
  \BibitemOpen
  \bibfield  {author} {\bibinfo {author} {\bibfnamefont {Lavinia}\ \bibnamefont
  {Heisenberg}},\ }\bibfield  {title} {\enquote {\bibinfo {title} {{More on
  effective composite metrics}},}\ }\href {\doibase 10.1103/PhysRevD.92.023525}
  {\bibfield  {journal} {\bibinfo  {journal} {Phys. Rev.}\ }\textbf {\bibinfo
  {volume} {D92}},\ \bibinfo {pages} {023525} (\bibinfo {year}
  {2015}{\natexlab{b}})},\ \Eprint {http://arxiv.org/abs/1505.02966}
  {arXiv:1505.02966 [hep-th]} \BibitemShut {NoStop}%
%%CITATION = ARXIV:1505.02966;%%
\bibitem [{\citenamefont
  {Heisenberg}(2015{\natexlab{c}})}]{Heisenberg:2015wja}%
  \BibitemOpen
  \bibfield  {author} {\bibinfo {author} {\bibfnamefont {Lavinia}\ \bibnamefont
  {Heisenberg}},\ }\bibfield  {title} {\enquote {\bibinfo {title} {{Non-minimal
  derivative couplings of the composite metric}},}\ }\href {\doibase
  10.1088/1475-7516/2015/11/005} {\bibfield  {journal} {\bibinfo  {journal}
  {JCAP}\ }\textbf {\bibinfo {volume} {1511}},\ \bibinfo {pages} {005}
  (\bibinfo {year} {2015}{\natexlab{c}})},\ \Eprint
  {http://arxiv.org/abs/1506.00580} {arXiv:1506.00580 [hep-th]} \BibitemShut
  {NoStop}%
%%CITATION = ARXIV:1506.00580;%%
\bibitem [{\citenamefont {Lagos}\ and\ \citenamefont
  {Noller}(2016)}]{Lagos:2015sya}%
  \BibitemOpen
  \bibfield  {author} {\bibinfo {author} {\bibfnamefont {Macarena}\
  \bibnamefont {Lagos}}\ and\ \bibinfo {author} {\bibfnamefont {Johannes}\
  \bibnamefont {Noller}},\ }\bibfield  {title} {\enquote {\bibinfo {title}
  {{New massive bigravity cosmologies with double matter coupling}},}\ }\href
  {\doibase 10.1088/1475-7516/2016/01/023} {\bibfield  {journal} {\bibinfo
  {journal} {JCAP}\ }\textbf {\bibinfo {volume} {1601}},\ \bibinfo {pages}
  {023} (\bibinfo {year} {2016})},\ \Eprint {http://arxiv.org/abs/1508.05864}
  {arXiv:1508.05864 [gr-qc]} \BibitemShut {NoStop}%
%%CITATION = ARXIV:1508.05864;%%
\bibitem [{\citenamefont {Melville}\ and\ \citenamefont
  {Noller}(2016)}]{Melville:2015dba}%
  \BibitemOpen
  \bibfield  {author} {\bibinfo {author} {\bibfnamefont {Scott}\ \bibnamefont
  {Melville}}\ and\ \bibinfo {author} {\bibfnamefont {Johannes}\ \bibnamefont
  {Noller}},\ }\bibfield  {title} {\enquote {\bibinfo {title} {{Generalised
  matter couplings in massive bigravity}},}\ }\href {\doibase
  10.1007/JHEP01(2016)094} {\bibfield  {journal} {\bibinfo  {journal} {JHEP}\
  }\textbf {\bibinfo {volume} {01}},\ \bibinfo {pages} {094} (\bibinfo {year}
  {2016})},\ \Eprint {http://arxiv.org/abs/1511.01485} {arXiv:1511.01485
  [hep-th]} \BibitemShut {NoStop}%
%%CITATION = ARXIV:1511.01485;%%
\bibitem [{\citenamefont {Nersisyan}\ \emph {et~al.}(2015)\citenamefont
  {Nersisyan}, \citenamefont {Akrami},\ and\ \citenamefont
  {Amendola}}]{Nersisyan:2015oha}%
  \BibitemOpen
  \bibfield  {author} {\bibinfo {author} {\bibfnamefont {Henrik}\ \bibnamefont
  {Nersisyan}}, \bibinfo {author} {\bibfnamefont {Yashar}\ \bibnamefont
  {Akrami}}, \ and\ \bibinfo {author} {\bibfnamefont {Luca}\ \bibnamefont
  {Amendola}},\ }\bibfield  {title} {\enquote {\bibinfo {title} {{Consistent
  metric combinations in cosmology of massive bigravity}},}\ }\href {\doibase
  10.1103/PhysRevD.92.104034} {\bibfield  {journal} {\bibinfo  {journal} {Phys.
  Rev.}\ }\textbf {\bibinfo {volume} {D92}},\ \bibinfo {pages} {104034}
  (\bibinfo {year} {2015})},\ \Eprint {http://arxiv.org/abs/1502.03988}
  {arXiv:1502.03988 [gr-qc]} \BibitemShut {NoStop}%
%%CITATION = ARXIV:1502.03988;%%
\bibitem [{\citenamefont {Cusin}\ \emph {et~al.}(2016)\citenamefont {Cusin},
  \citenamefont {Durrer}, \citenamefont {Guarato},\ and\ \citenamefont
  {Motta}}]{Cusin:2015tmf}%
  \BibitemOpen
  \bibfield  {author} {\bibinfo {author} {\bibfnamefont {Giulia}\ \bibnamefont
  {Cusin}}, \bibinfo {author} {\bibfnamefont {Ruth}\ \bibnamefont {Durrer}},
  \bibinfo {author} {\bibfnamefont {Pietro}\ \bibnamefont {Guarato}}, \ and\
  \bibinfo {author} {\bibfnamefont {Mariele}\ \bibnamefont {Motta}},\
  }\bibfield  {title} {\enquote {\bibinfo {title} {{A general mass term for
  bigravity}},}\ }\href {\doibase 10.1088/1475-7516/2016/04/051} {\bibfield
  {journal} {\bibinfo  {journal} {JCAP}\ }\textbf {\bibinfo {volume} {1604}},\
  \bibinfo {pages} {051} (\bibinfo {year} {2016})},\ \Eprint
  {http://arxiv.org/abs/1512.02131} {arXiv:1512.02131 [astro-ph.CO]}
  \BibitemShut {NoStop}%
%%CITATION = ARXIV:1512.02131;%%
\bibitem [{\citenamefont {Doran}\ and\ \citenamefont
  {Robbers}(2006)}]{Doran:2006kp}%
  \BibitemOpen
  \bibfield  {author} {\bibinfo {author} {\bibfnamefont {Michael}\ \bibnamefont
  {Doran}}\ and\ \bibinfo {author} {\bibfnamefont {Georg}\ \bibnamefont
  {Robbers}},\ }\bibfield  {title} {\enquote {\bibinfo {title} {{Early dark
  energy cosmologies}},}\ }\href {\doibase 10.1088/1475-7516/2006/06/026}
  {\bibfield  {journal} {\bibinfo  {journal} {JCAP}\ }\textbf {\bibinfo
  {volume} {0606}},\ \bibinfo {pages} {026} (\bibinfo {year} {2006})},\ \Eprint
  {http://arxiv.org/abs/astro-ph/0601544} {arXiv:astro-ph/0601544 [astro-ph]}
  \BibitemShut {NoStop}%
%%CITATION = ASTRO-PH/0601544;%%
\bibitem [{\citenamefont {Pettorino}\ \emph {et~al.}(2013)\citenamefont
  {Pettorino}, \citenamefont {Amendola},\ and\ \citenamefont
  {Wetterich}}]{Pettorino:2013ia}%
  \BibitemOpen
  \bibfield  {author} {\bibinfo {author} {\bibfnamefont {Valeria}\ \bibnamefont
  {Pettorino}}, \bibinfo {author} {\bibfnamefont {Luca}\ \bibnamefont
  {Amendola}}, \ and\ \bibinfo {author} {\bibfnamefont {Christof}\ \bibnamefont
  {Wetterich}},\ }\bibfield  {title} {\enquote {\bibinfo {title} {{How early is
  early dark energy?}}}\ }\href {\doibase 10.1103/PhysRevD.87.083009}
  {\bibfield  {journal} {\bibinfo  {journal} {Phys. Rev.}\ }\textbf {\bibinfo
  {volume} {D87}},\ \bibinfo {pages} {083009} (\bibinfo {year} {2013})},\
  \Eprint {http://arxiv.org/abs/1301.5279} {arXiv:1301.5279 [astro-ph.CO]}
  \BibitemShut {NoStop}%
%%CITATION = ARXIV:1301.5279;%%
\bibitem [{\citenamefont {Gomes}\ and\ \citenamefont
  {Amendola}(2014)}]{Gomes:2013ema}%
  \BibitemOpen
  \bibfield  {author} {\bibinfo {author} {\bibfnamefont {A.~R.}\ \bibnamefont
  {Gomes}}\ and\ \bibinfo {author} {\bibfnamefont {Luca}\ \bibnamefont
  {Amendola}},\ }\bibfield  {title} {\enquote {\bibinfo {title} {{Towards
  scaling cosmological solutions with full coupled Horndeski Lagrangian: the
  KGB model}},}\ }\href {\doibase 10.1088/1475-7516/2014/03/041} {\bibfield
  {journal} {\bibinfo  {journal} {JCAP}\ }\textbf {\bibinfo {volume} {1403}},\
  \bibinfo {pages} {041} (\bibinfo {year} {2014})},\ \Eprint
  {http://arxiv.org/abs/1306.3593} {arXiv:1306.3593 [astro-ph.CO]} \BibitemShut
  {NoStop}%
%%CITATION = ARXIV:1306.3593;%%
\bibitem [{\citenamefont {Gomes}\ and\ \citenamefont
  {Amendola}(2016)}]{Gomes:2015dhl}%
  \BibitemOpen
  \bibfield  {author} {\bibinfo {author} {\bibfnamefont {Adalto~R.}\
  \bibnamefont {Gomes}}\ and\ \bibinfo {author} {\bibfnamefont {Luca}\
  \bibnamefont {Amendola}},\ }\bibfield  {title} {\enquote {\bibinfo {title}
  {{The general form of the coupled Horndeski Lagrangian that allows
  cosmological scaling solutions}},}\ }\href {\doibase
  10.1088/1475-7516/2016/02/035} {\bibfield  {journal} {\bibinfo  {journal}
  {JCAP}\ }\textbf {\bibinfo {volume} {1602}},\ \bibinfo {pages} {035}
  (\bibinfo {year} {2016})},\ \Eprint {http://arxiv.org/abs/1511.01004}
  {arXiv:1511.01004 [gr-qc]} \BibitemShut {NoStop}%
%%CITATION = ARXIV:1511.01004;%%
\bibitem [{\citenamefont {Caldwell}(2002)}]{Caldwell:1999ew}%
  \BibitemOpen
  \bibfield  {author} {\bibinfo {author} {\bibfnamefont {R.~R.}\ \bibnamefont
  {Caldwell}},\ }\bibfield  {title} {\enquote {\bibinfo {title} {{A Phantom
  menace?}}}\ }\href {\doibase 10.1016/S0370-2693(02)02589-3} {\bibfield
  {journal} {\bibinfo  {journal} {Phys. Lett.}\ }\textbf {\bibinfo {volume}
  {B545}},\ \bibinfo {pages} {23--29} (\bibinfo {year} {2002})},\ \Eprint
  {http://arxiv.org/abs/astro-ph/9908168} {arXiv:astro-ph/9908168 [astro-ph]}
  \BibitemShut {NoStop}%
%%CITATION = ASTRO-PH/9908168;%%
\bibitem [{\citenamefont {Caldwell}\ \emph {et~al.}(2003)\citenamefont
  {Caldwell}, \citenamefont {Kamionkowski},\ and\ \citenamefont
  {Weinberg}}]{Caldwell:2003vq}%
  \BibitemOpen
  \bibfield  {author} {\bibinfo {author} {\bibfnamefont {Robert~R.}\
  \bibnamefont {Caldwell}}, \bibinfo {author} {\bibfnamefont {Marc}\
  \bibnamefont {Kamionkowski}}, \ and\ \bibinfo {author} {\bibfnamefont
  {Nevin~N.}\ \bibnamefont {Weinberg}},\ }\bibfield  {title} {\enquote
  {\bibinfo {title} {{Phantom energy and cosmic doomsday}},}\ }\href {\doibase
  10.1103/PhysRevLett.91.071301} {\bibfield  {journal} {\bibinfo  {journal}
  {Phys. Rev. Lett.}\ }\textbf {\bibinfo {volume} {91}},\ \bibinfo {pages}
  {071301} (\bibinfo {year} {2003})},\ \Eprint
  {http://arxiv.org/abs/astro-ph/0302506} {arXiv:astro-ph/0302506 [astro-ph]}
  \BibitemShut {NoStop}%
%%CITATION = ASTRO-PH/0302506;%%
\end{thebibliography}%

\end{document}